\documentclass[journal]{IEEEtran}
\usepackage{graphicx}
\usepackage{cite}
\usepackage{caption}
\usepackage{float}
\usepackage{subcaption}
\usepackage{fixltx2e}
\usepackage{gensymb}
\usepackage{stfloats}
\usepackage{amsmath}
\usepackage{authblk}
\begin{document}
\title{Deep Reformulated Laplacian Tone Mapping}

\author[2]{Jie Yang  \thanks{Corresponding author e-mail: yangjie@westlake.edu.cn}}
\author[1]{Ziyi Liu} 
\author[1]{Mengchen Lin}
\author[3]{Svetlana Yanushkevich}
\author[1]{Orly Yadid-Pecht}

\affil[1]{I2Sense lab, University of Calgary, Calgary T2N 1N4, Canada}
\affil[2]{Westlake University, Hangzhou  310024, China}
\affil[3]{Biometric Technologies Laboratory, Schulich School of Engineering, University of Calgary, Calgary T2N 1N4, Canada}

\maketitle
\begin{abstract}
Wide dynamic range (WDR) images contain more scene details and contrast when compared to common images. However, it requires tone mapping to process the pixel values in order to display properly. The details of WDR images can diminish during the tone mapping process.  In this work, we address the problem by combining a novel reformulated Laplacian pyramid and deep learning. The reformulated Laplacian pyramid always decompose a WDR image into two frequency bands where the low-frequency band is global feature-oriented, and the high frequency band is local feature-oriented. The reformulation preserves the local features in its original resolution and condenses the global features into a low-resolution image.   The generated frequency bands are reconstructed and fine-tuned to output the final tone mapped image that can display on the screen with minimum detail and contrast loss. The experimental results demonstrate that the proposed method outperforms state-of-the-art WDR image tone mapping methods. The code is made publicly available at
https://github.com/linmc86/Deep-Reformulated-Laplacian-Tone-Mapping.
\end{abstract}

\begin{IEEEkeywords}
Tone mapping, wide dynamic range image, image processing, machine learning.
\end{IEEEkeywords}

%
\IEEEpeerreviewmaketitle

\section{Introduction}
Wide dynamic range (WDR) imaging plays an important role in many imaging-related applications including photography, machine vision, medical imaging, and self-driving cars. Unlike traditional images that may suffer from under- and over-exposure, WDR images are obtained with WDR sensors with huge dynamic range \cite{yadid1997wide,spivak2009wide} or radiance recovery algorithm such as \cite{debevec1997recovering} that take multiple exposure images to compensate the under and over exposure regions.
WDR images greatly avoid the detail and contrast loss issues of conventional low dynamic range (LDR) images that often affect the human visual experience. However, unlike most conventional LDR images where the pixel values range from $0$ to $255$, the range of the pixels of WDR images are distributed over a much wider range. They may range from $0$ to $1,000,000$ or $1 \times 10^{-6}$ to $1.0$ based on the way that they are acquired. Although WDR display devices do exist in the commercial market for direct WDR display, they are still far away from representing all available luminance levels. In fact, the absolute majority of displays are, and in the foreseeable future will most likely be, of a very limited dynamic range. Therefore, to show WDR images on commonly used displays, additional tone mapping is still needed to convert the WDR images to a standard displayable level. To avoid any misunderstanding, we call the displayable image that is tone mapped from WDR as WDR-LDR to distinguish from conventional LDR images.


Previous methods of tone mapping employ various gradient reduction methods to compress the dynamic range \cite{fattal2002gradient,durand2002fast,reinhard2005dynamic,paris2015local,gu2013local}. Unfortunately, their WDR-LDR output often inevitably loses some details and contrast that are preferred by the human visual system (HVS). 
This is because tone mapping is not only a gradient reduction problem, but rather an in-depth topic involving human perception. A good WDR tone mapping algorithm could not only compress the large gradient but also enhance the local details of WDR images. In this paper, we propose to directly learn the global compression and local detail manipulation functionalities between the WDR images and the WDR-LDR images. Our method takes a WDR image as input, and tone maps it to WDR-LDR automatically, compressing the global dynamic range while enhancing local details. This work has the following key contributions. First, we present the reformulated Laplacian method to decompose the global feature and local features from the original WDR image. The reformulated Laplacian method condenses the global features into a low-resolution image which facilitate global feature extraction during convolution operations. Secondly, we present a two-stage network architecture and a full end-to-end learning approach which can directly tone map WDR images to WDR-LDR images. The entire network is a joint of three sub-networks that focuses on global compression, local manipulation, and fine tuning, respectively. The three sub-networks work cooperatively to produce the final WDR-LDR image. Code and model are available on our project page.

\section{Related Work}
In this section, we discuss works relevant to our research. The works include image-to-image transformation, conventional approaches that tone map WDR image to WDR-LDR image and reverse tone mapping that reconstructs a WDR image from an LDR image.


\textbf{Image-to-image transformation} Generally speaking, WDR tone mapping is an image-to-image transformation task. In recent years, many image-to-image transformation tasks are tackled by training deep convolutional neural networks. For example, deep neural networks for denoising, colorization, semantic segmentation, and super-resolution applications are massively proposed and show great performance improvement when compared with traditional methods \cite{zhang2017beyond,dong2016image,ronneberger2015u,zhang2016colorful}. Style transfer methods that adopt perceptual loss can produce certain artistic rendered counterpart of an input image and preserve the original image content \cite{johnson2016perceptual,gatys2016image}. Perceptual loss measures the high-level image feature representations extracted from pre-trained convolutional neural networks. These features are more sensitive to HVS than simple pixel values. 
In-network encoder-decoder architectures are also widely used in image transformation works where the original image is encoded into a low dimensional latent representation and then decoded to reconstruct the required image \cite{ronneberger2015u,noh2015learning,long2015fully,tsai2017deep}.

\textbf{LDR to WDR} The most well-known approach to generate a WDR image is merge
multiple LDR photographs that were taken with different exposures \cite{debevec1997recovering}. It is still widely used in many applications. To remove ghost artifacts caused by misalignment between images of different exposures, many effective techniques were proposed \cite{heo2010ghost,lee2014ghost,sen2012robust} including CNN-based solution \cite{wu2018deep}. Unlike WDR radiance recovery that fuses all available information of a bracketed of images, reverse tone mapping simply generates the missing information from a single LDR image. In recent years, with the growing popularity of machine learning and abundant WDR sources, traditional reverse tone mapping methods \cite{banterle2006inverse,rempel2007ldr2hdr,kovaleski2009high,wang2015pseudo} were overperformed by machine learning based approaches. Endo et al. \cite{endo2017deep} proposed a convolutional neural network architecture that is able to create a serial of bracketed images of different exposures from a single LDR image. A WDR image is then generated from these bracketed images. 
 Eilertsen et al. \cite{eilertsen2017hdr} proposed encoder-decoder architecture that is able to reconstruct WDR image from arbitrary single exposed LDR image with unknown camera response functions and post-processing.
Marnerides et al. \cite{marnerides2018expandnet} proposed ExpandNet which consists of three different branches to reconstruct the missed information of an LDR image. A generative adversarial network is also proposed to carry out reverse tone mapping which could generate images with wider dynamic range \cite{lee2018deep}. 

\textbf{WDR to WDR-LDR} 
The research of tone mapping a WDR image to WDR-LDR image has been lasting for decades. The simplest approach to tone mapping a WDR image is to use a global tone mapping operator (TMO). Global TMO applies a single global function to all pixels in the image where identical pixels will be given an identical output value within the range of the display device.Tumblin and Rushmeier \cite{tumblin1993tone} and Ward \cite{ward1994contrast} were the early researchers who developed global operators
for tone mapping. 

Recently, Khan et al. \cite{khan2018tone} proposed a global TMO that uses a sensitivity model of the human visual system. In general, global TMOs are easy to implement and mostly artifacts-free, and they have unique advantages in hardware implementations. However, the tone mapped images mostly suffer from low brightness, low contrast and loss of details
due to the global compression of the dynamic range. Different from global TMOs, local TMOs are able to tone map a pixel based on local statistics and reveal more detail and contrast. Some early local TMOs were inspired by certain kind of features of human visual system \cite{reinhard2005dynamic,van2006encoding,spitzer2003biological}. Some local TMOs solve the WDR image compression as a constrained optimization problem \cite{mantiuk2008display,ma2014high}. 
In recent years, various edge preserving filters based TMOs were developed \cite{durand2002fast,farbman2008edge,gu2013local,paris2015local}, and showed unprecedented results when compared with the aforementioned methods. A comprehensive review and classification of tone mapping algorithms can be found in \cite{eilertsen2017comparative}. Recently, a machine learning method that can effectively calculate the coefficients of a locally-affine model in bilateral space was reported \cite{gharbi2017deep}. It shows the great potential and performance that machine learning can provide for WDR tone mapping. 


\begin{figure*}[t]
\begin{center}
   \includegraphics[scale = 0.23]{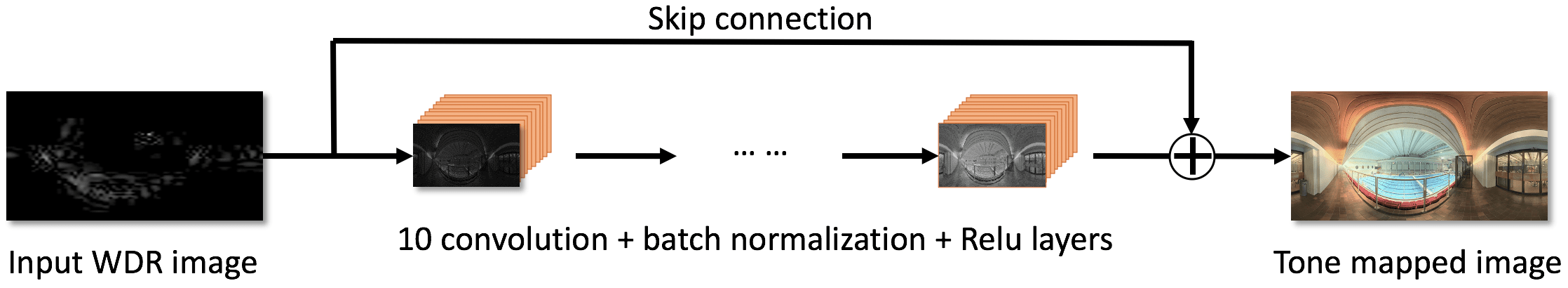}
\end{center}	
   \caption{An overview of the 10 layers CNN architecture.}
\label{fig:failed_structure}
\end{figure*}

\begin{figure*}[tb]
        \centering
        \begin{subfigure}[b]{0.49\textwidth}  
            \centering 
            \includegraphics[width=\textwidth]{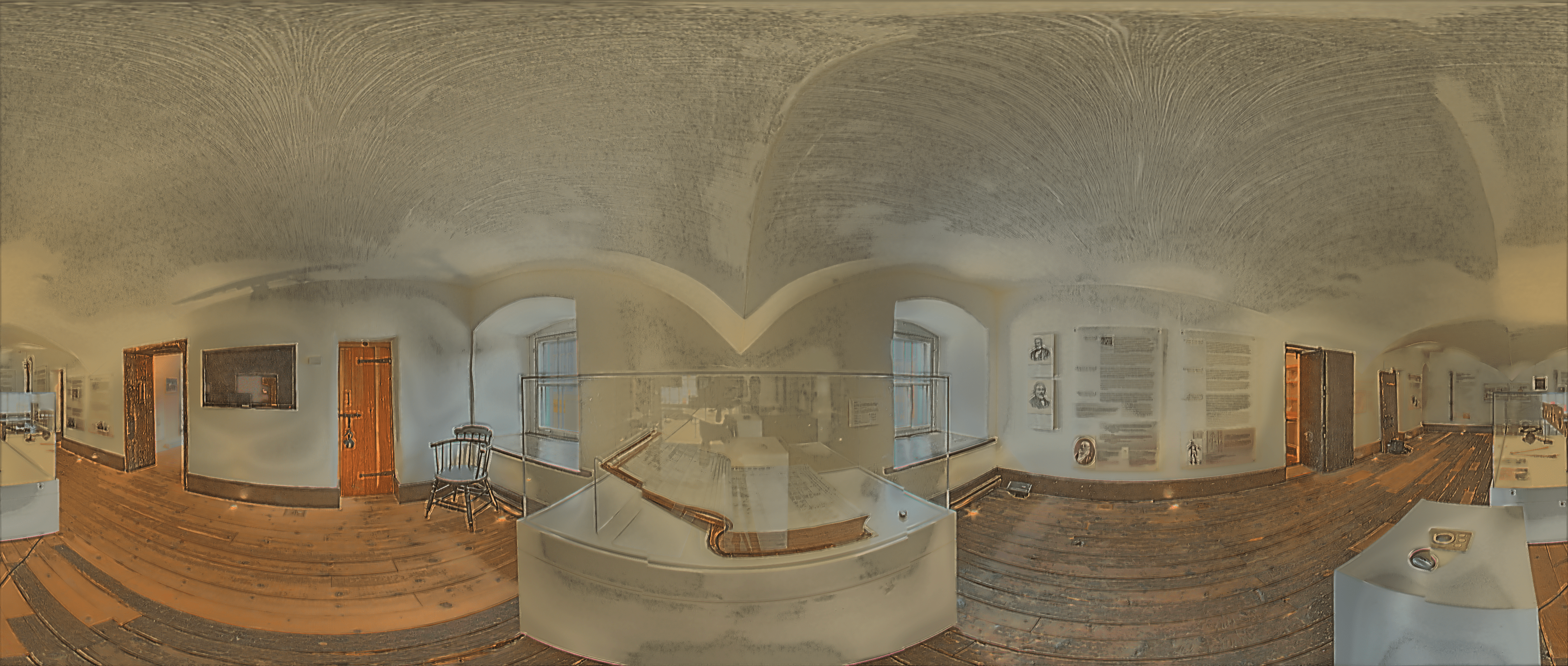}
            \caption[]%
            {{\small }}    
        \end{subfigure}
        \begin{subfigure}[b]{0.49\textwidth}   
            \centering 
            \includegraphics[width=\textwidth]{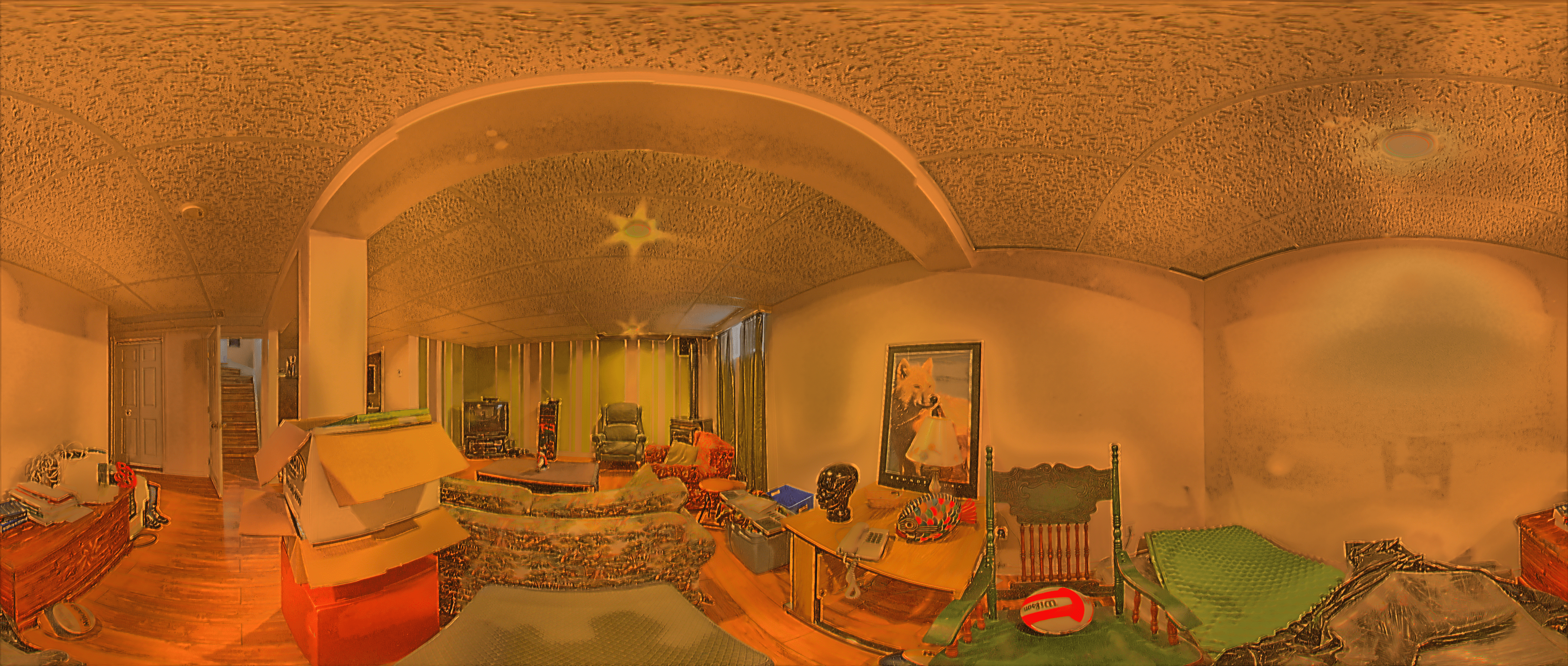}
            \caption[]%
            {{\small }}
        \end{subfigure}
        \centering
        \begin{subfigure}[b]{0.49\textwidth}  
            \centering 
            \includegraphics[width=\textwidth]{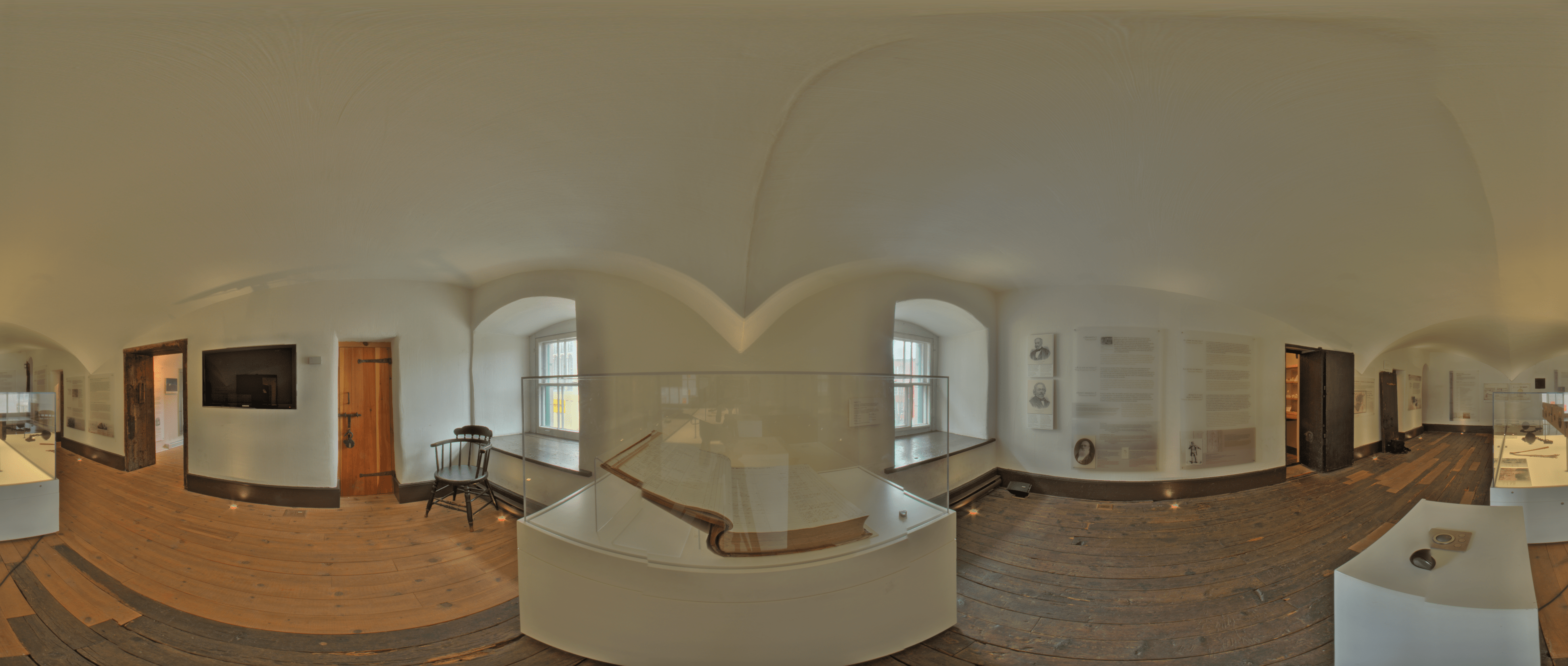}
            \caption[]%
            {{\small }}
        \end{subfigure}
        \begin{subfigure}[b]{0.49\textwidth}   
            \centering 
            \includegraphics[width=\textwidth]{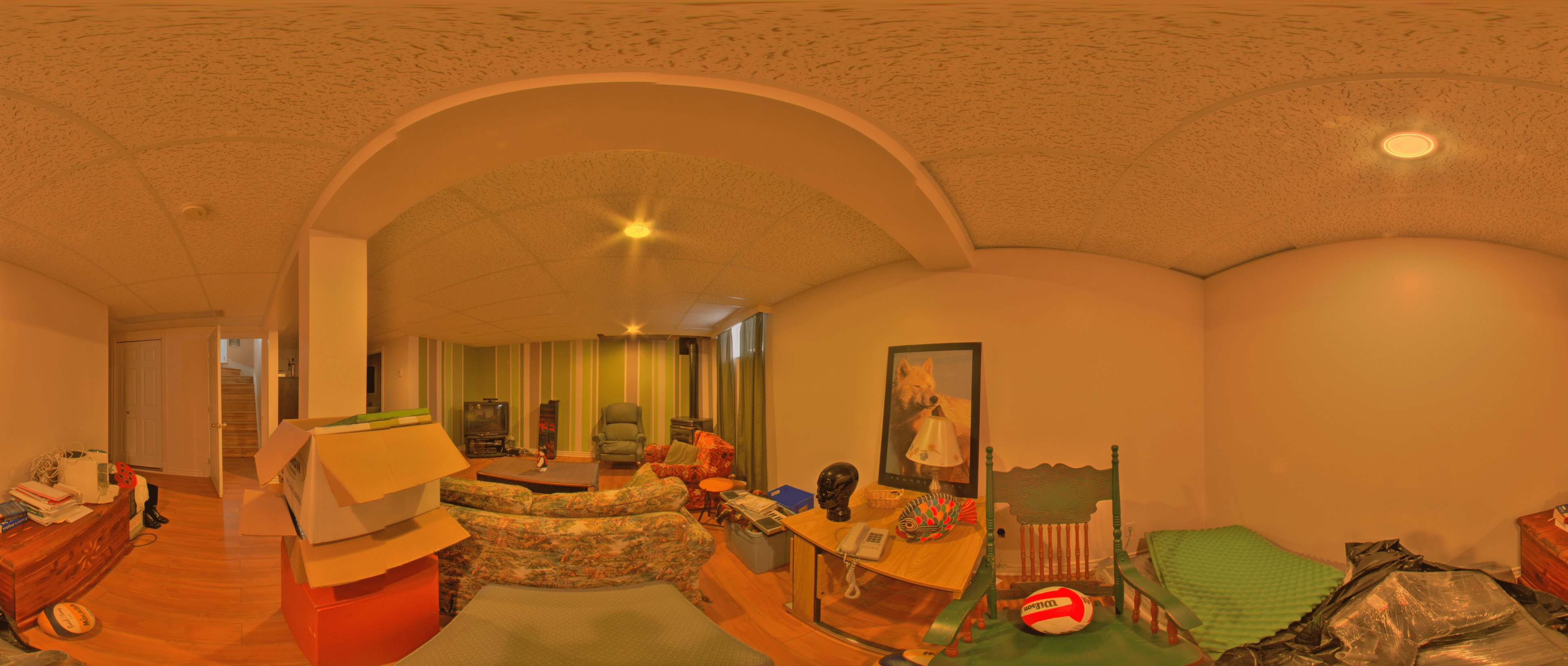}
            \caption[]%
            {{\small }}
        \end{subfigure}
        \caption{The tone mapped results of the CNN with flat architecture (a and b) and reformulated Laplacian pyramid architecture (c and d). Images (a) and (c), and the images (b) and (d) exhibit the same scene, respectively.}
        \label{fig:failed_structure_results}
\end{figure*}

\section{Approach}
To train a learning-based TMO to learn the mapping from a WDR image to a WDR-LDR image, we originally design our CNN model to be 10 layers flat with skip connection architecture shown in Figure \ref{fig:failed_structure}.  We used a combination of the well-known $\ell_1$-norm loss, Structual dissimilarity (DSSIM) \cite{loza2006structural} loss, and feature loss \cite{johnson2016perceptual} to train our network.  The $\ell_1$-norm can be formulated as:

\begin{equation}
    \ell_1 = ||f(x;\theta) - y||_1
\end{equation}

where $f(.)$ represents the CNN network that takes the input image $x$ and the weight $\theta$.  $y$ is the ground truth. 

The DSSIM loss is a variation of the Structual Similarity index (SSIM) \cite{Wang2004Image} that reflects the distance between two images. The DSSIM can be formulated as:

\begin{equation}
    DSSIM = 1 - ssim(f(x;\theta) - y)
\end{equation}

Feature loss, $\ell_{feat}(x,y)$ as a part of the perceptual loss, was proposed by \cite{johnson2016perceptual}.  It uses 16-layer VGG network pre-trained on ImageNet to measure the semantic differences between two images.  Unlike the $\ell_1$-norm pushes the output image to exactly match the label in each pixel,  $l_{feat}(x,y)$ encourages them to increase the similarities in different feature levels. Suppose $\phi_i(x)$ is the output of the feature loss network at $i$-th activation layer, and the activation map is a shape of $W_i \times H_i \times C_i$. We adopted 5 convolutional layers of VGG-16. The feature loss function is formulated as:
\begin{equation}
    \ell_{feat}(x_g,y_g) = \sum_{i=1}^{5}\frac{1}{W_iH_iC_i}||(\phi_i(x_g)-\phi_i(y_g)||_1
\end{equation}

\begin{figure*}[t]
\begin{center}
   \includegraphics[scale = 0.30]{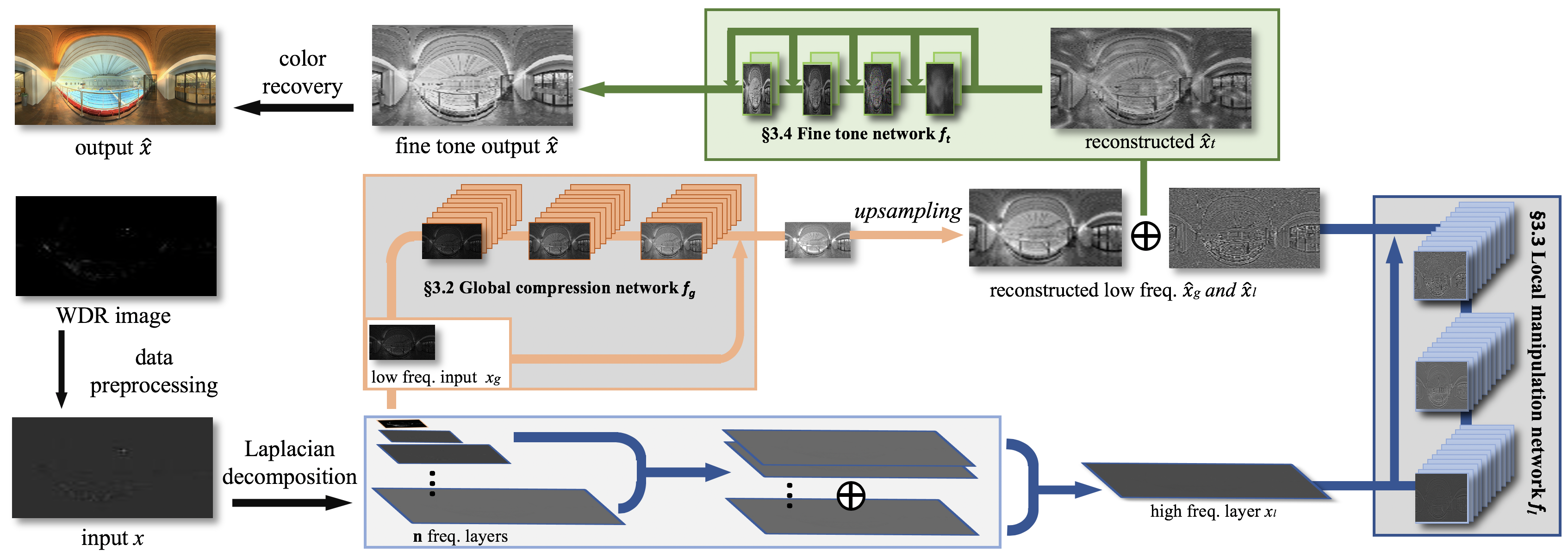}
\end{center}	
   \caption{An overview of the proposed deep multi-bands tone mapping architecture. It decompose an input WDR image into multiple frequency bands with Laplacian pyramid, every band is mapped to the WDR-LDR domain with a specific deep neural network.}
\label{fig:architecture}
\end{figure*}

After many experimental attempts, the result directly generated by such CNN with skip connection architecture shows unpleasing.  The tone mapped images exhibit in severe contrast loss and color distortion.  Figure \ref{fig:failed_structure_results} shows two examples of the tone mapped results comparing to our novel TMO that will be introduced in the next section. These cases demonstrate that CNN, in general, can be used to compress the gradient of a WDR image.  However, CNN with this architecture lacks the ability to generate a WDR-LDR image with a smooth texture and high contrast.  It also fails to preserve the details in the overexposed regions such as the scenery outside the window in (a).  In addition, lots of halo artifacts can also be visually observed in high gradient areas.  This is likely because CNN with ordinary architecture has difficulty extracting the high-frequency feature of a WDR image. Frequency means the rate of change of intensity per pixel. If you have an area in your image changing from black to white, which takes many pixels to represent that intensity variation, it is called low-frequency, and vice versa. For that reason, we came up with the idea of redesigning our CNN to operate on the different image frequencies. One network can focus on the gradient compression in the high-frequency layer, while the other network focuses on the compression of the naturalness.  In the end, the result of the two image frequency bands will be reconstructed back to generate the tone mapped WDR-LDR image.  We combine our CNN with the reformulated Laplacian pyramid to complete this task.

Figure \ref{fig:architecture} presents an overview of the novel architecture. The objective of our work is to find the weight $\theta$, which tone map the input image $x$ to an output image $\hat{x}$, i.e. $\hat{x} = f(x;\theta)$. The input WDR image $x$ is first decomposed into $n$ different frequency bands $x_0$, $x_1$, ..., $x_{n-2}$, $x_g$ with Laplacian pyramid decomposition where $x_0$ is the highest frequency band and $x_{g}$ is the lowest frequency band. The high frequency bands from $x_0$ to $x_{n-2}$ are further Laplacian reconstructed to a single image $x_l$ which has the original resolution of the WDR image.  The entire network $f$ is composed of three sub-networks, global compression network $f_g$,  local manipulation network $f_l$, and fine tune network $f_t$. $f_g$ is used to generate the low frequency Laplacian decomposition of $\hat{x}_g$, i.e.  $\hat{x}_g = f_g(x_g;\theta_g)$. Network $f_l$ is used to generate the high frequency components of $\hat{x}_l$, $\hat{x_l} = f_l(x_l;\theta_l)$. 
Network $f_g$ handles global features while network $f_l$ deals with the high frequency local features. The generated images of $\hat{x}_l$ and $\hat{x}_g$ are reconstructed and fine toned through network $f_t$ to output the final WDR-LDR image $\hat{x}$.

\subsection{Laplacian Pyramid Reformulation}
\label{sect:lap_py}
Laplacian pyramid condenses the global luminance information of an image to lower resolution without sacrificing the detail since the traditional Laplacian pyramid reconstruction operation will nevertheless restore the image back. On the other hand, applying convolutional operation over the image with lower resolution can effectively decrease the computational complexity, thus reduces the requirement of the computing device.

\begin{figure*}[t]
\begin{center}
   \includegraphics[scale = 0.35]{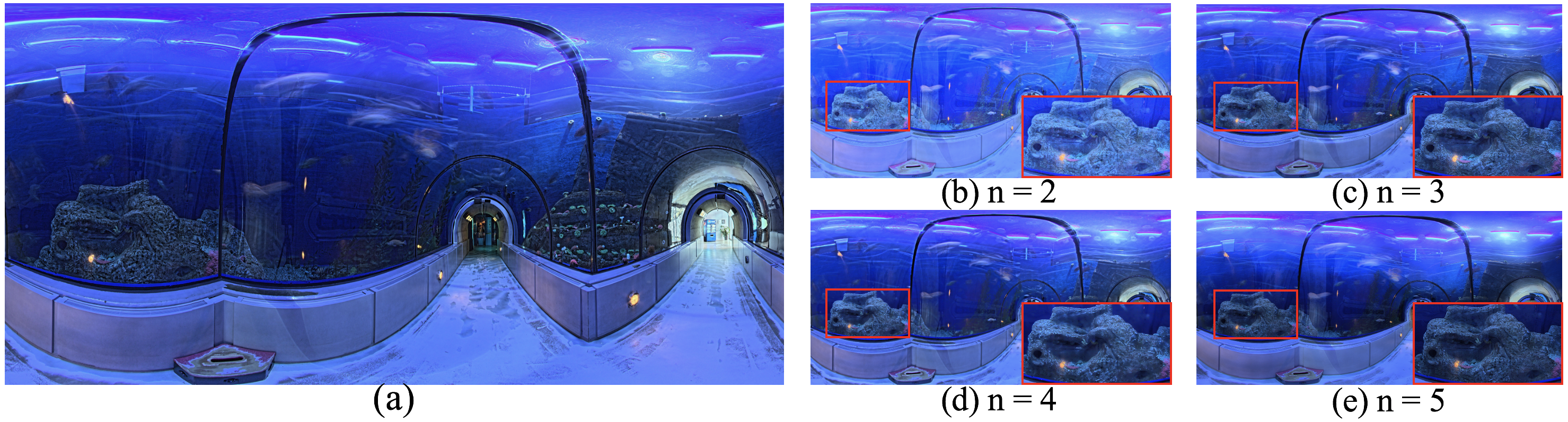}
\end{center}	
   \caption{Visual comparison of the resulting images in different frequency bands. (a) is the ground truth image. (b), (c), (d) and (e) are the images with the frequency band $n=2$, $3$, $4$ and $5$, respectively.}
\label{fig:level_conparison}
\end{figure*}

A WDR image can be segmented into $n$ different frequency bands with a Laplacian pyramid. The lowest frequency band contains the global luminance terrain of the original image and the higher frequencies contain local detail and textural information which varies fast in space.  The advantage of using Laplacian decomposition is apparent. 
\begin{enumerate} 
\item Taking the lowest frequency layer $x_g$ as an example, its resolution is reduced by $2^{n-1}$ times in both width and height when compared to the original image, moreover, the global luminance terrain is well preserved in $x_g$. 
\item In subsequent processing, even a small kernel in the neural network can process a large receptive field of the original image. 
\item Additionally, the low-resolution input of $x_g$ can significantly reduce the required computation for training the parameter $\theta_g$.  Figure \ref{fig:level_conparison} shows the visual comparison with different choices of the number of the frequency band $n$.
\end{enumerate}

However, the generated Laplacian pyramid also has certain drawbacks. Firstly, there are $n$ layers of images and each contains different frequency components of the original WDR image. It would be difficult to process all different layers with a single neural network because the low frequency layer needs to be compressed greatly while the high frequency layers only need to be manipulated locally.  Furthermore, if the $n$ layers are processed with $n$ different neural networks, the complexity of tone mapping model will also grow and make it not feasible to fit in hardware devices. 

\begin{figure*}[tb]
\begin{center}
    \centering
    \includegraphics[scale = 0.37]{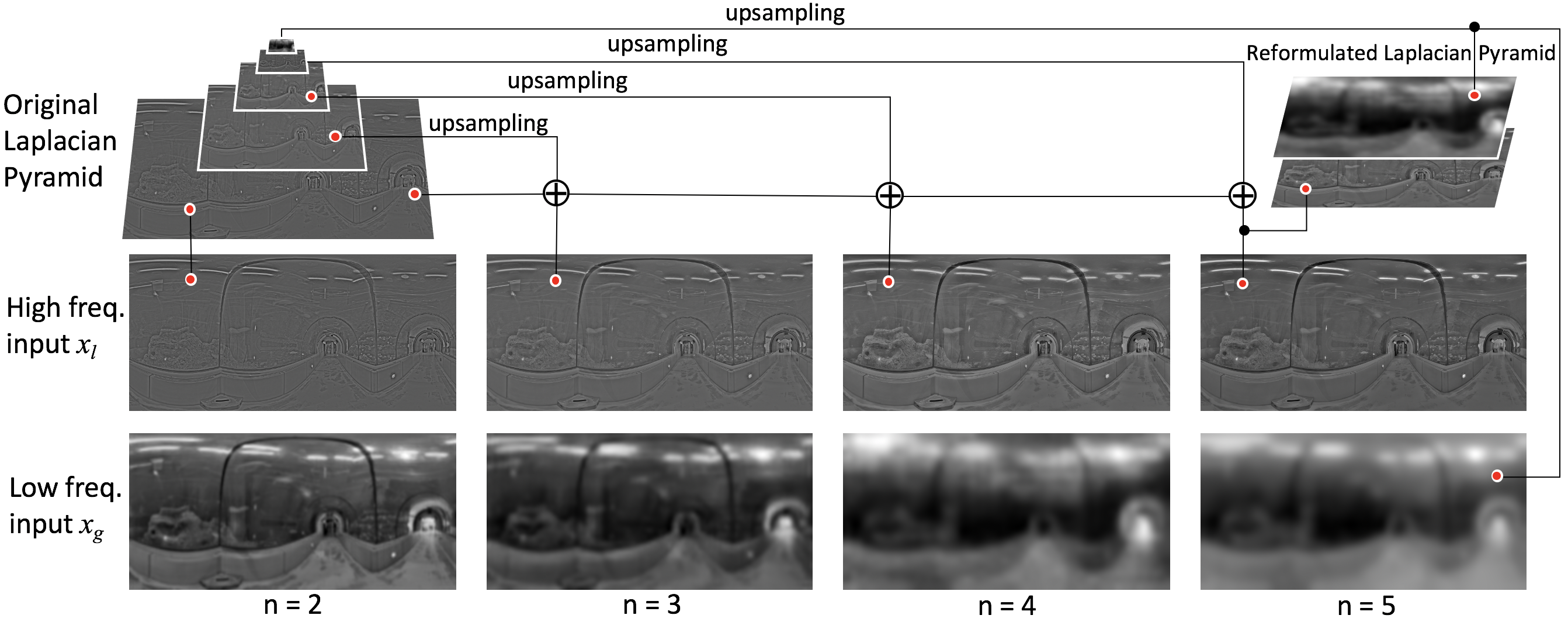}
    \captionof{figure}{Illustration of the relationship and the differences between reformulated Laplacian pyramid and the original Laplacian pyramid. Unlike the original pyramid shape image decomposition (top-left, $n=5$), the reformulated structure (top-right, $n=5$) always contains two layers. Layers at the same frequency are connected with red dots. The high frequency layer of the reformulated pyramid in each $n$ can be reconstructed by adding the high frequency layer and its upsampled previous high frequency layer in the original Laplacian pyramid. }
    \label{fig:level_explain}
\end{center}
\end{figure*}

To overcome the mentioned drawbacks caused by the Laplacian pyramid, we reconstruct an image from the entire Laplacian pyramid without the lowest frequency layer. The generated $x_l$ is a single image that has the same resolution as the original WDR image. Now, the Laplacian pyramid is reformulated into two layers with $x_l$ representing all high frequency components and $x_g$ representing low frequency global luminance terrain.  Figure \ref{fig:level_explain} intuitively shows the relationship and the difference between the reformulated Laplacian pyramid and the original one.  Using two layers in the Laplacian pyramid structure has the following advantages. First, it reduces the original Laplacian pyramid model from $n$ layers to $2$ layers, hence the computation complexity of subsequent processing is significantly reduced. Secondly, the segmentation of high and low frequency components of the WDR image leads network $f_g$ and $f_l$ to focus on simple tasks, namely, global compression and local detail manipulation, respectively.


\subsection{Global Compression Network}
The global compression network of $f_g$ focuses on the compression of the global dynamic range of the WDR image, namely $x_g$. After the decomposition of the Laplacian pyramid, $x_g$ is a low resolution image and only contains global luminance information of the original WDR image. Unlike many image transformation works \cite{yang2018image,lee2018deep,cai2018learning} that employ encode-decode architecture to avoid the loss of the global feature during convolution, our architecture is able to achieve the same effect with the help of the low resolution representation $x_g$. A small $k \times k$ kernel is able to cover $(2^{n-1}*k)^2$ pixels of the original WDR image if the WDR image is decomposed to $n$ layers.
Therefore, we adopt a simple CNN architecture to do the compression. The detail of the proposed global compression network is summarized Table \ref{table:globalnetwork}.

$W$ and $H$ are the width and height of the input image $x_g$, respectively. 
Given an input image $x$, and the ground truth WDR-LDR image $y$, 
we use the $\ell_1$-norm, feature loss $\ell_{feat}$ and $\ell_2$ regularization as the loss function:
\begin{equation}
    \ell_{global} = \alpha \ell_1 +\beta \ell_{feat}(x_g, y_g) + \gamma R(\theta_g)
\end{equation}
where $\alpha$ = 0.5, $\beta$ = 0.5 and $\gamma$ = 0.2.  The $\ell_1$-norm can be formulated as:
\begin{equation}
    \ell_1 = ||f_g(x_g;\theta_g) - y_g||_1
\end{equation}
where $y_g$ is the lower frequency part of the corresponding reformulated Laplacian pyramid of ground truth $y$. As our dataset is not ample comparing to all WDR image representations, we implement $\ell_2$ regularization loss $R(\theta_g)$ to all our neural networks to prevent over-fitting.

\subsection{Local Manipulation Network}
The purpose of the local manipulation network $f_l$ is to manipulate the high frequency part of the WDR image, namely $x_l$. Unlike $x_g$, the high frequency features contained in $x_l$ are mostly local. For simplicity, we adopt the same architecture in Table \ref{table:globalnetwork} to do the local manipulation. This is because $x_l$ has the same resolution as the WDR image, the kernels in Table \ref{table:globalnetwork} will only cover a local image patch instead of a global area. The same network can serve two different goals when cooperated with $x_g$ and $x_l$, respectively. 
The learning objective of the local manipulation network and the global compression network are the same. We use the same set of parameters and loss function:
\begin{equation}
    \ell_{local} = \alpha \ell_1 +\beta \ell_{feat}(x_l, y_l) + \gamma R(\theta_l)
\end{equation}

$y_l$ is the high frequency part of the corresponding reformulated Laplacian pyramid of the ground truth $y$.

\begin{table}
\footnotesize
\begin{center}
\caption{Detail of Global Compression Network and Local Manipulation Network.}
\begin{tabular}{l|c|c|c|c}
\hline
 Layers & Input Size & Kernel Size & Stride & Kernel Num.\\
\hline
 Input  & $ W \times H$ & - & - & - \\
 Conv\_1 & $ W \times H$ & $3 \times 3$ & $1$ & $32$ \\
 Batch\_norm\_1 & $W \times H \times 32$ & - & - & - \\
 Conv\_2 & $ W \times H \times 32$ & $3 \times 3$ & $1$ & $32$ \\
 Batch\_norm\_2 & $W \times H \times 32$ & - & - & - \\
 Conv\_3 & $ W \times H \times 32$ & $3 \times 3$ & $1$ & $32$ \\
 Batch\_norm\_3 & $W \times H \times 32$ & - & - & - \\
 Conv\_4 & $ W \times H \times 32$ & $3 \times 3$ & $1$ & $32$ \\
 Batch\_norm\_4 & $W \times H \times 32$ & - & - & - \\
 Conv1$\times$1 & $ W \times H \times 32$ & - & - & $1$ \\
 \hline
 Output & $W \times H$ &\multicolumn{3}{c}{Input + Conv1$\times$1}\\
\hline
\end{tabular}
\label{table:globalnetwork}
\end{center}
\end{table}

\subsection{Fine Tune Network}
The global compression network and the local manipulation network are able to generate the corresponding reformulated Laplacian layer $\hat{x}_g$ and $\hat{x}_l$, respectively. The Laplacian pyramid requires additional operations to add all frequency layers. That is, $\hat{x}_t$ = upsampling($\hat{x}_g$) + $\hat{x}_l$.  However, image $\hat{x}$ cannot guarantee overall visual quality since $\hat{x}_g$ and $\hat{x}_l$ are produced with separate neural networks. Moreover, color shifts, regional blurry and other artifacts may also occur in $\hat{x}$.

To overcome these possible issues, we utilize a fine tune network $f_t$ to further refine the reconstructed image $\hat{x}_t$ to the desired ground truth image. 
$f_t$ is a \textit{ResNet} architecture with large feature maps and small depth since the main feature of the image has been learned from the  previous two neural networks. 

The \textit{ResNet} contains 4 residual blocks.  Each residual block consists of 2 convolutional layers.  We use $3 \times 3 \times 32$ kernels for every layer with stride of 1, and we use a batch normalization layer after each convolutional layer.  At the end of this \textit{ResNet}, a $1 \times 1$ convolution layer is applied to condense all extracted features from the 32-channel receptive field to 1. 
The loss function of fine tune network is slightly different than the previous network.  We adopted feature loss $l_{feat}$ and $\ell_2$-norm:
\begin{equation}
    \ell_t = \alpha_t \ell_2 + \beta_t \ell_{feat}(\hat{x}_t, y)
\end{equation}
where $\alpha_t$ = 0.6 and $\beta_t$ = 0.4.


\section{Experiments}
In this section, we first present the experimental setup and then analyze the effects of the proposed Laplacian pyramid reformation. We then compare the proposed model with the state-of-the-art methods on two databases.

\subsection{Training Data Generation}
We trained the proposed network for WDR tone mapping on Laval indoor dataset \cite{gardner2017learning}. This dataset contained 2,233 high-resolution ($7768 \times 3884$), high dynamic range indoor panoramas WDR images captured by Canon 5D Mark III camera. 
In the Laval indoor dataset, some images contain watermarks of different scale in the bottom region. 
We discard the bottom 15\% of the panoramas to remove watermarks on the original images. 
After this cropping, the image resolution became $7768 \times 3301$.  And the total number of images used in the experiment was 2,125.
These images are further down-sampled to one-quarter of its original resolution and transferred to luminance image. The luminance image is generated and recovered using methods described in \cite{gu2013local}.
We generate 20 sub-images from each training sample. 
The size of the sub-images are drawn uniformly from the range [20\%, 60\%] of the size of an input WDR image and re-sampled to $512 \times 512$ pixels 
The ground truth WDR-LDR images are generated using various tools including Luminance HDR \footnote{https://github.com/LuminanceHDR/LuminanceHDR}, HDR tool box provided by \cite{Banterle2017}, Photoshop with human tuning and supervision. 

\begin{figure*}[t]
        \centering
        \begin{subfigure}[b]{0.245\textwidth}
            \centering
            \includegraphics[width=\textwidth]{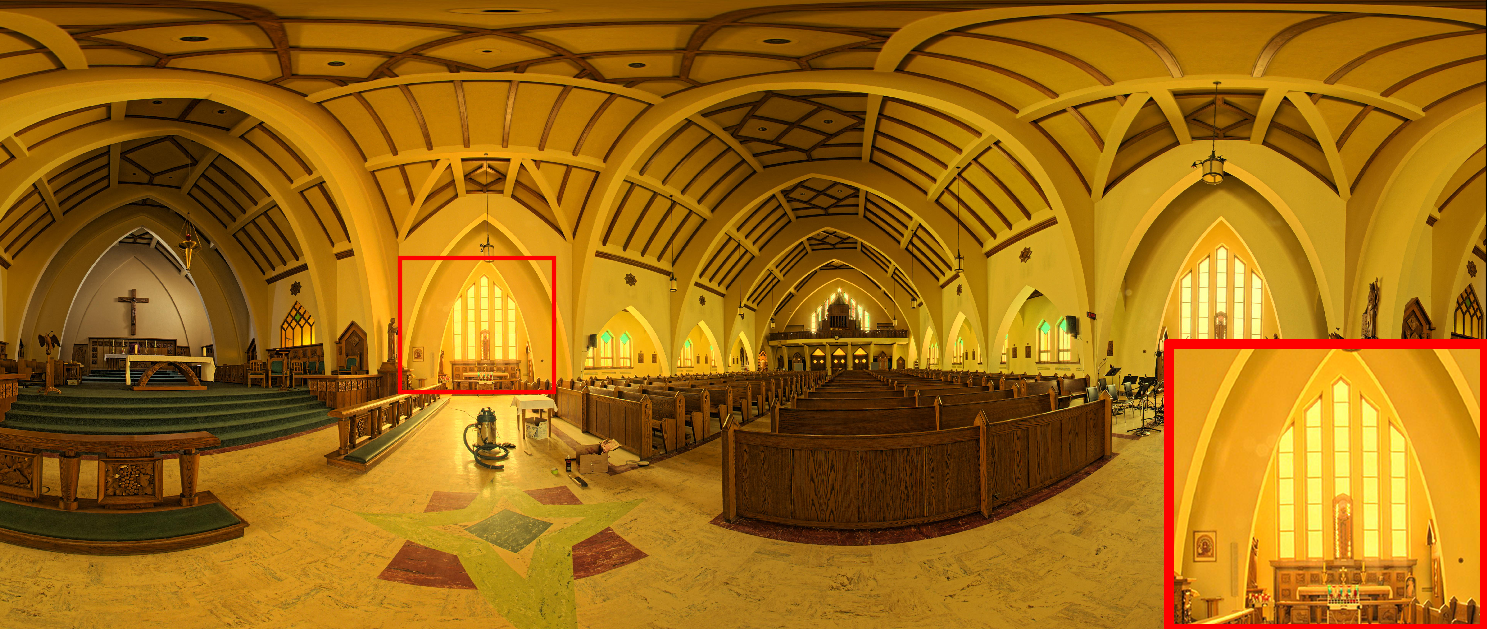}
            \caption[]%
            {{\small Reference }}    
            \label{fig:mean and std of net14}
        \end{subfigure}
        \begin{subfigure}[b]{0.245\textwidth}  
            \centering 
            \includegraphics[width=\textwidth]{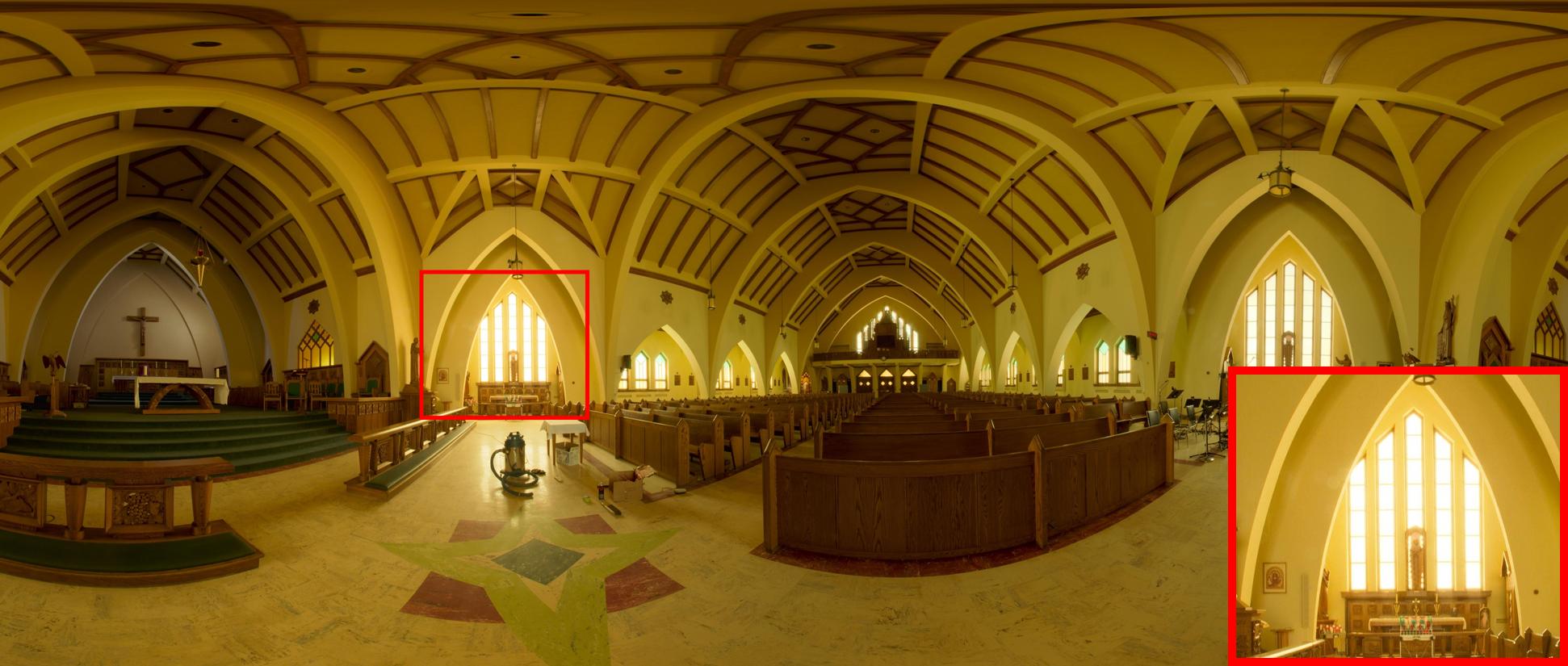}
            \caption[]%
            {{\small Mantiuk \cite{mantiuk2008display} }}    
            \label{fig:mean and std of net24}
        \end{subfigure}
        \begin{subfigure}[b]{0.245\textwidth}   
            \centering 
            \includegraphics[width=\textwidth]{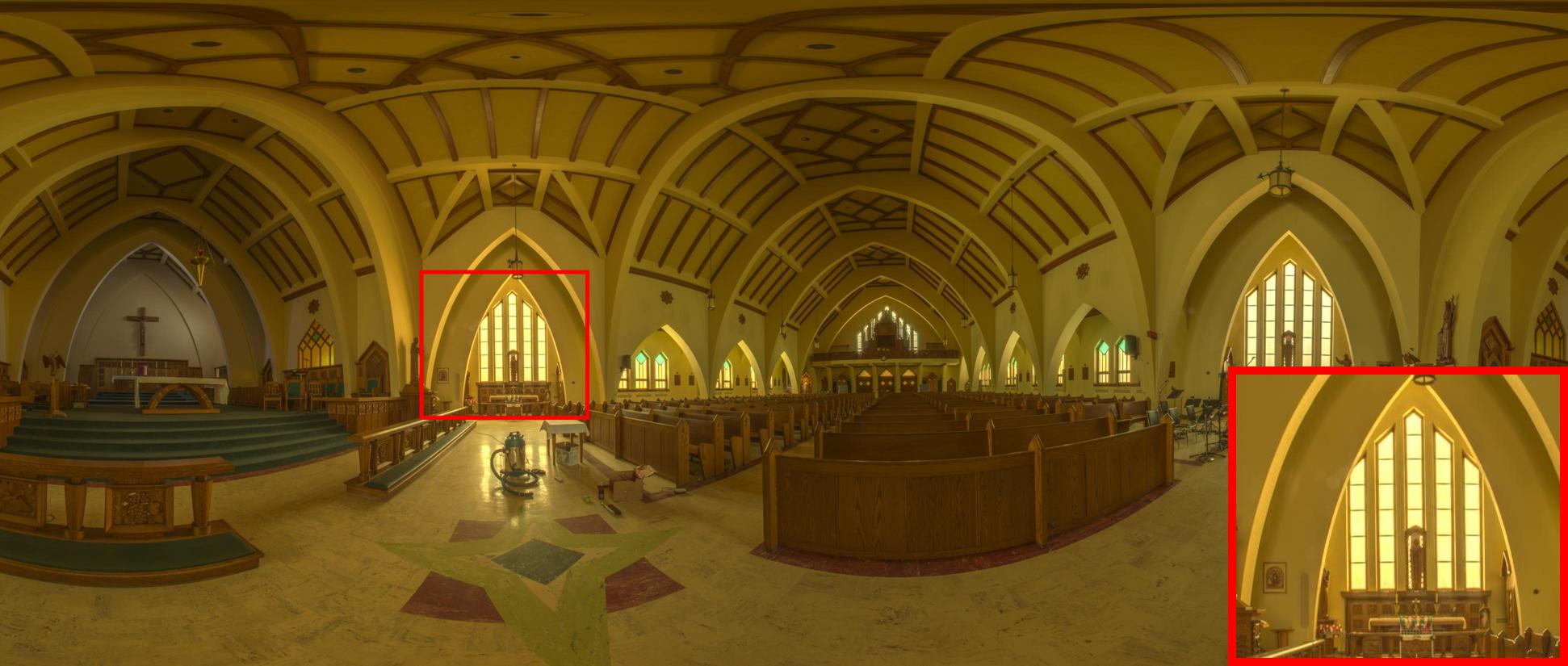}
            \caption[]%
            {{\small Paris \cite{paris2015local} }}    
            \label{fig:mean and std of net34}
        \end{subfigure}
        \begin{subfigure}[b]{0.245\textwidth}   
            \centering 
            \includegraphics[width=\textwidth]{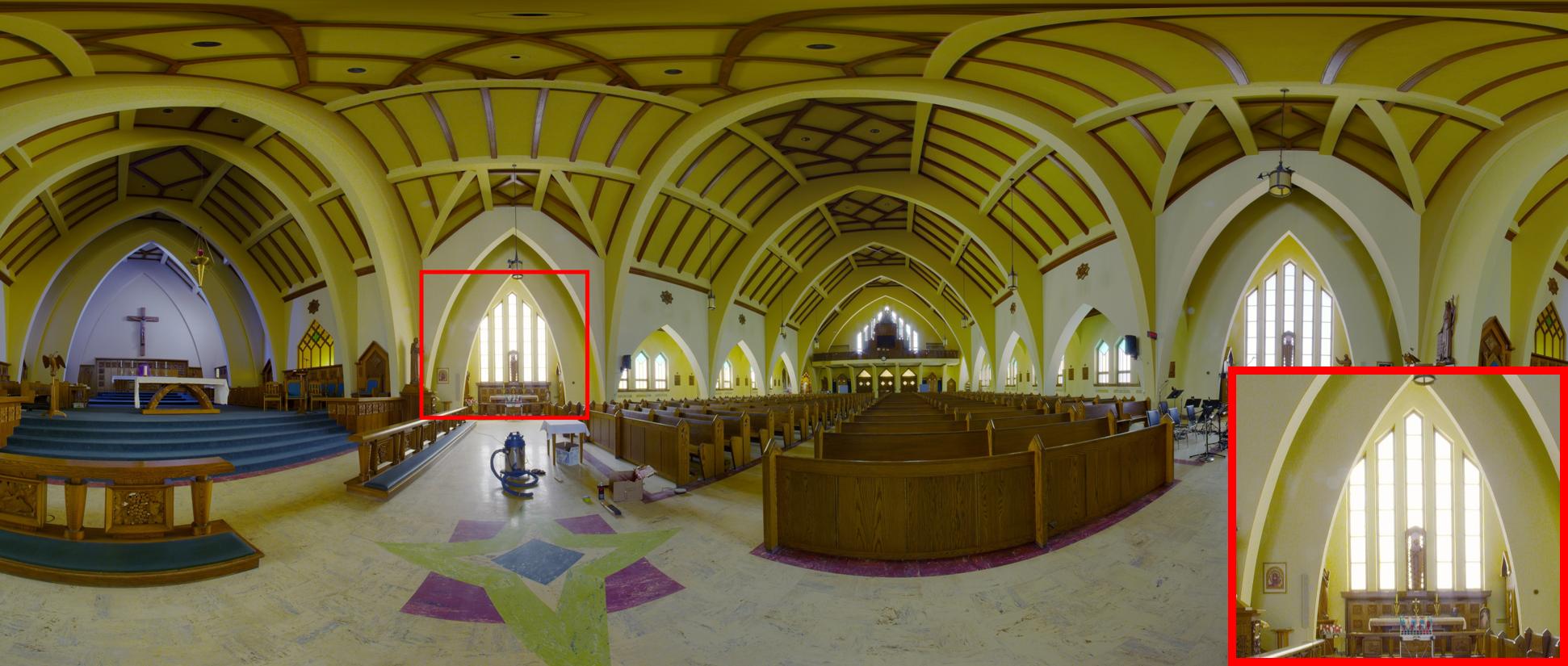}
            \caption[]%
            {{\small Ferradans \cite{ferradans2011analysis}     }}    
            \label{fig:mean and std of net44}
        \end{subfigure}
        \label{fig:mean and std of nets}
        \centering
        \begin{subfigure}[b]{0.245\textwidth}
            \centering
            \includegraphics[width=\textwidth]{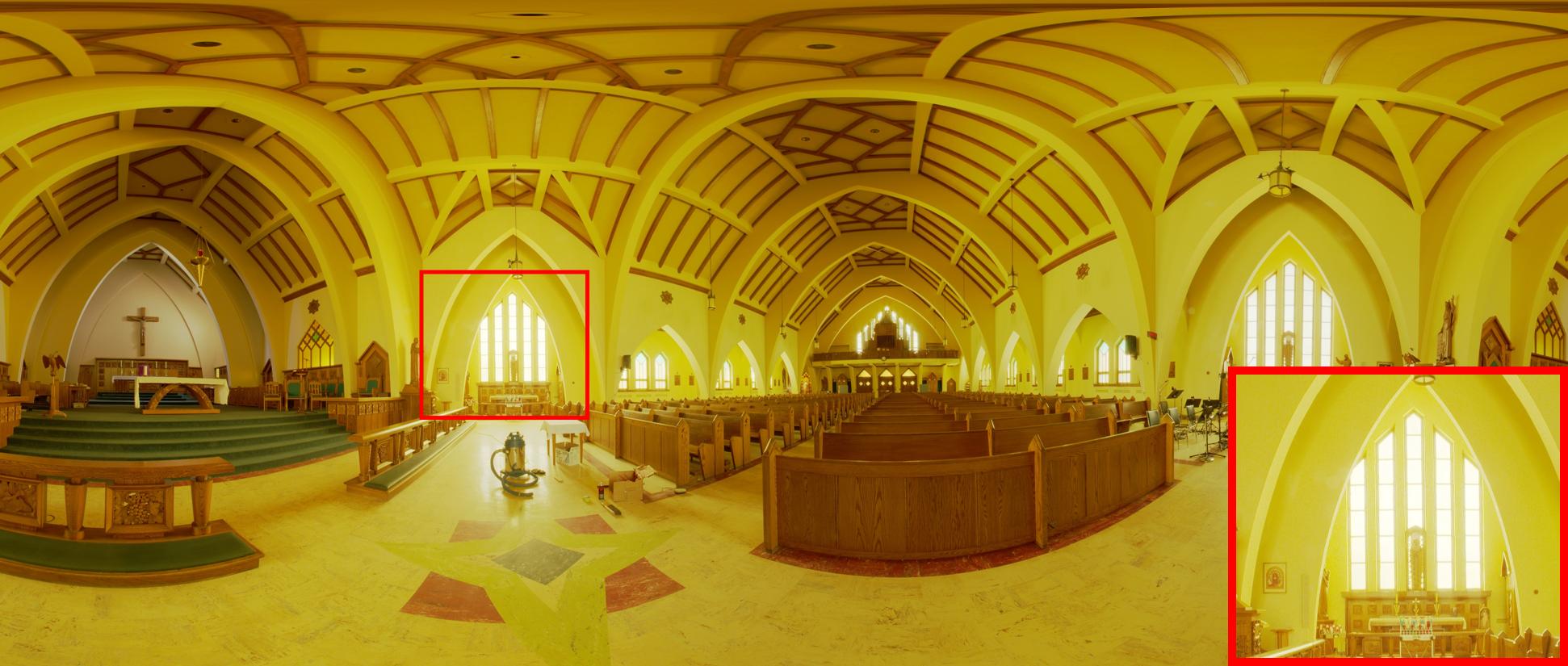}
            \caption[]%
            {{\small Mai \cite{mai2011optimizing} }}    
            \label{fig:mean and std of net14}
        \end{subfigure}
        \begin{subfigure}[b]{0.245\textwidth}  
            \centering 
            \includegraphics[width=\textwidth]{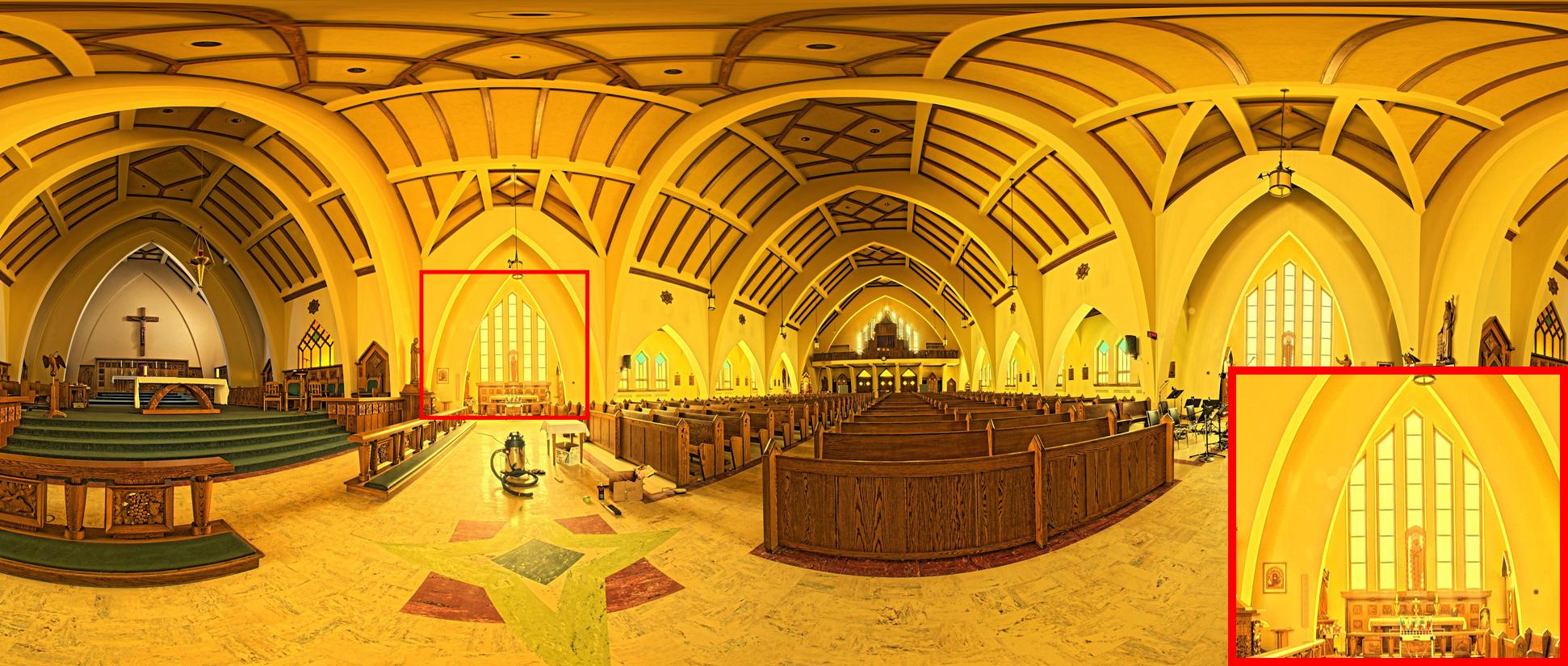}
            \caption[]%
            {{\small Gu \cite{gu2013local} }}    
            \label{fig:mean and std of net24}
        \end{subfigure}
        \begin{subfigure}[b]{0.245\textwidth}   
            \centering 
            \includegraphics[width=\textwidth]{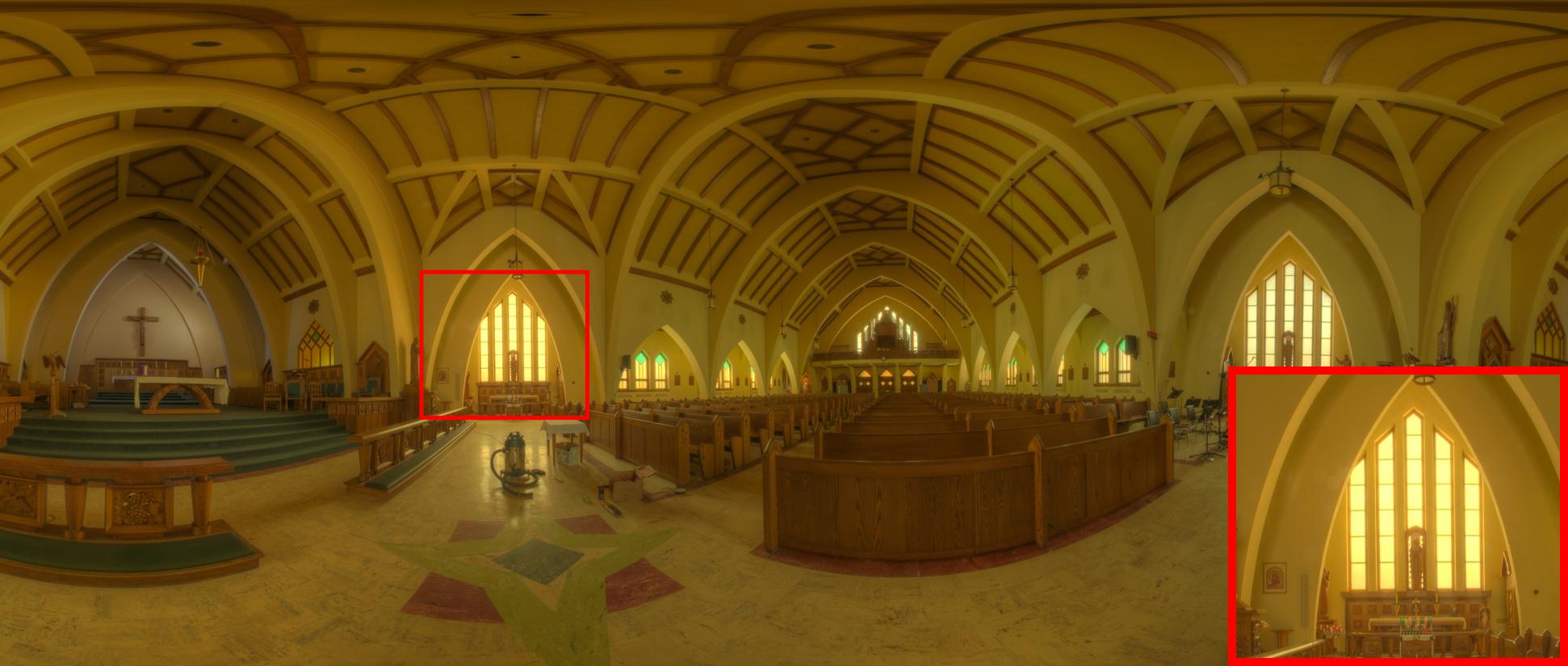}
            \caption[]%
            {{\small Photomatix \cite{photomatrix} }}    
            \label{fig:mean and std of net34}
        \end{subfigure}
        \begin{subfigure}[b]{0.245\textwidth}   
            \centering 
            \includegraphics[width=\textwidth]{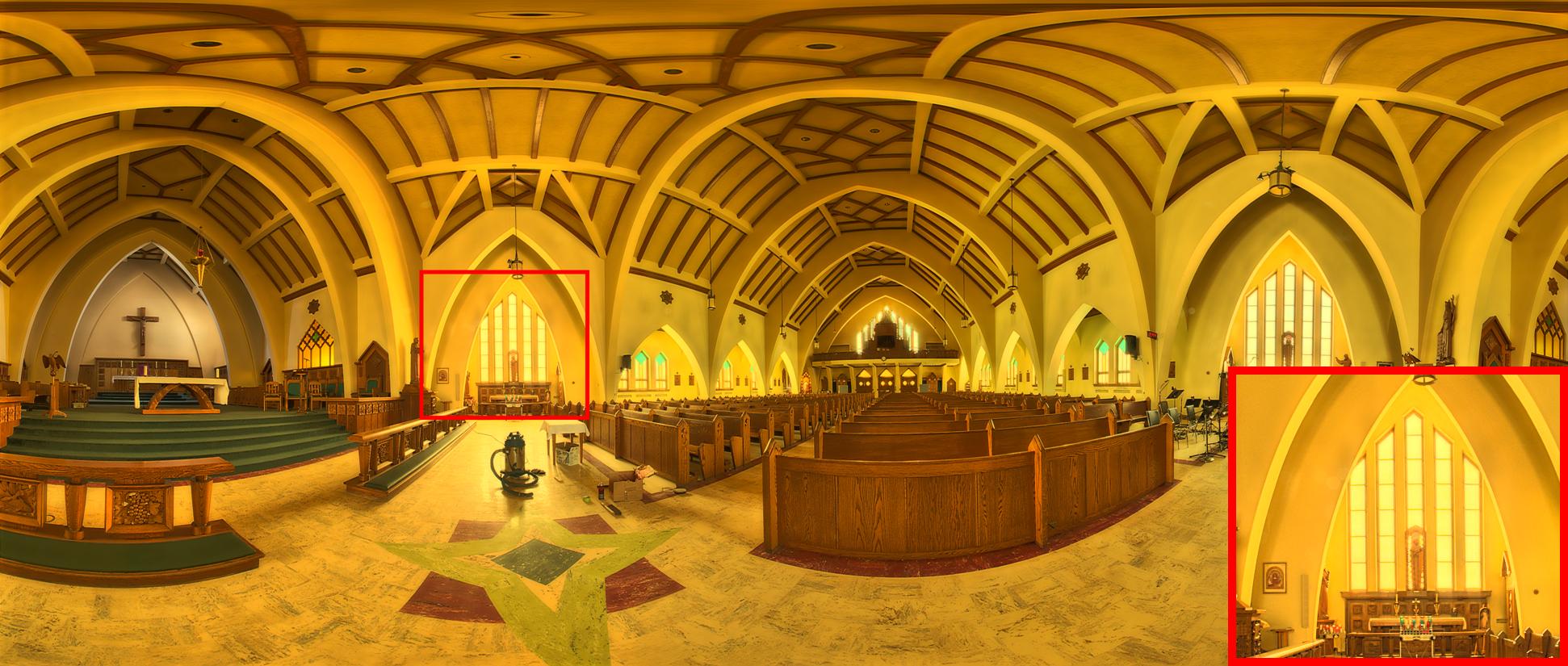}
            \caption[]%
            {{\small Proposed  }}    
            \label{fig:mean and std of net44}
        \end{subfigure}
        \caption{Visual comparison on the test set. The proposed method can effectively compress the global dynamic range while preserving local detail and contrast.}
        \label{fig:laval_global_compress}
\end{figure*}

\subsection{Ground Truth Generation}
Similar to  Cai’s \cite{Cai2018deep} reference image generation method, we generated high-quality ground truth images using several TMOs and human tuning. We used 6 TMOs in this process including Fattal \cite{fattal2002gradient}, Ferradans \cite{ferradans2011analysis}, Mantiuk \cite{mantiuk2008display}, Drago \cite{drago2003adaptive}, Durand \cite{durand2002fast}, and Reinhard \cite{reinhard2010high} from Luminance HDR and HDR tool box.  Then we employed 4 volunteers and 2 photographers in this process. The two photographers first picked out the images they thought were unsatisfactory (such as too dark, too bright, or exists distortion), and used Photoshop to fix them according to their own preferences. The volunteers performed the random pairwise comparison independently in the 7 sets of tone mapped images by given instruction:

\begin{itemize}
  \item Select one image of two that best suits your visual preferences.
  \item Spend no more than 5 sec for each pair.
\end{itemize}

Images with the same vote or that couldn't be selected within 5 seconds will be circulated back to photographers to modify, and then send to volunteers in the next round until all images have been selected.

\subsection{Implementation Details}
We randomly selected 70\% images for training our model and use the remaining 10\% for validation and 20\% for testing.  
The network parameters are initialized using the truncated normal initializer. 
All training experiments are performed
using the TensorFlow\footnote{https://www.tensorflow.org/?hl=zh-cn} deep learning library. 
We adopt the ADAM optimizer for loss minimization with the learning rate is $10^{-3}$, momentum $\beta_1$ = 0.9 and $\beta_2$ = 0.999, $\epsilon = 10^{-8}$. We use mini-batch gradient descent with batch size 8 for local manipulation, 64 for global compression and 4 for fine tuning. 
The forgoing networks are trained in multiple steps. The network $f_l$ and $f_g$ are trained first. 
And then, we use the loss function of $f_t$ to jointly train the entire system containing $f_l$, $f_g$ and $f_t$.  The proposed model is trained in an end-to-end fashion.

\subsection{Parameter Setting}
The process of Laplacian pyramid reformulation described in Section \ref{sect:lap_py} has a hidden parameter $n$ which indicates the number of layers during the original Laplacian decomposition. A different $n$ value will certainly affect the training and lead to different results. In order to evaluate the effect of this parameter on the final trained model, we trained our model with $n=2, 3,..., 7$ and evaluated the average PSNR, SSIM \cite{Wang2004Image} and FSITM \cite{nafchi2014fsitm} on the test data set.  The result is summarized in Table \ref{table:PSNR}.  It is not surprising that the median $n$ values achieve average higher metrics. Actually, a smaller $n$ value will assign most information to the $x_g$ image while a larger $n$ value will move more frequency bands to $x_l$. Suppose $n$ is so large that $x_g$ has only one pixel, then the final image will solely be determined by $f_l$. On the other hand, if $n$ is too small, then $x_l$ will contain limited information which deteriorates the desired functionality of $f_l$.  The model with $n = 6$ gives the highest metric values. In the rest of this paper, we set $n = 6$ for all remaining experiments.

\begin{figure*}[tb]
        \centering
        \begin{subfigure}[b]{0.245\textwidth}  
            \centering 
            \includegraphics[width=\textwidth]{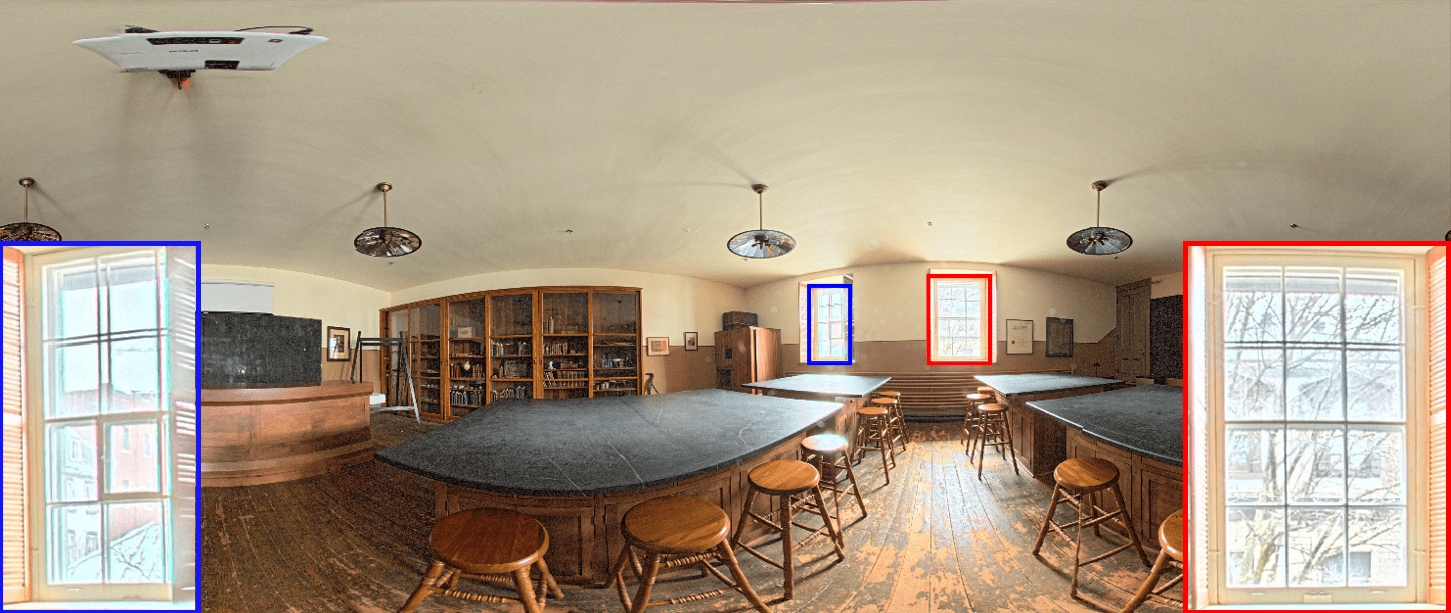}
            \caption[]%
            {{\small Reference }}    
            \label{fig:mean and std of net24}
        \end{subfigure}
        \begin{subfigure}[b]{0.245\textwidth}  
            \centering 
            \includegraphics[width=\textwidth]{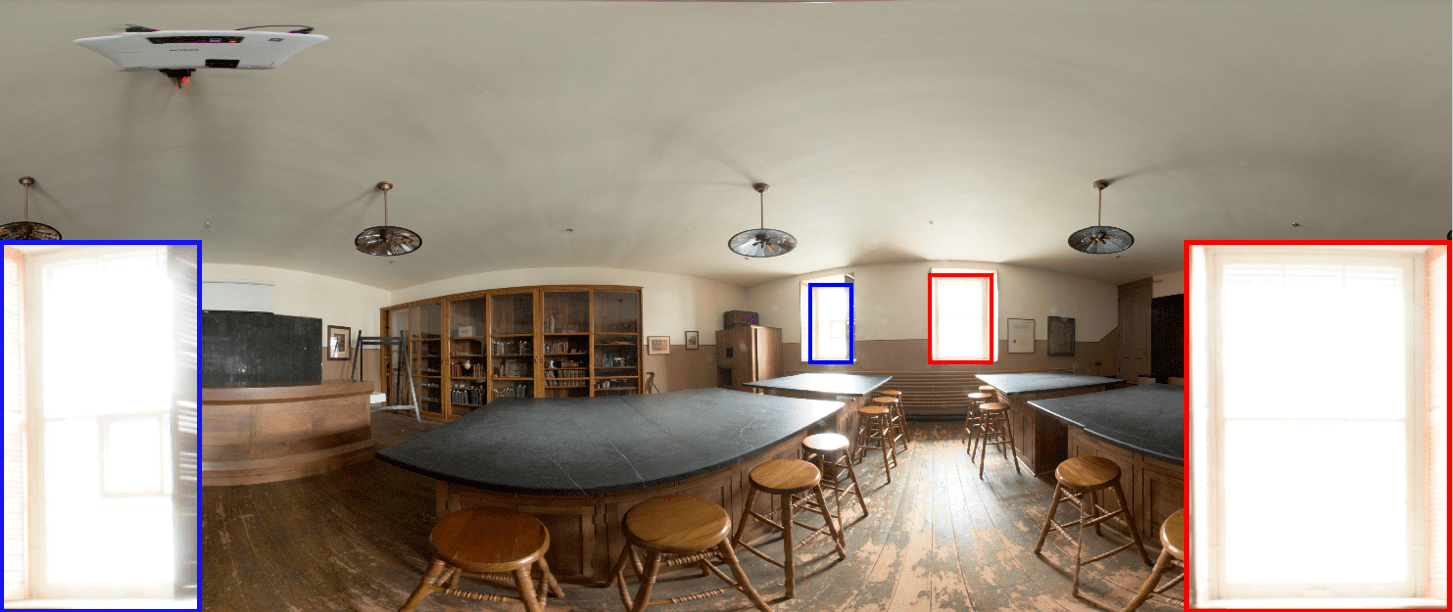}
            \caption[]%
            {{\small Mantiuk \cite{mantiuk2008display} }}    
            \label{fig:mean and std of net24}
        \end{subfigure}
        \begin{subfigure}[b]{0.245\textwidth}   
            \centering 
            \includegraphics[width=\textwidth]{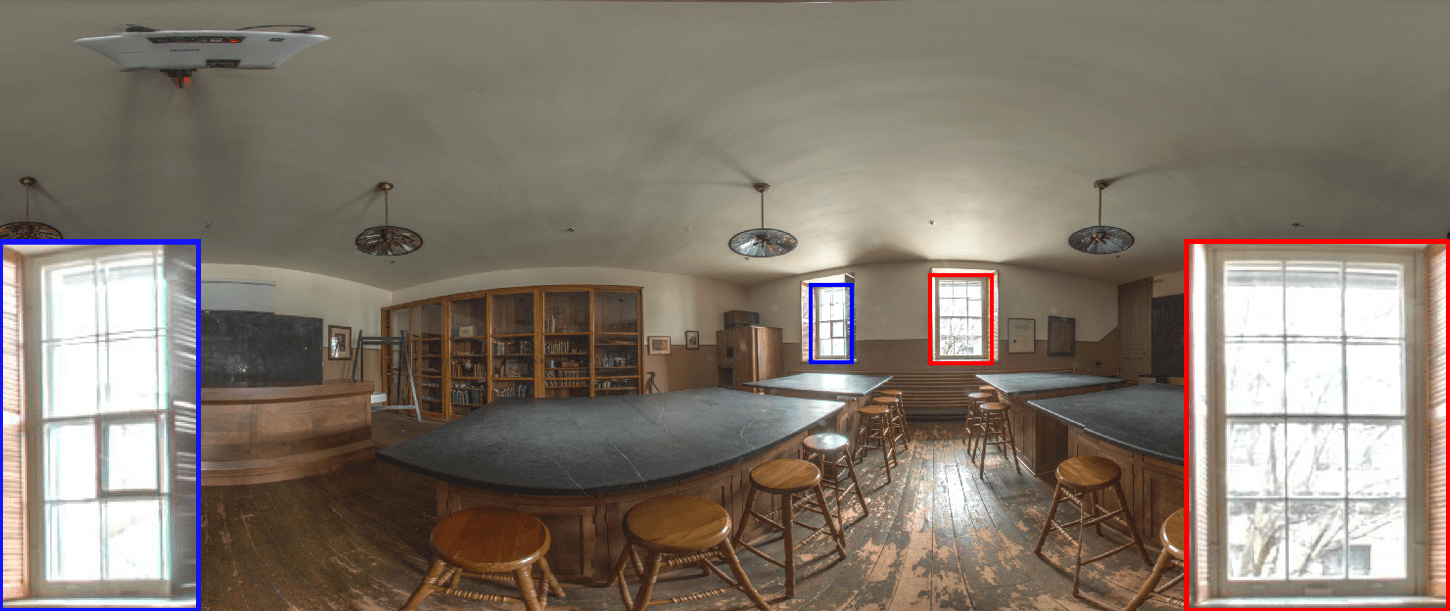}
            \caption[]%
            {{\small Paris \cite{paris2015local} }}    
            \label{fig:mean and std of net34}
        \end{subfigure}
        \begin{subfigure}[b]{0.245\textwidth}   
            \centering 
            \includegraphics[width=\textwidth]{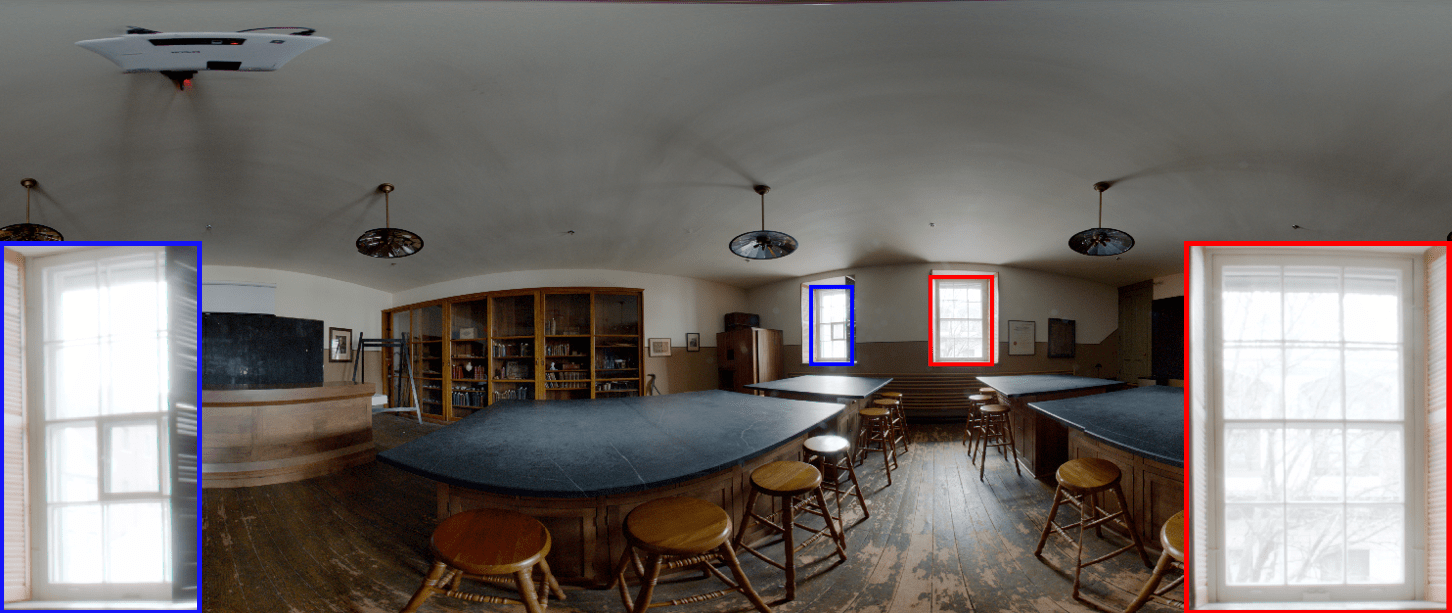}
            \caption[]%
            {{\small Ferradans \cite{ferradans2011analysis}     }}    
            \label{fig:mean and std of net44}
        \end{subfigure}
        \label{fig:mean and std of nets}
        \centering
        \begin{subfigure}[b]{0.245\textwidth}
            \centering
            \includegraphics[width=\textwidth]{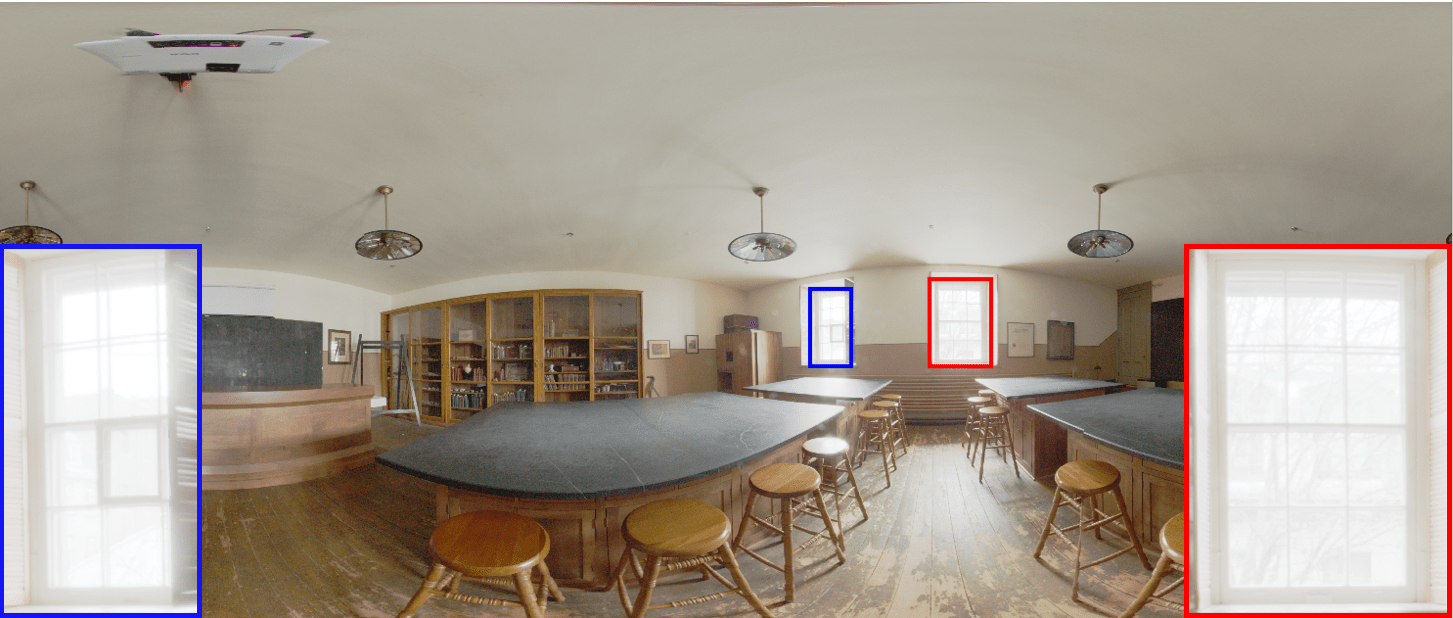}
            \caption[]%
            {{\small Mai \cite{mai2011optimizing} }}    
            \label{fig:mean and std of net14}
        \end{subfigure}
        \begin{subfigure}[b]{0.245\textwidth}
            \centering
            \includegraphics[width=\textwidth]{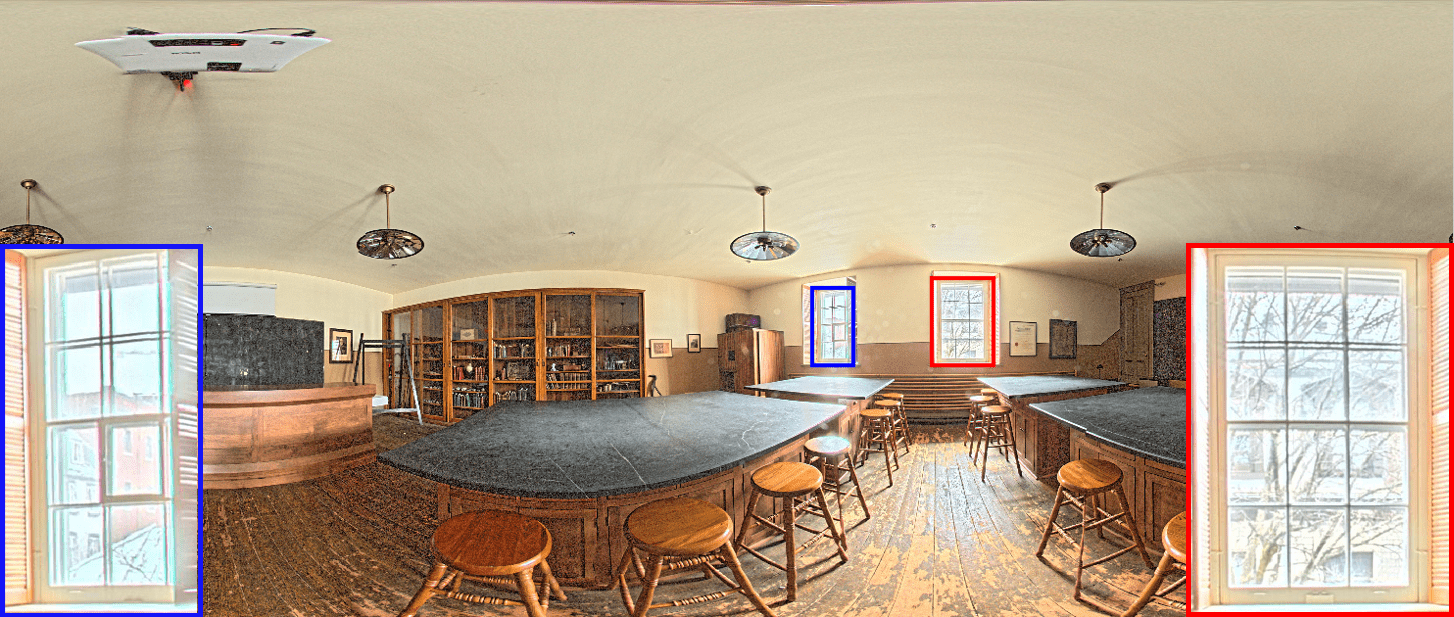}
            \caption[]%
            {{\small Gu \cite{gu2013local} }}    
            \label{fig:mean and std of net14}
        \end{subfigure}
        \begin{subfigure}[b]{0.245\textwidth}   
            \centering 
            \includegraphics[width=\textwidth]{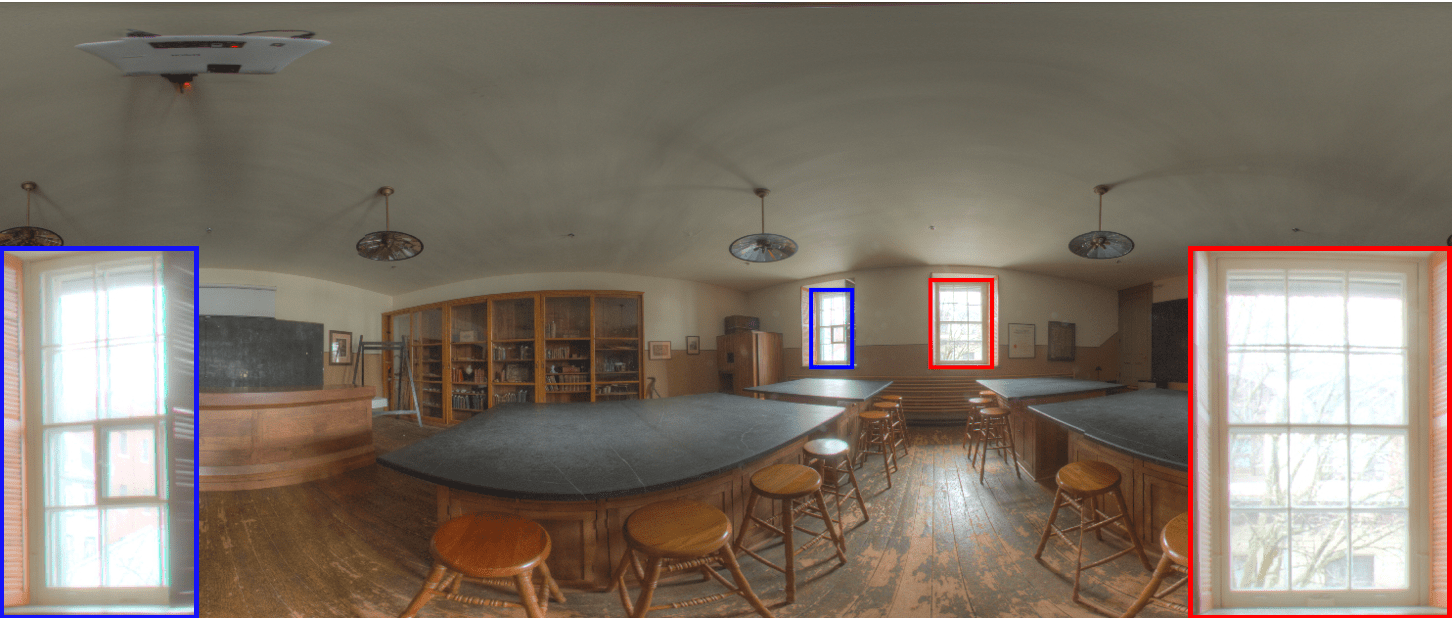}
            \caption[]%
            {{\small Photomatrix \cite{photomatrix} }}    
            \label{fig:mean and std of net34}
        \end{subfigure}
        \begin{subfigure}[b]{0.245\textwidth}   
            \centering 
            \includegraphics[width=\textwidth]{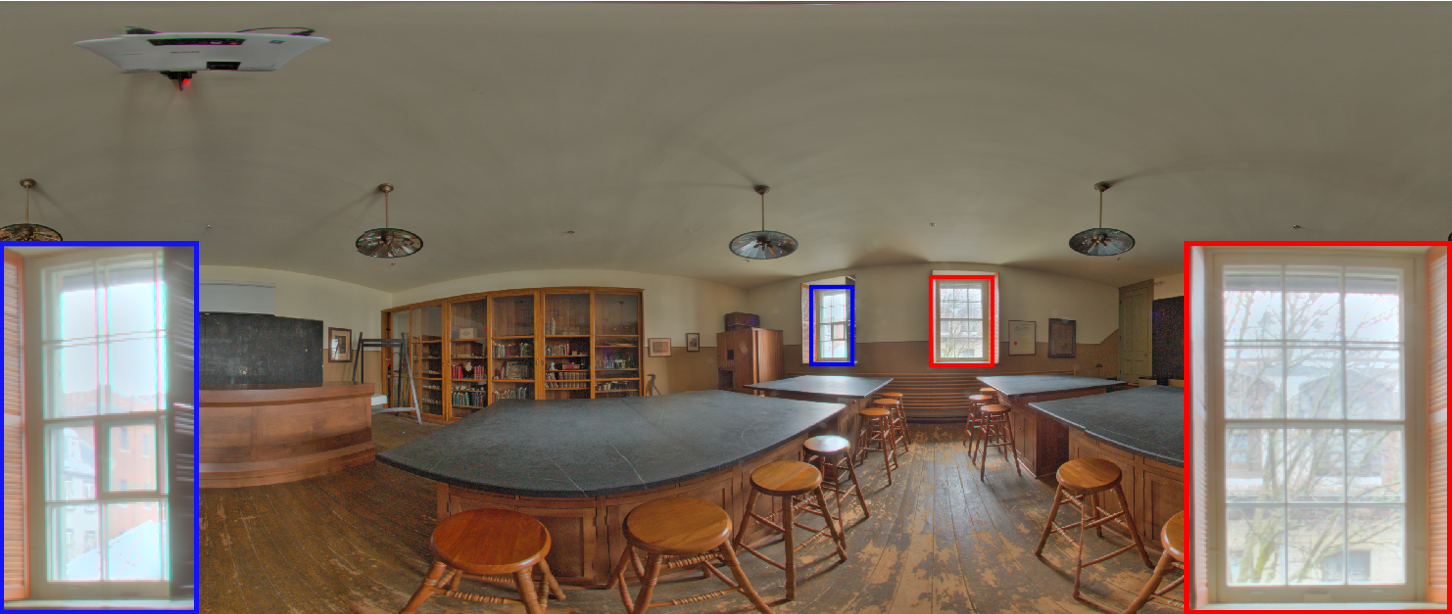}
            \caption[]%
            {{\small Proposed  }}    
            \label{fig:mean and std of net44}
        \end{subfigure}
         \centering
        \begin{subfigure}[b]{0.245\textwidth}  
            \centering 
            \includegraphics[width=\textwidth]{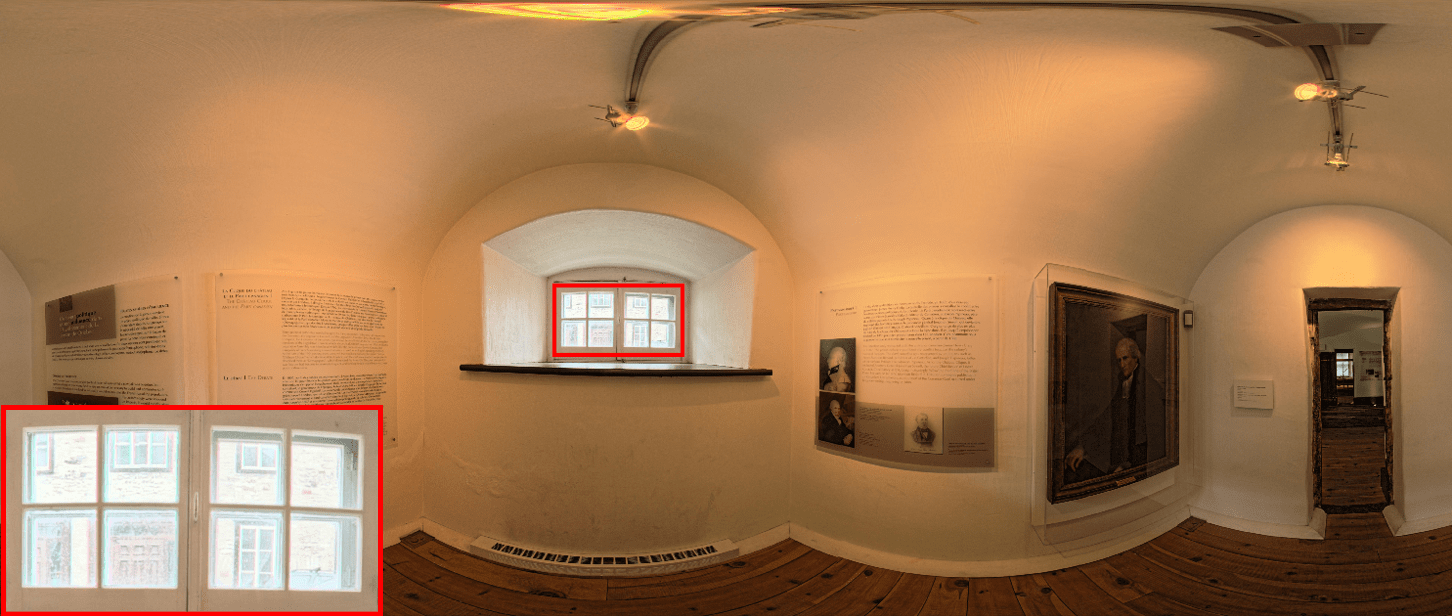}
            \caption[]%
            {{\small Reference }}    
            \label{fig:mean and std of net24}
        \end{subfigure}
        \begin{subfigure}[b]{0.245\textwidth}  
            \centering 
            \includegraphics[width=\textwidth]{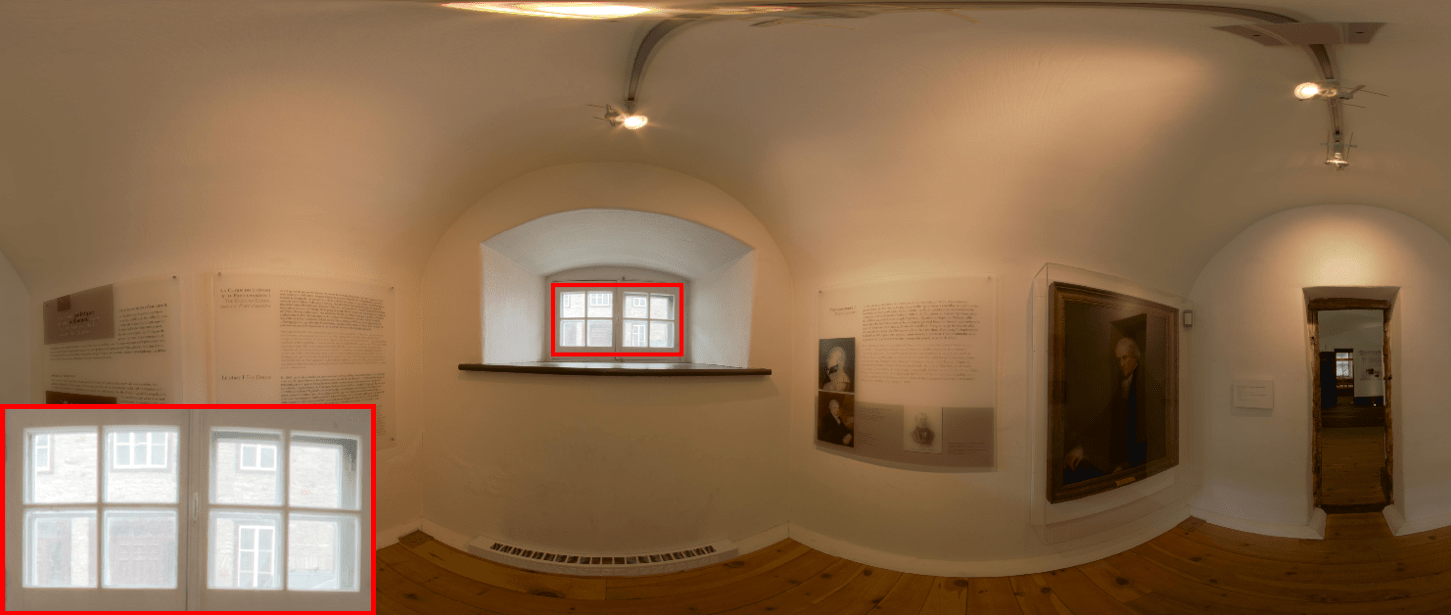}
            \caption[]%
            {{\small Mantiuk \cite{mantiuk2008display} }}    
            \label{fig:mean and std of net24}
        \end{subfigure}
        \begin{subfigure}[b]{0.245\textwidth}   
            \centering 
            \includegraphics[width=\textwidth]{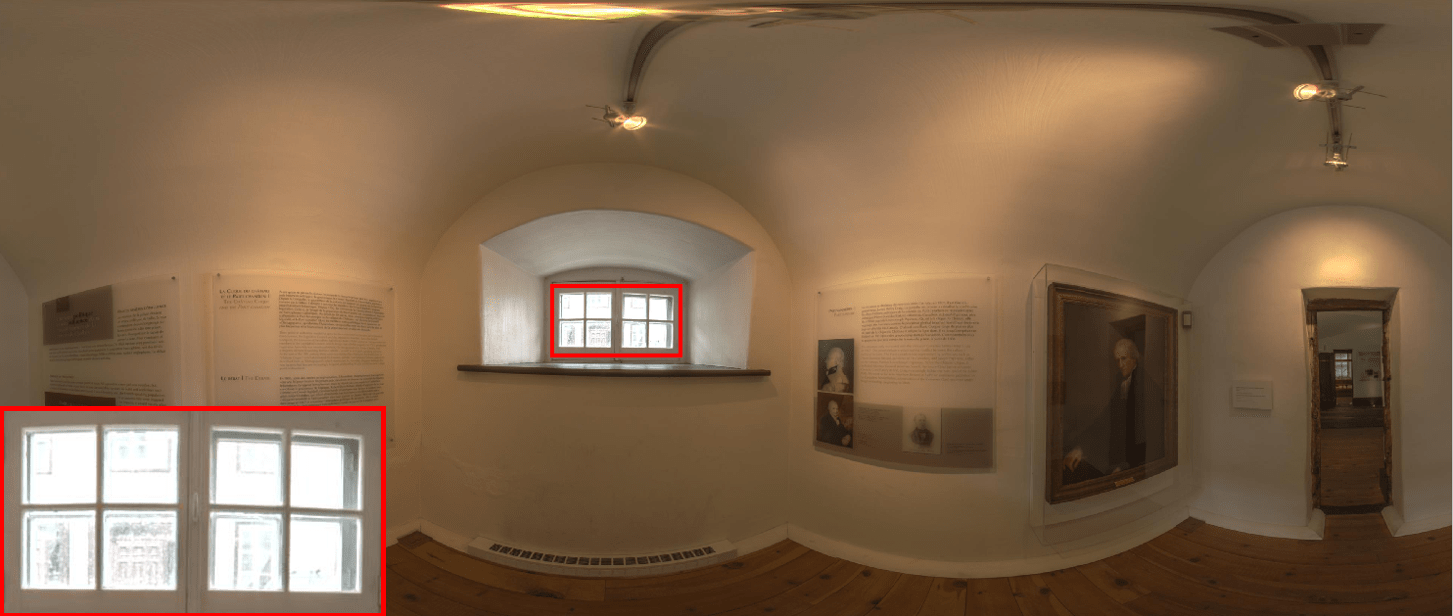}
            \caption[]%
            {{\small Paris \cite{paris2015local} }}    
            \label{fig:mean and std of net34}
        \end{subfigure}
        \begin{subfigure}[b]{0.245\textwidth}   
            \centering 
            \includegraphics[width=\textwidth]{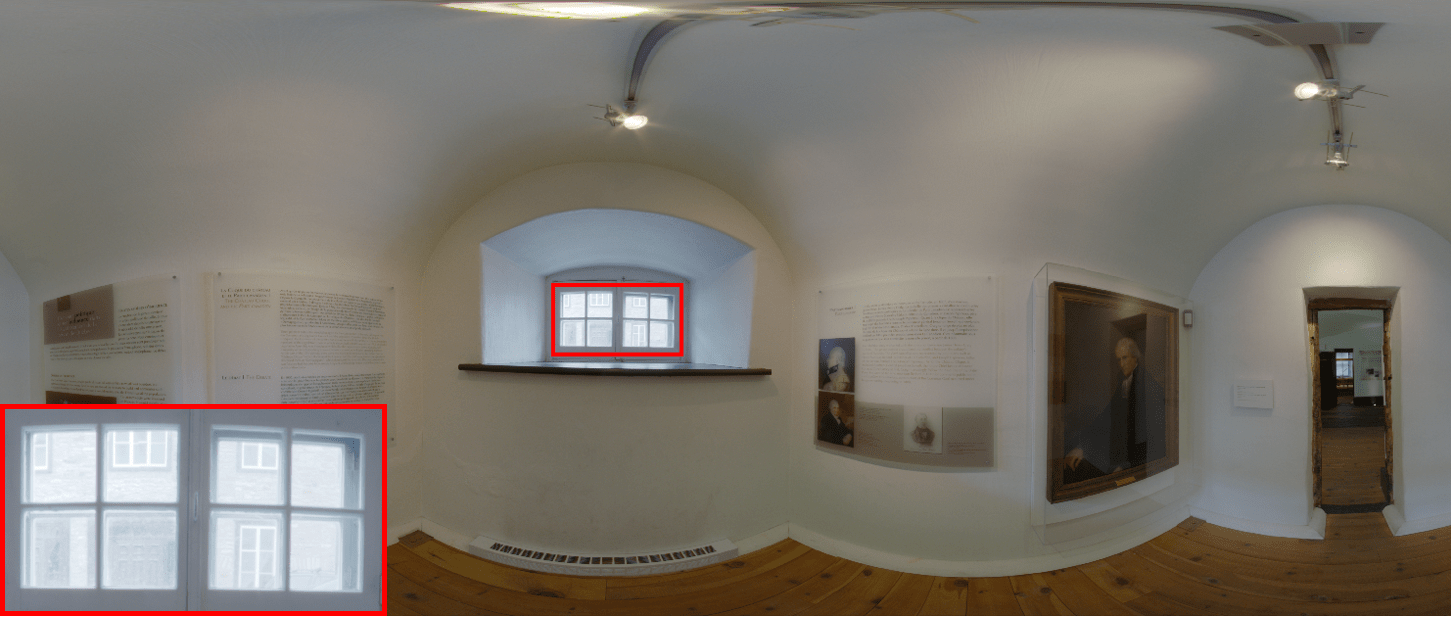}
            \caption[]%
            {{\small Ferradans \cite{ferradans2011analysis}     }}    
            \label{fig:mean and std of net44}
        \end{subfigure}
        \label{fig:mean and std of nets}
        \centering
        \begin{subfigure}[b]{0.245\textwidth}
            \centering
            \includegraphics[width=\textwidth]{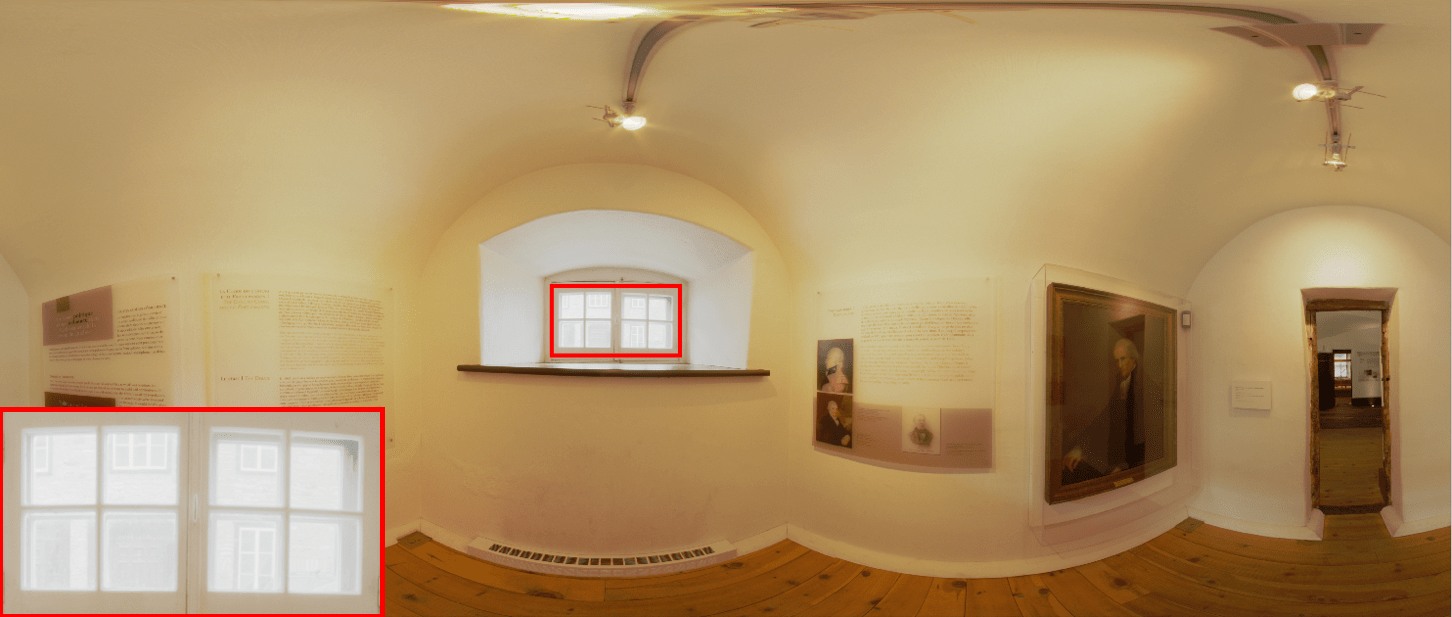}
            \caption[]%
            {{\small Mai \cite{mai2011optimizing} }}    
            \label{fig:mean and std of net14}
        \end{subfigure}
        \begin{subfigure}[b]{0.245\textwidth}
            \centering
            \includegraphics[width=\textwidth]{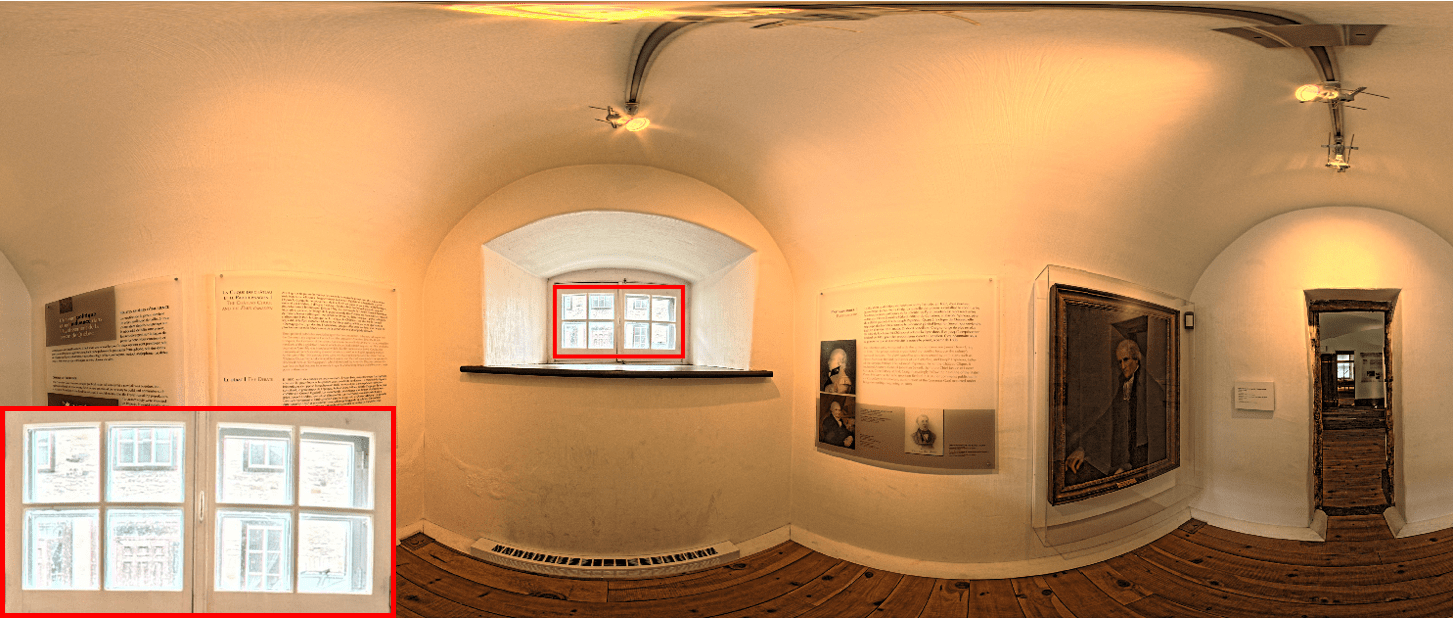}
            \caption[]%
            {{\small Gu \cite{gu2013local} }}    
            \label{fig:mean and std of net14}
        \end{subfigure}
        \begin{subfigure}[b]{0.245\textwidth}   
            \centering 
            \includegraphics[width=\textwidth]{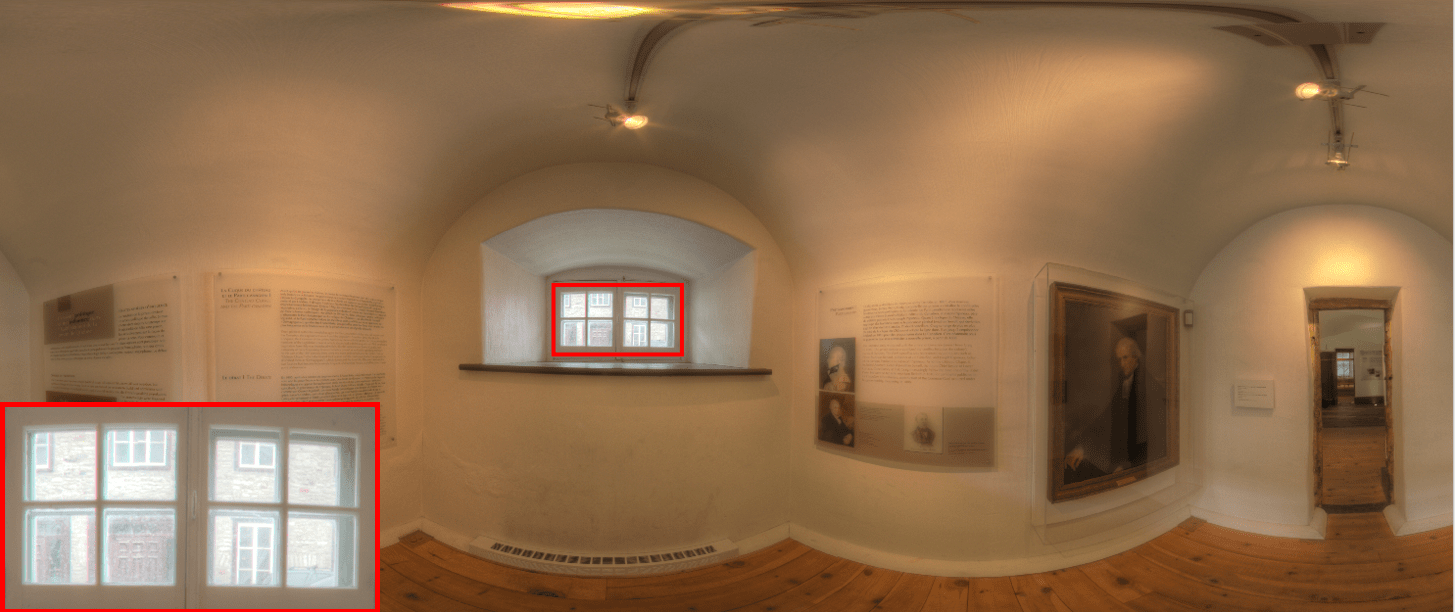}
            \caption[]%
            {{\small Photomatrix \cite{photomatrix} }}    
            \label{fig:mean and std of net34}
        \end{subfigure}
        \begin{subfigure}[b]{0.245\textwidth}   
            \centering 
            \includegraphics[width=\textwidth]{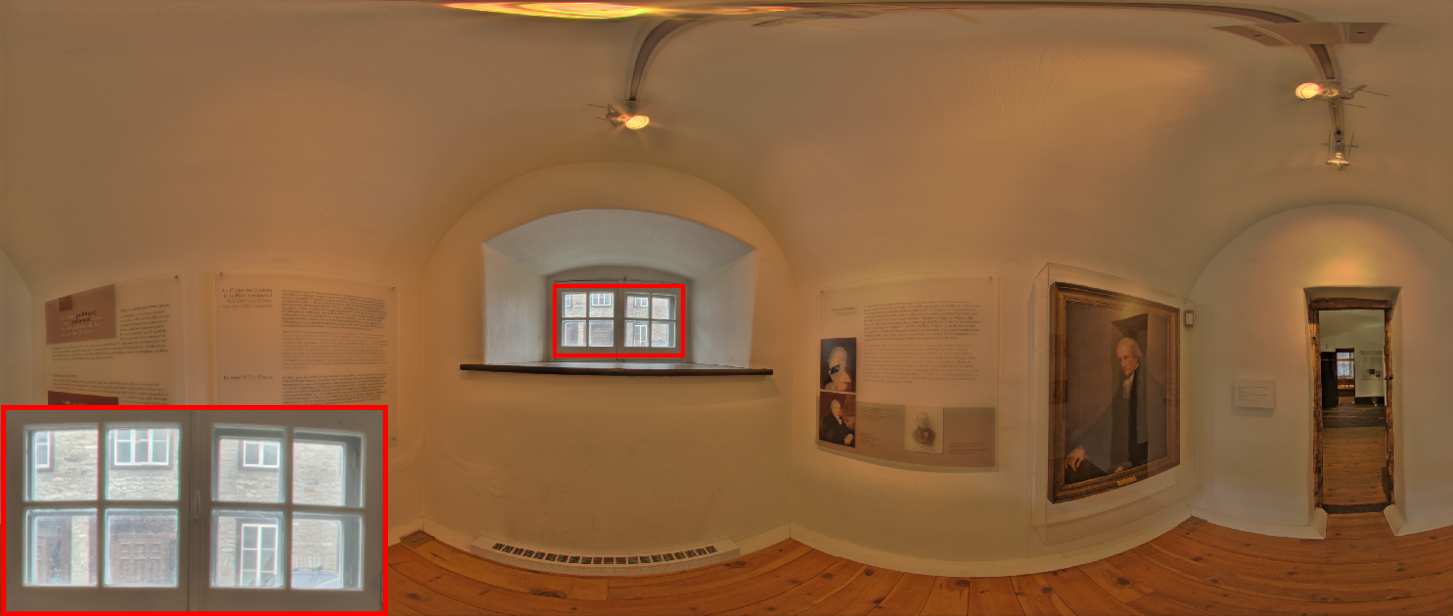}
            \caption[]%
            {{\small Proposed  }}    
            \label{fig:mean and std of net44}
        \end{subfigure}
        \caption{Visual comparison on the test set. The proposed method is able to enhance and recover local details that cannot be seen with other algorithms.}
        \label{fig:Laval_database_result}
\end{figure*}

\begin{table}[tb]
\footnotesize
\begin{center}
\caption{Average PSNR, SSIM and FSITM values computed for models with different $n$ values. Median $n$ values are able to achieve relatively higher indices when compared with the two end values.}
\begin{tabular}{c|c|c|c}
\hline
Parameter & PSNR (dB) & SSIM  &FSITM\\ \hline
     $n  = 2$       &  17.4065  & 0.8586  & 0.8983\\
     $n  = 3$       &  17.2129  & 0.8475  & 0.9209\\
     $n  = 4$       &  19.2187  & 0.8417  & 0.9212\\
     $n  = 5$       &  19.6627  & 0.8782  & 0.9321\\
     $n  = 6$       &  \textbf{20.0335}  & \textbf{0.8948}  & \textbf{0.9378}\\
     $n  = 7$       &  16.7248  & 0.7996  & 0.9195\\ \hline
\end{tabular}
\label{table:PSNR}
\end{center}
\end{table}

\begin{table}[tb]
\footnotesize
\begin{center}
\caption{Comparison of the average running time on an image with size $3884\times1650$. \cite{paris2015local} requires more than one hour of processing time.}
\begin{tabular}{l|c}
\hline
Methods & Time(s) \\ \hline
\cite{gu2013local} & 2.4 \\ \hline
\cite{mantiuk2008display} & 1.7 \\ \hline
\cite{ferradans2011analysis} & 8.2 \\ \hline
\cite{mai2011optimizing} & 0.7 \\ \hline
\cite{photomatrix} & 3.7 \\ \hline
Ours & 0.61(with GPU)  \\ \hline
\end{tabular}
\label{table:running_time}
\end{center}
\end{table}

\subsection{Running Time}
We report the processing time of each algorithm in Table \ref{table:running_time}. We evaluated all methods on a PC with Intel(R) Core(TM) i5-8600 CPU 3.10GHz, 16G memory.  We used one HDR image of size $3884\times1650$ as input.  Note that learning-based TM solutions are designed under GPU environment by convention since they run often significantly slower than other TM approaches under CPU environment.  Our model runs 24.95 seconds with CPU and 0.61 seconds with GPU (Nvidia Titan Xp).

\begin{figure*}[tb]
        \centering
        \begin{subfigure}[b]{0.245\textwidth}
            \centering
            \includegraphics[width=\textwidth]{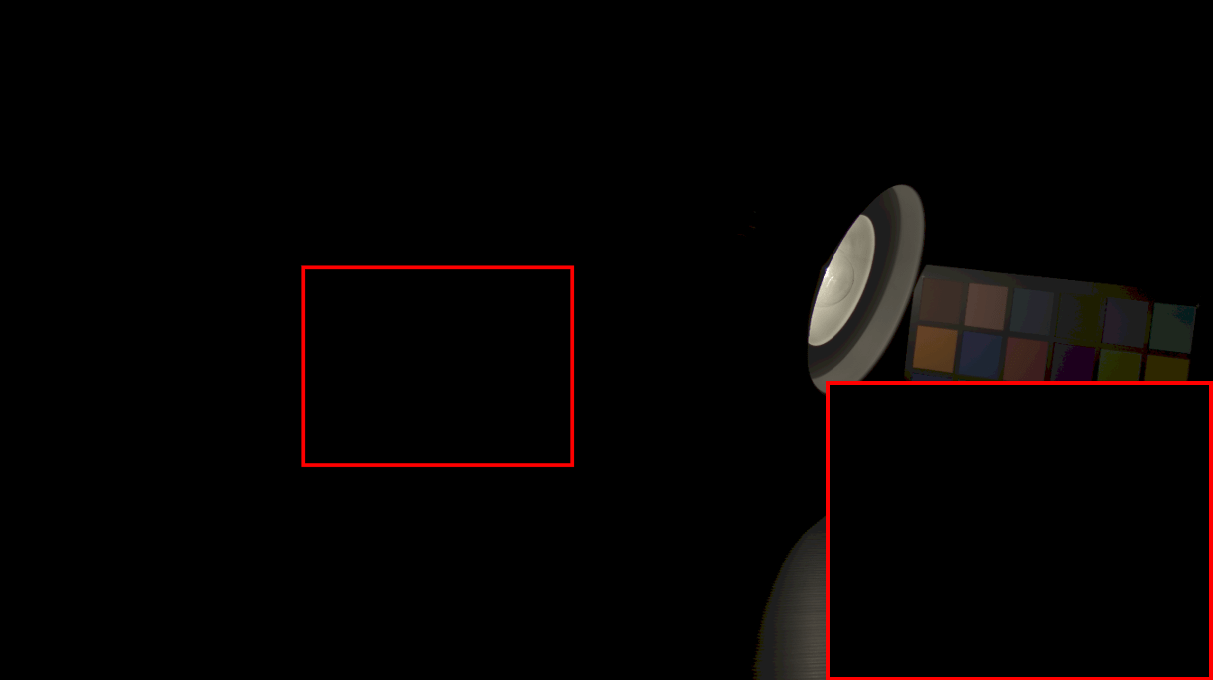}
            \caption[]%
            {{\small Gamma correction }}    
            \label{fig:mean and std of net14}
        \end{subfigure}
        \begin{subfigure}[b]{0.245\textwidth}  
            \centering 
            \includegraphics[width=\textwidth]{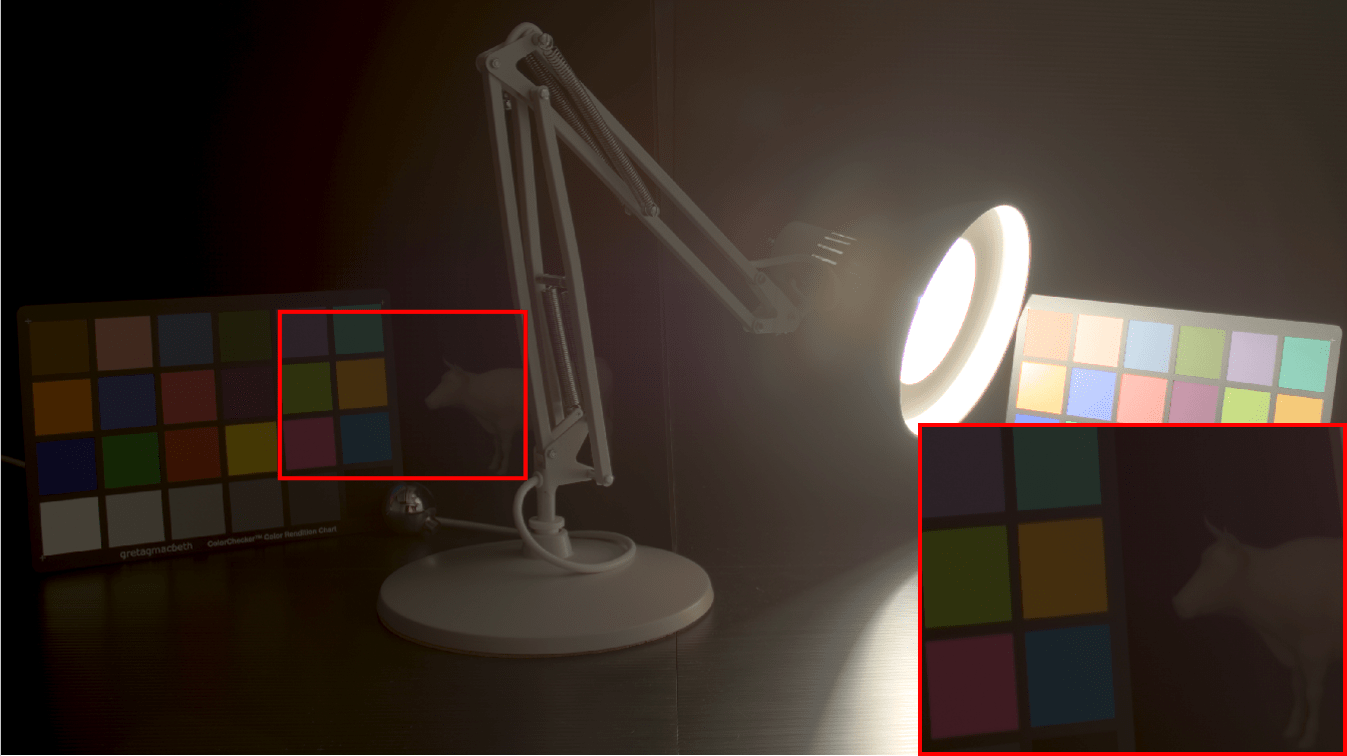}
            \caption[]%
            {{\small Mantuik \cite{mantiuk2008display} }}    
            \label{fig:mean and std of net24}
        \end{subfigure}
        \begin{subfigure}[b]{0.245\textwidth}   
            \centering 
            \includegraphics[width=\textwidth]{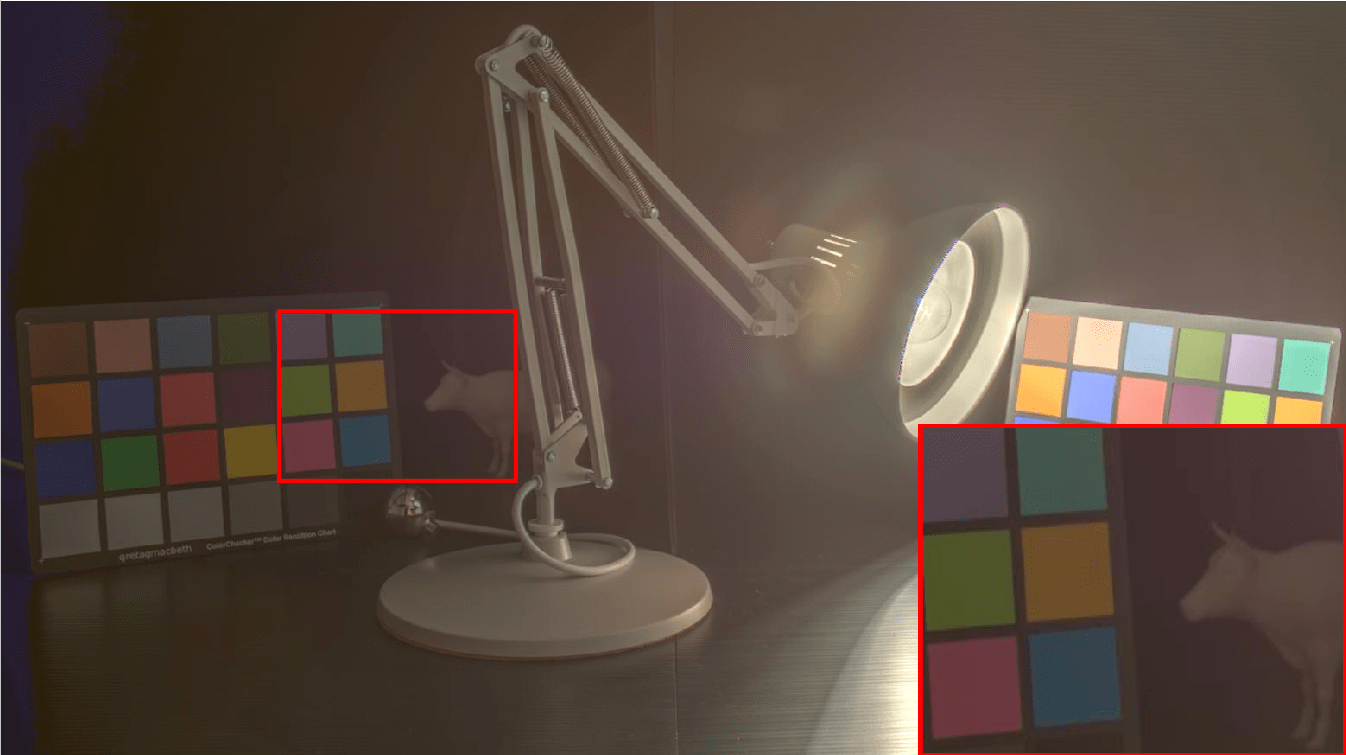}
            \caption[]%
            {{\small Paris\cite{paris2015local} }}    
            \label{fig:mean and std of net34}
        \end{subfigure}
        \begin{subfigure}[b]{0.245\textwidth}   
            \centering 
            \includegraphics[width=\textwidth]{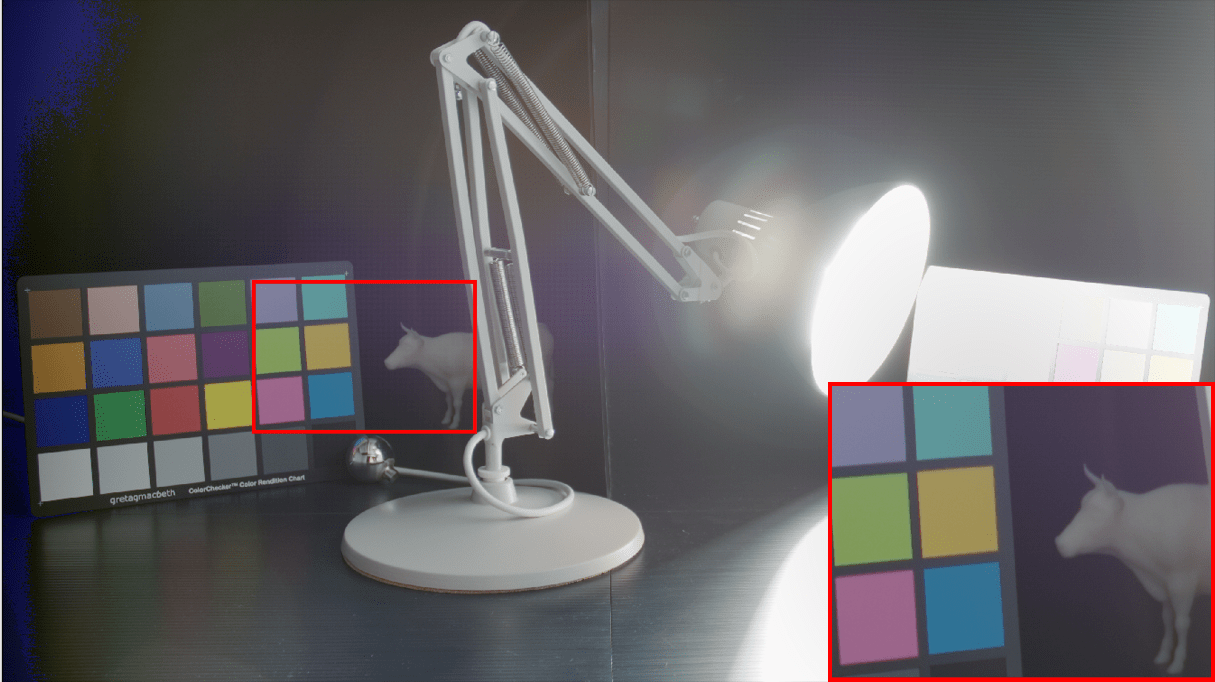}
            \caption[]%
            {{\small Ferradans \cite{ferradans2011analysis} }}    
            \label{fig:mean and std of net44}
        \end{subfigure}
      
        \label{fig:mean and std of nets}
        \centering
        \begin{subfigure}[b]{0.245\textwidth}
            \centering
            \includegraphics[width=\textwidth]{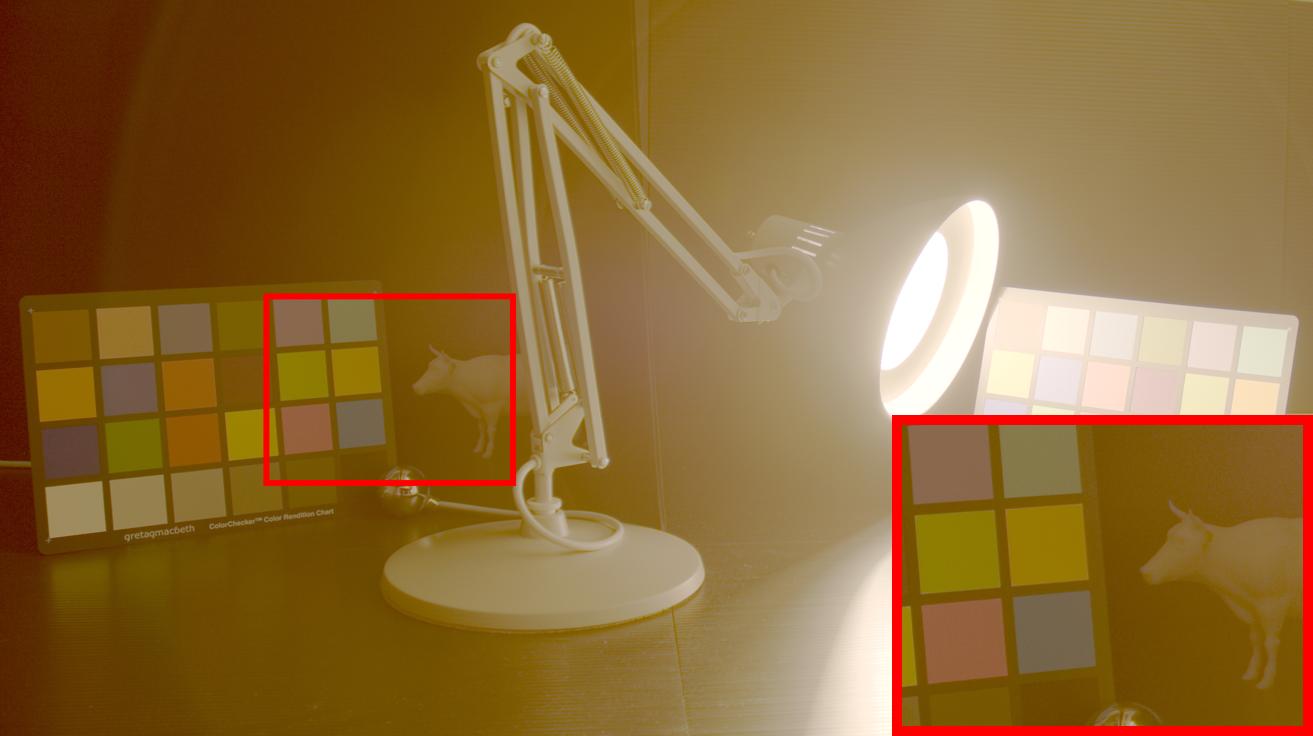}
            \caption[]%
            {{\small Mai \cite{mai2011optimizing} }}    
            \label{fig:mean and std of net14}
        \end{subfigure}
        \begin{subfigure}[b]{0.245\textwidth}  
            \centering 
            \includegraphics[width=\textwidth]{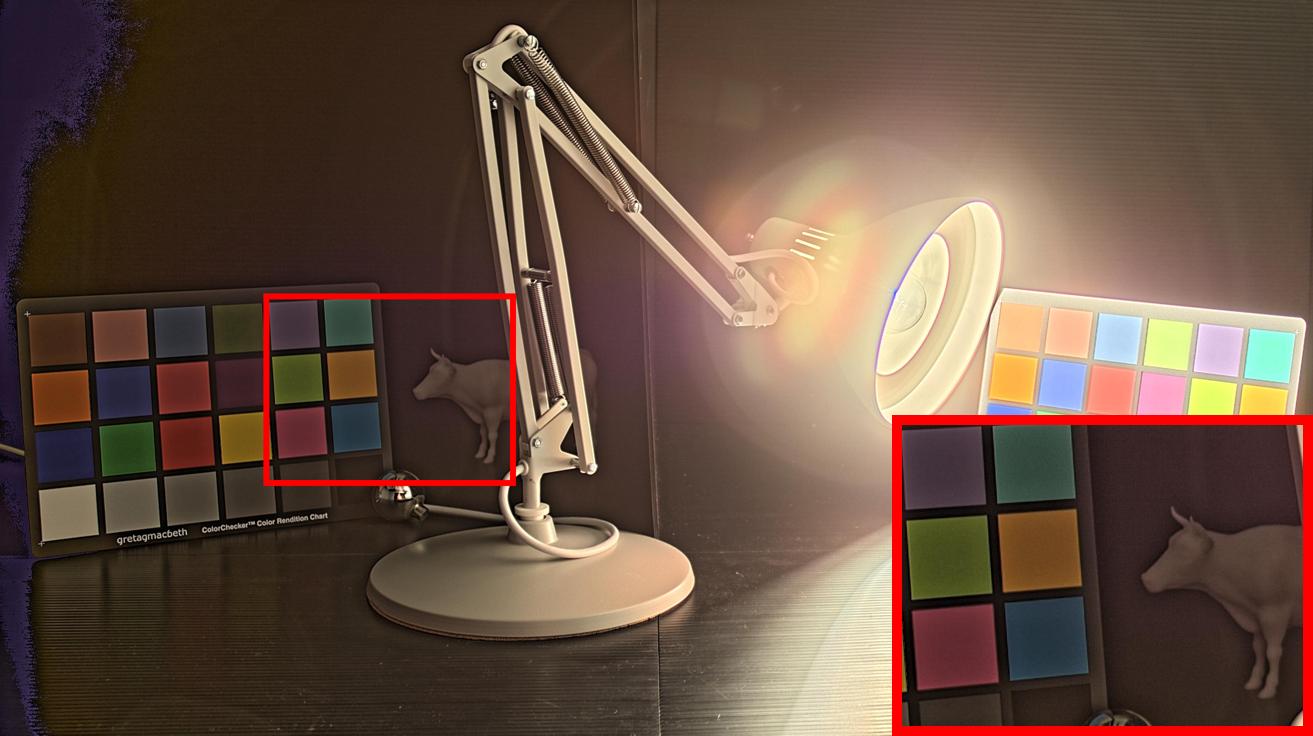}
            \caption[]%
            {{\small Gu \cite{gu2013local} }}    
            \label{fig:mean and std of net24}
        \end{subfigure}
        \begin{subfigure}[b]{0.245\textwidth}   
            \centering 
            \includegraphics[width=\textwidth]{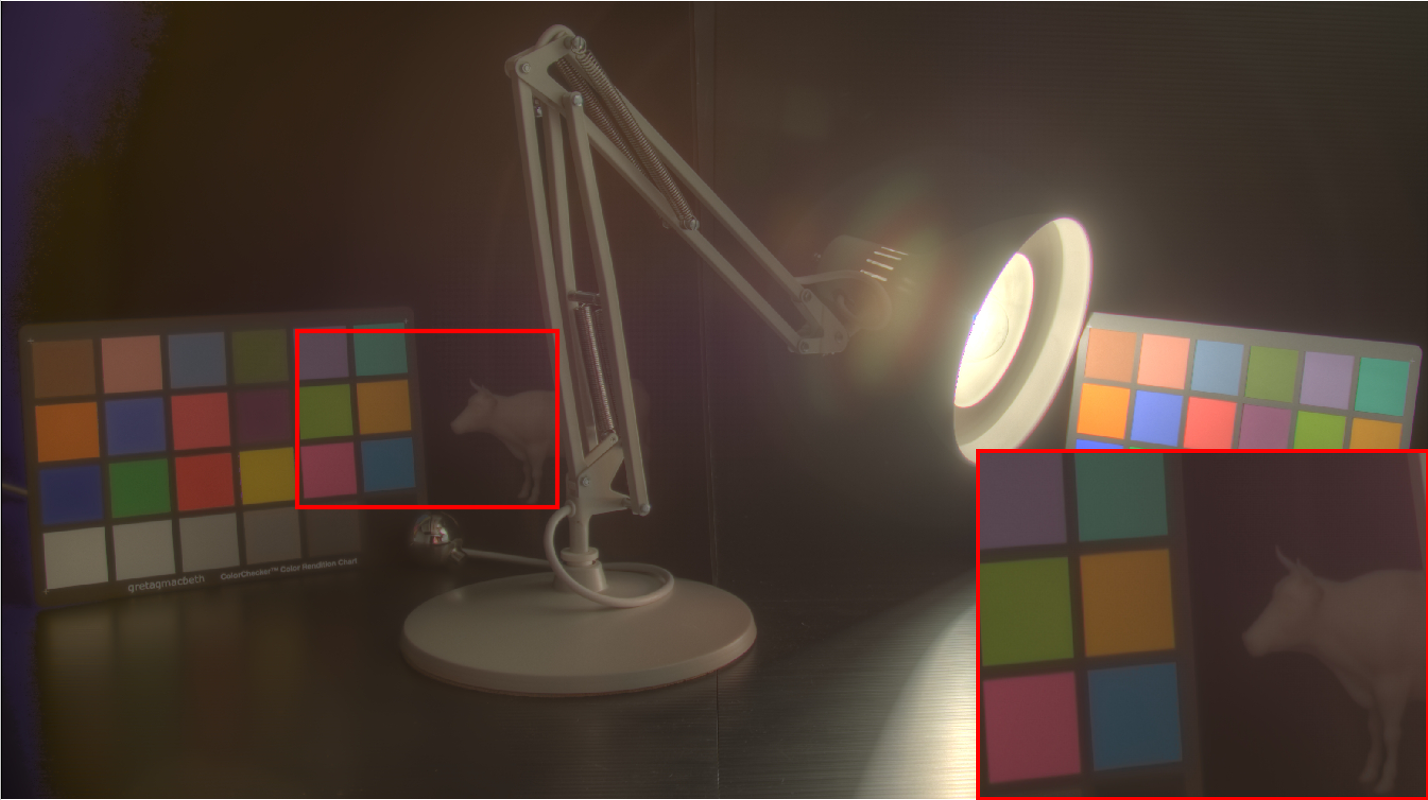}
            \caption[]%
            {{\small Photomatix \cite{photomatrix} }}    
            \label{fig:mean and std of net34}
        \end{subfigure}
        \begin{subfigure}[b]{0.245\textwidth}   
            \centering 
            \includegraphics[width=\textwidth]{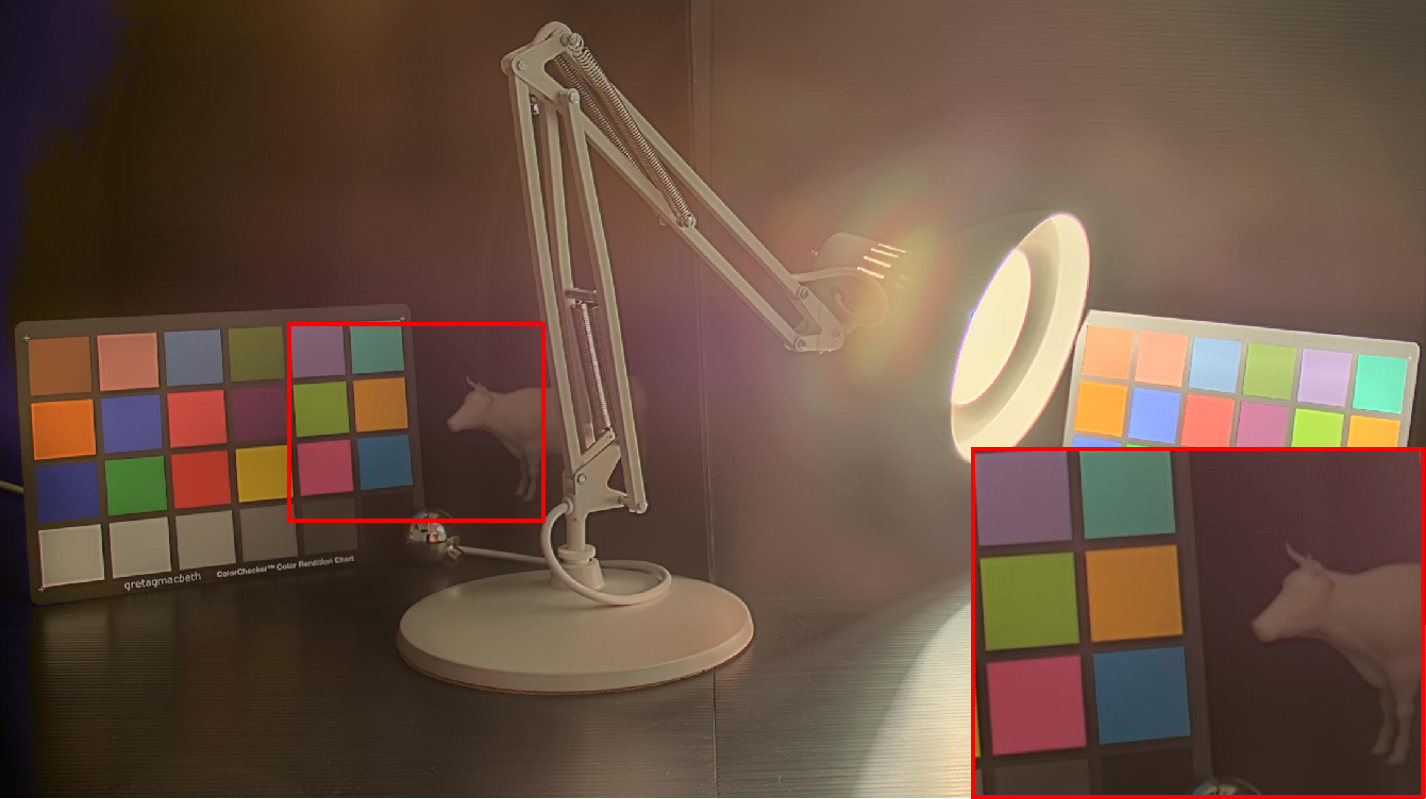}
            \caption[]%
            {{\small Proposed  }}    
            \label{fig:mean and std of net44}
        \end{subfigure}
        \label{fig:mean and std of nets}

        
        \centering
        \begin{subfigure}[b]{0.245\textwidth}
            \centering
            \includegraphics[width=\textwidth]{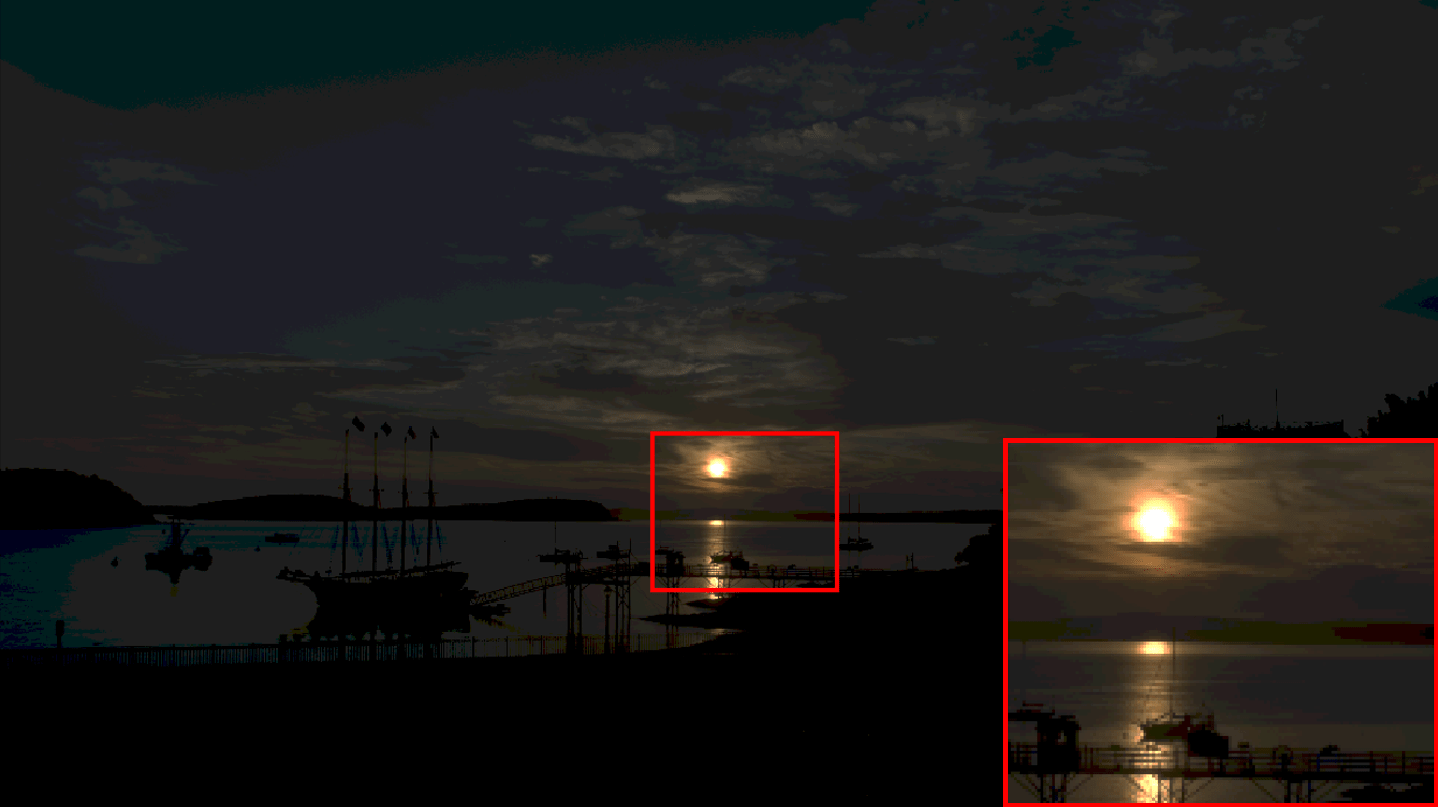}
            \caption[]%
            {{\small Gamma correction }}    
            \label{fig:mean and std of net14}
        \end{subfigure}
        \begin{subfigure}[b]{0.245\textwidth}  
            \centering 
            \includegraphics[width=\textwidth]{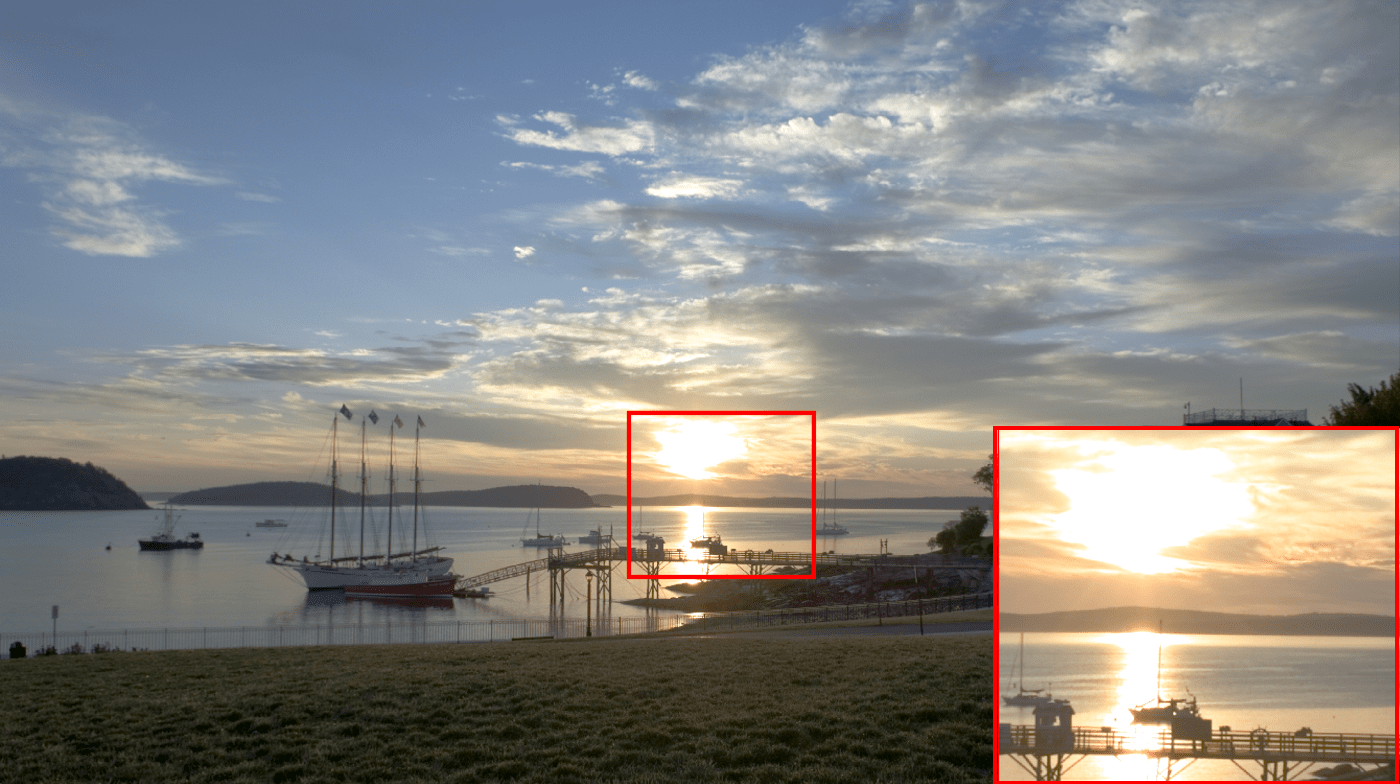}
            \caption[]%
            {{\small Mantiuk \cite{mantiuk2008display} }}    
            \label{fig:mean and std of net24}
        \end{subfigure}
        \begin{subfigure}[b]{0.245\textwidth}   
            \centering 
            \includegraphics[width=\textwidth]{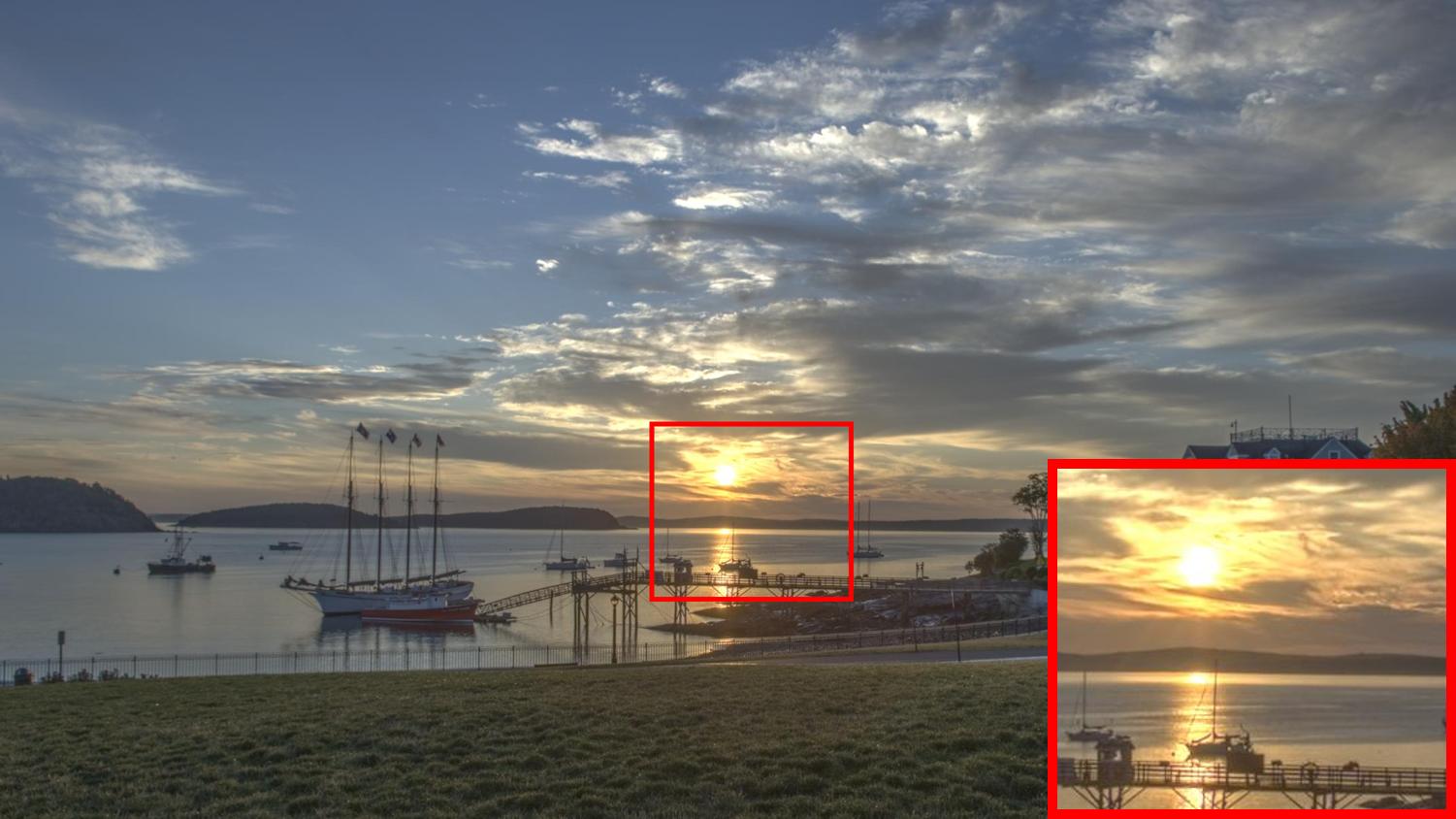}
            \caption[]%
            {{\small Paris\cite{paris2015local} }}    
            \label{fig:mean and std of net34}
        \end{subfigure}
        \begin{subfigure}[b]{0.245\textwidth}   
            \centering 
            \includegraphics[width=\textwidth]{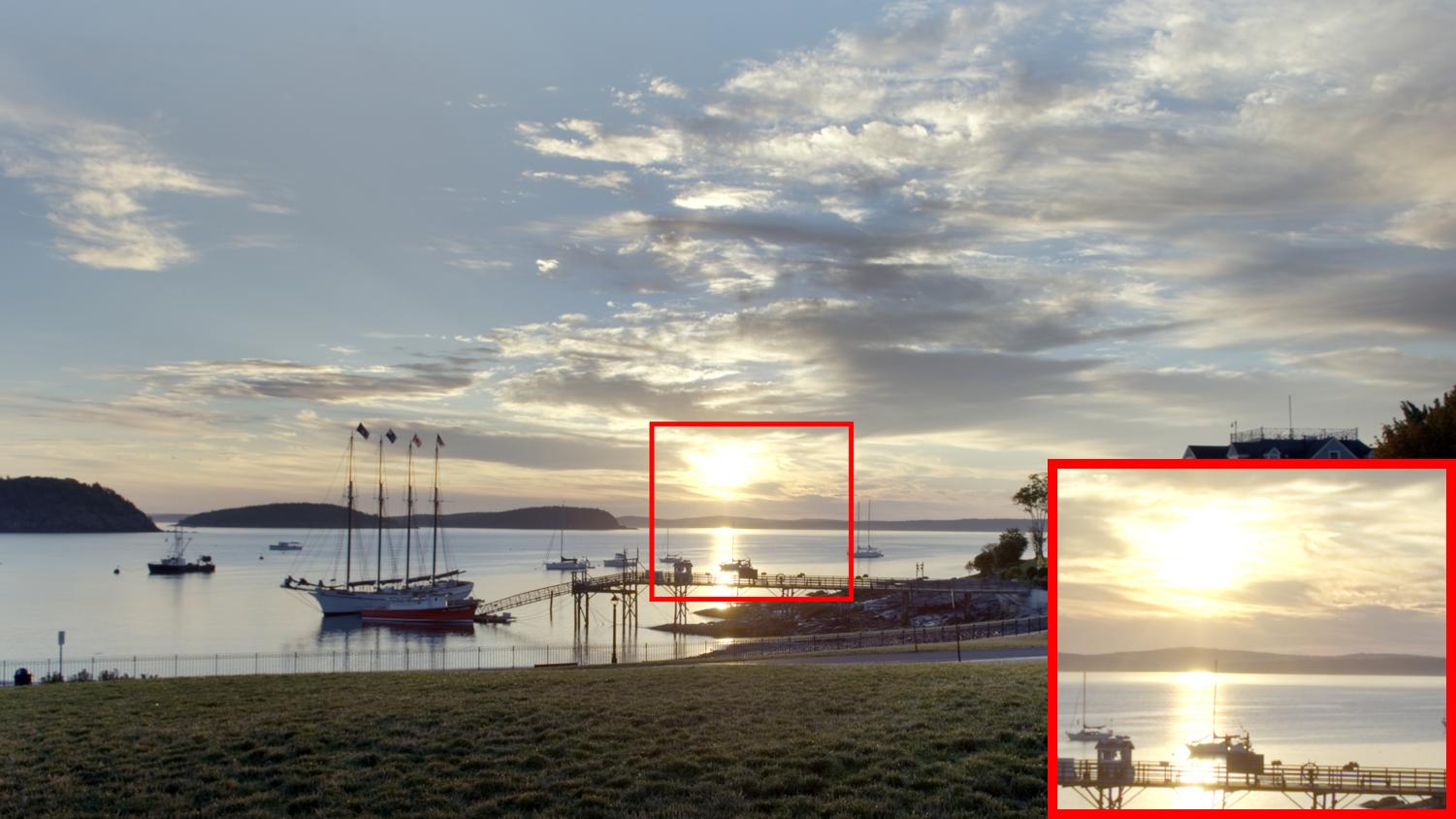}
            \caption[]%
            {{\small Ferradans \cite{ferradans2011analysis}     }}    
            \label{fig:mean and std of net44}
        \end{subfigure}
        \label{fig:mean and std of nets}
        \centering
        \begin{subfigure}[b]{0.245\textwidth}
            \centering
            \includegraphics[width=\textwidth]{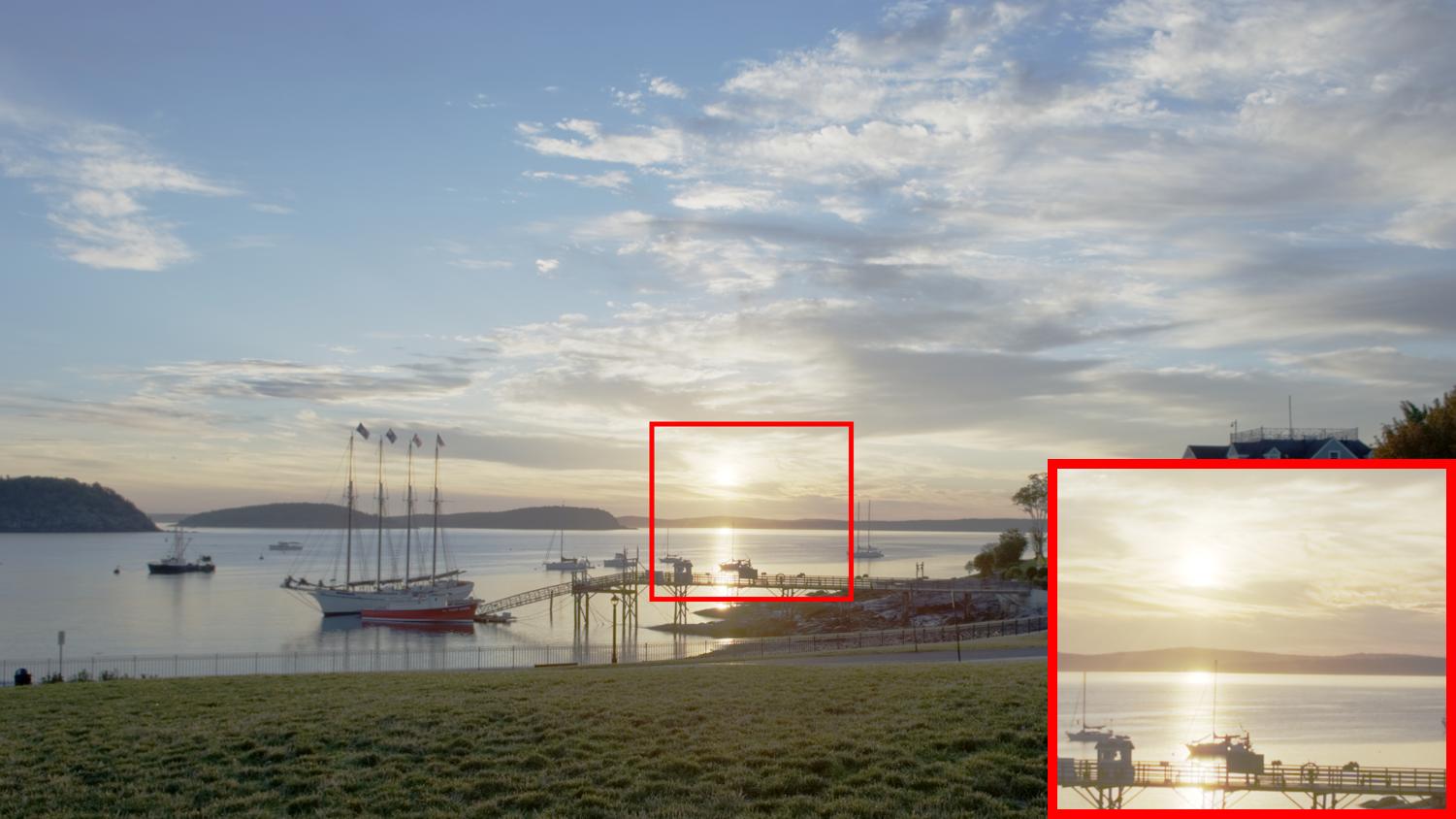} 
            \caption[]%
            {{\small Mai \cite{mai2011optimizing} }}    
            \label{fig:mean and std of net14}
        \end{subfigure}
        \begin{subfigure}[b]{0.245\textwidth}  
            \centering 
            \includegraphics[width=\textwidth]{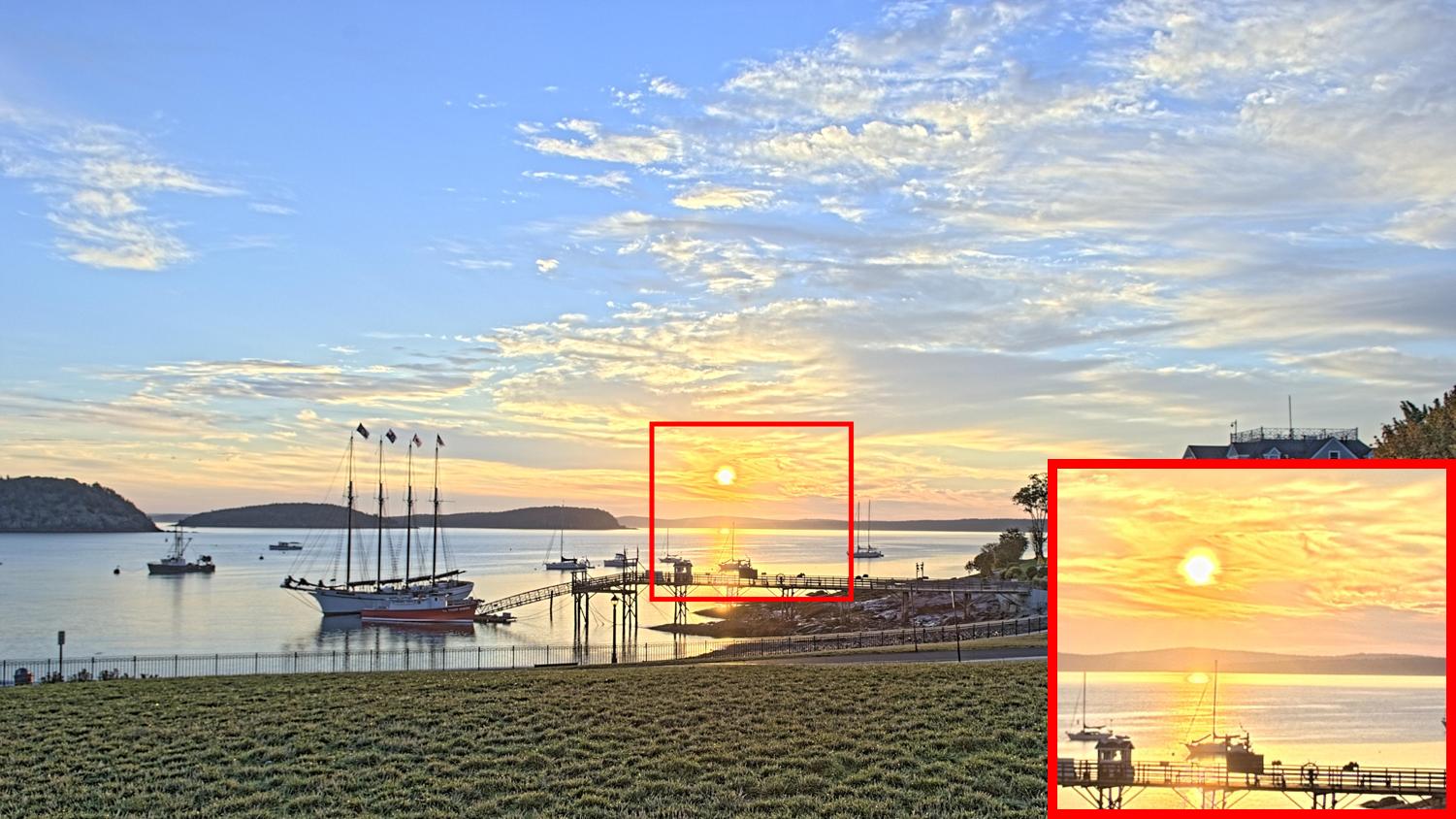}
            \caption[]%
            {{\small Gu \cite{gu2013local} }}    
            \label{fig:mean and std of net24}
        \end{subfigure}
        \begin{subfigure}[b]{0.245\textwidth}   
            \centering 
            \includegraphics[width=\textwidth]{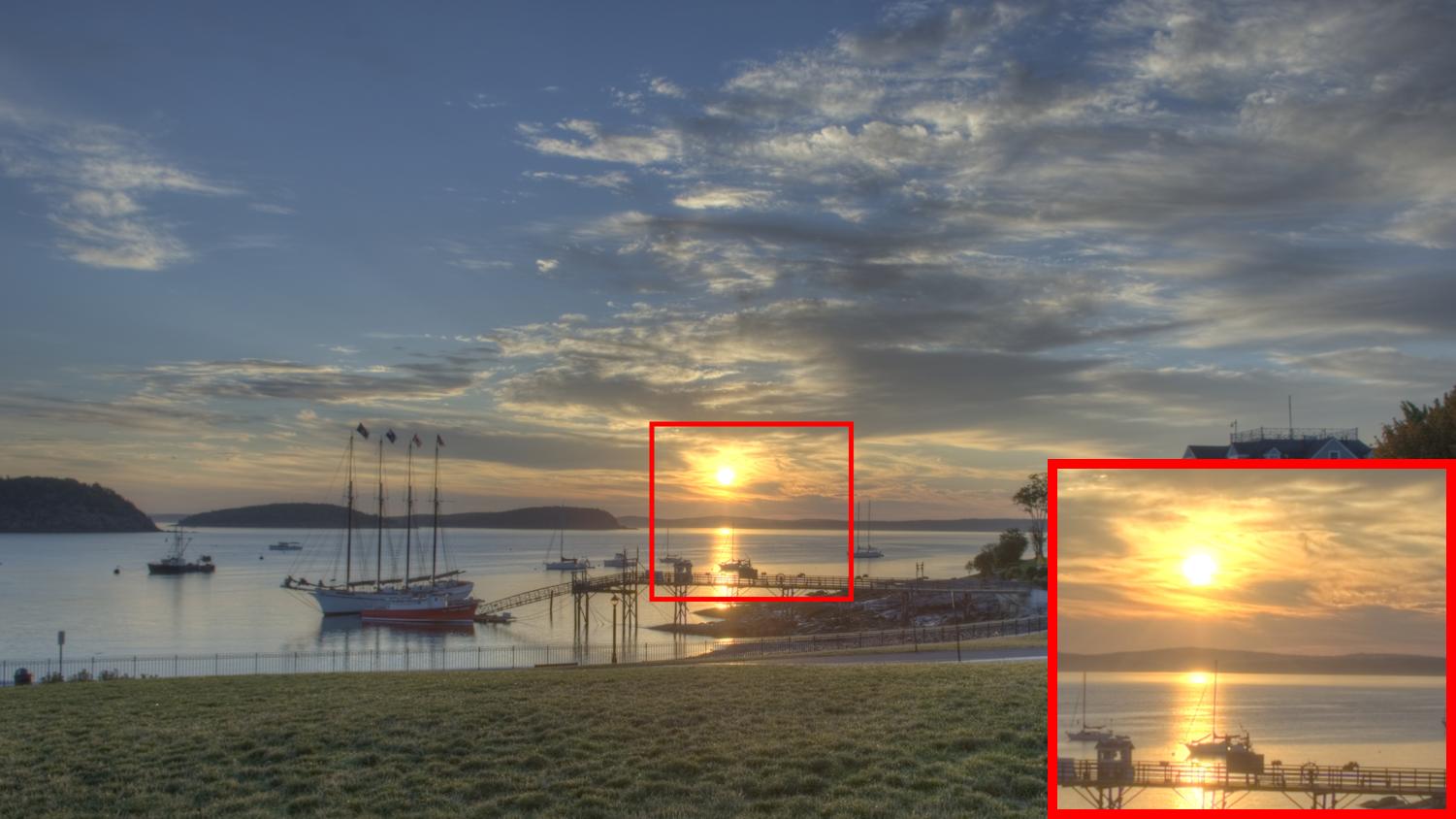}
            \caption[]%
            {{\small Photomatrix\cite{photomatrix} }}    
            \label{fig:mean and std of net34}
        \end{subfigure}
        \begin{subfigure}[b]{0.245\textwidth}   
            \centering 
            \includegraphics[width=\textwidth]{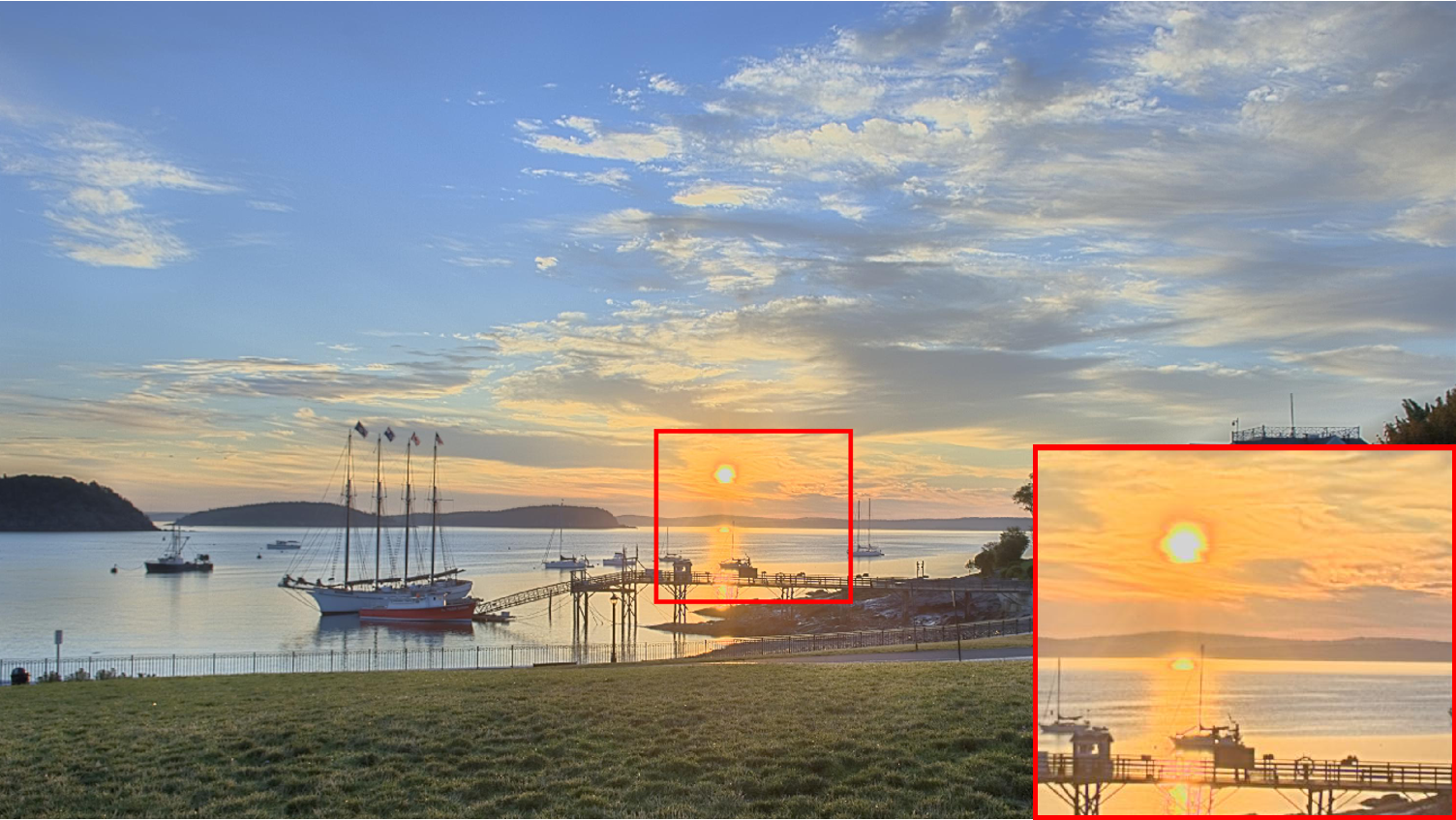}
            \caption[]%
            {{\small Proposed  }}    
            \label{fig:mean and std of net44}
        \end{subfigure}
        \caption{Visual comparison on the Fairchild database. The proposed model renders more detail in saturated regions with less artifacts when compared with other state-of-the-art approaches.}
        \label{fig:fairchild_result}
\end{figure*} 

\subsection{Comparison With State-of-the-art Methods}

We compare the proposed model with other 6 state-of-the-art image tone mapping methods, namely, Gu TMO \cite{gu2013local}, Mantiuk TMO \cite{mantiuk2008display}, Paris TMO \cite{paris2015local}, Ferradans TMO \cite{ferradans2011analysis}, Mai TMO \cite{mai2011optimizing}, and Photomatix TMO \cite{photomatrix}. Among them, Gu TMO \cite{gu2013local} is based on edge preserving filter theory, images tone mapped with this TMO usually exhibit more detail. Paris TMO \cite{paris2015local} is based on local Laplacian operator and it is good at preserving details from the WDR image. Mantiuk TMO \cite{mantiuk2008display} regards tone mapping as an optimization problem, though it has some difficulty in preserving detail, it can give very natural looking image results.  Ferradans TMO \cite{ferradans2011analysis} and  Mai TMO \cite{mai2011optimizing} are popular tone mapping methods and are used in open source applications. Photomatix is a commercial software dedicated to WDR image tone mapping. 

The following results of the mentioned algorithms are obtained from their online websites or open source projects with default parameter settings. We evaluate the model on the test dataset first and then on a totally different database.

\subsubsection{Objective Quality Assessment} 
We first compared our algorithm with these methods on images from the test set of the Laval database. The results on one image are shown in Figure \ref{fig:laval_global_compress}. All methods are able to produce acceptable images. However, they also have different problems. 

For example, images obtained with Mantiuk TMO, Paris TMO, Ferradans TMO, and Photomatix TMO are generally darker than other images which give them a disadvantage for screening and human observing. Mai TMO generates the brightest image but it also saturates the area within the red rectangle. Both Gu TMO and the proposed method are able to generate images that are similar to the reference image. However, the image from the proposed TMO looks more natural than the image of Gu TMO because of the global luminance distribution. Moreover, it can display more local detail in the floor when compared with the Gu TMO.

Actually, the proposed model is able to extract local detail even under some extreme conditions. Two examples are given in Figure \ref{fig:Laval_database_result}. The two images show a commonly seen WDR scenario where there is extreme luminance difference between inside and outside window area. The proposed model can still display the scenes outside the window more clearly than other results including the reference images.

\begin{table}[tb]
\footnotesize
\begin{center}
\caption{PSNR, SSIM, FSITM and HDR-VDP2 indices for different tone mapping methods. The values are obtained using the test data set.}
\begin{tabular}{l|c|c|c|c}
\hline
 Methods & PSNR & SSIM & FSITM & HDR-VDP2 \\ \hline
Gu \cite{gu2013local} & 16.5024 & 0.7755 & 0.830 & 35.165 \\
Mantiuk \cite{mantiuk2008display} & 14.8641 & 0.7563 & \textbf{0.876} & 38.864 \\
Paris \cite{paris2015local} & 18.3214 & 0.8443 & 0.853  & 39.598 \\
Ferradans \cite{ferradans2011analysis} & 16.7242 & 0.8756 & 0.863 & 39.219 \\
Mai \cite{mai2011optimizing} & 19.4638 & 0.8842 & 0.858 & 39.346 \\
Photomatrix \cite{photomatrix} & 17.0730 & 0.8565 & 0.856 & 42.194 \\
 Ours & \textbf{20.0335} & \textbf{0.8948} & 0.864 & \textbf{42.215} \\ \hline
\end{tabular}
\label{table:laval_database_PSNR}
\end{center}
\end{table}

\begin{table}[tb]
\footnotesize
\begin{center}
\caption{Average TMQI, BTMQI, FSITM and HDR-VDP2 indices for different tone mapping methods. The values are obtained using the Fairchild database. }
\begin{tabular}{l|c|c|c|c}
\hline
 Methods & TMQI & BTMQI & FSITM & HDR-VDP2 \\ \hline
Gu \cite{gu2013local} & 0.8300 & 3.6683 &  0.823 & 26.161 \\
Mantiuk \cite{mantiuk2008display} & 0.9194 & 3.7474 & \textbf{0.872} & 27.686 \\
Paris \cite{paris2015local} & 0.9228 & 2.8849 & 0.857 & 26.747 \\
Ferradans \cite{ferradans2011analysis} & 0.8418 & 4.4286 & 0.857 & 27.35 \\
Mai \cite{mai2011optimizing} & 0.8894 & 3.8959 & 0.859 & 27.464 \\
Photomatrix \cite{photomatrix} & 0.8978 & 3.3978 & 0.869 & 27.226 \\
 Ours & \textbf{0.9257} & \textbf{4.5110} & 0.868 & \textbf{29.323} \\ \hline
\end{tabular}
\label{table:fairchild_metrices}
\end{center}
\end{table}

PSNR, SSIM, FSITM and HDR-VDP2 \cite{mantiuk2011hdr} are employed to assess these algorithms quantitatively. FSITM is designed to evaluate the feature similarity index for tone-mapped images.  We measured the quality of the images using HDR-VDP2, the visual metric that mimics the anatomy of the HVS to evaluate the quality of HDR images.  The average indices obtained from the test set of Laval database is summarized in Table \ref{table:laval_database_PSNR}. In all four metrics, the proposed algorithm is able to achieve the highest scores for PSNR, SSIM and HDR-VDP2. For the FSITM, the algorithm achieves the second highest score.   


To further demonstrate the robustness of our method, we evaluate our method on images outside the test set. We choose Fairchild database \cite{fairchild2007hdr} which contains 105 WDR images containing various situations. It is a commonly used benchmark for measuring tone mapping methods. Two result images are shown in Figure \ref{fig:fairchild_result}. The first image has a very wide dynamic range in front of and behind the lamp. Mantiuk TMO, Ferrandans TMO, Mai TMO, and the proposed models are able to generate images while other algorithms show various color artifacts in the top left dim region. In the four images without artifacts, the proposed model is able to show the color boards clearly under both dark and bright lighting conditions. In the second image, only Gu TMO and the proposed model are able to show clearly the shape of the sun. Images obtained with other methods cannot show this detail. Since there are no reference images for Fairchild database, we use the blind quality indexes alone with TMQI \cite{Hojatollah2013Objective} to quantitatively measure the performance of different methods.  Blind quality assessment of tone mapped images (BTMQI) \cite{Ke2016Blind} are two blind metrics that do not require a reference image to compute a metric score.  The computed indices are summarized in Table \ref{table:fairchild_metrices}. The proposed model is able to achieve the highest TMQI and BTMQI scores.


\begin{table}[tb]
\footnotesize
\begin{center}
\caption{The summary of the subjective experimental results. $Sum$ indicates the total number of each TMO being selected in the \textit{Comparative Selection} Section.  $Ave_{selection}$ represents the average number of each TMO being selected in the \textit{Comparative Selection} Section.  $Ave_{rating}$ represents the average rating score of each TMO in the \textit{Image Quality Rating} Section.  The computational and statistical details are elaborated in the Supplementary material.}
\begin{tabular}{l|c|c|c}\hline
 & $Sum$ & $Ave_{selection}$ & $Ave_{rating}$ \\ \hline
Ferradans \cite{ferradans2011analysis} & 69 & 8.625 & 6.184 \\ \hline
Gu \cite{gu2013local} & 111 & 13.875 & 6.403 \\ \hline
Mai \cite{mai2011optimizing} & 65 & 8.125 & 6.571 \\ \hline
Mantiuk \cite{mantiuk2008display} & 54 & 6.75 & 6.079 \\ \hline
Paris \cite{paris2015local} & 52 & 6.5 & 5.586 \\ \hline
Photomatrix \cite{photomatrix} & 39 & 4.875 & 5.642 \\ \hline
Ours & \textbf{112} & \textbf{14} & \textbf{6.580} \\ \hline

\end{tabular}
\label{table:subjective_survey}
\end{center}
\end{table}

\subsubsection{Subjective Preference Assessment} 
We used human-preferred tuning to generate our ground truth images. In this process, no metrics were employed because no golden standard metrics exist to evaluate the quality of a tone mapped image nor do the metrics truly reflect the observers' preference. These metrics can only be used as limited references. To assess the actual visual experience of our tone mapped LDR images, we also carried out a subjective preference experiment beyond the set of objective indices measurement to assess the visual quality of the image generated from our deep learning solution. 

The subjective experiment has two sections: \textit{Comparative Selection} Section and \textit{Image Quality Rating} Section. Each section contains 8 groups of tone mapped LDR images.  In the Comparative Selection Section, each group contains the LDR images generated by all 7 algorithms from the same scene for visual comparison. Participants were asked to select one image of their visual preference.  In the Image Quality Rating Section, participants were asked to rate each LDR image from $1$ to $10$ based on the degree of visual comfort and the degree of details revealed in the image.   Details refer to the brightness, contrast, and the extent to which overexposure and underexposure details are revealed.  $1$ represents ``dislike" or ``fuzzy details" and $10$ stands for ``most favorite" or ``clear and rich details".  $5$ is somewhat in the middle. All images were randomly selected from Laval HDR dataset.

We used SurveyHero\footnote{https://www.surveyhero.com}  website to build our subjective experiment project.  We send the survey invitation randomly via email, social networking websites, and application (see appendix for the experiment details).  The survey result is shown in Table \ref{table:subjective_survey}.  In terms of visual experience that meant to be tested in the Comparative Selection Section, the voting results of our algorithm and Gu's algorithm are significantly better than other approaches. The good visual experience comes from the low frequency layer that our Global Compression Network can effectively learn from the global luminance terrain of the human-tuned ground truth image. In the Image Quality Rating Section, the Local Manipulation Network in our model can extract and enhance the local high frequency details, therefore avoiding the lack of details in the overexposed and underexposed areas of the image after tone mapping. Our result achieves the highest voting score in both sections.  

\section{Conclusion}
In this work, we have proposed a new tone mapping method that can perform high-resolution WDR image tone mapping. 
To preserve the global low frequency feature as well as maintain local high frequency detail, we have proposed a novel reformulated Laplacian method to decompose a WDR image into a low-resolution image which contains the low frequency component of the WDR image and a high-resolution image which contains the remaining higher frequencies of the WDR image. The two images are processed by a dedicated global compression network and a local manipulation neural network, respectively. The global compression network learns how to compress the global scale gradient of a WDR image and the local manipulation network learns to manipulate local features. The generated images from the two networks are further merged together to produce the final output image for screen display. We visually and quantitatively compared our model with other state-of-the-art tone mapping methods with images from and outside the targeted database. The results showed that the proposed method outperforms other methods, and sometimes even shows better results than the ground truth.

\ifCLASSOPTIONcaptionsoff
  \newpage
\fi



%

\bibliographystyle{IEEEtran}
\bibliography{references}

\clearpage
\appendix

\subsection{Comparative Selection}
\begin{figure*}[!h]
    \centering
    \includegraphics[scale = 0.32]{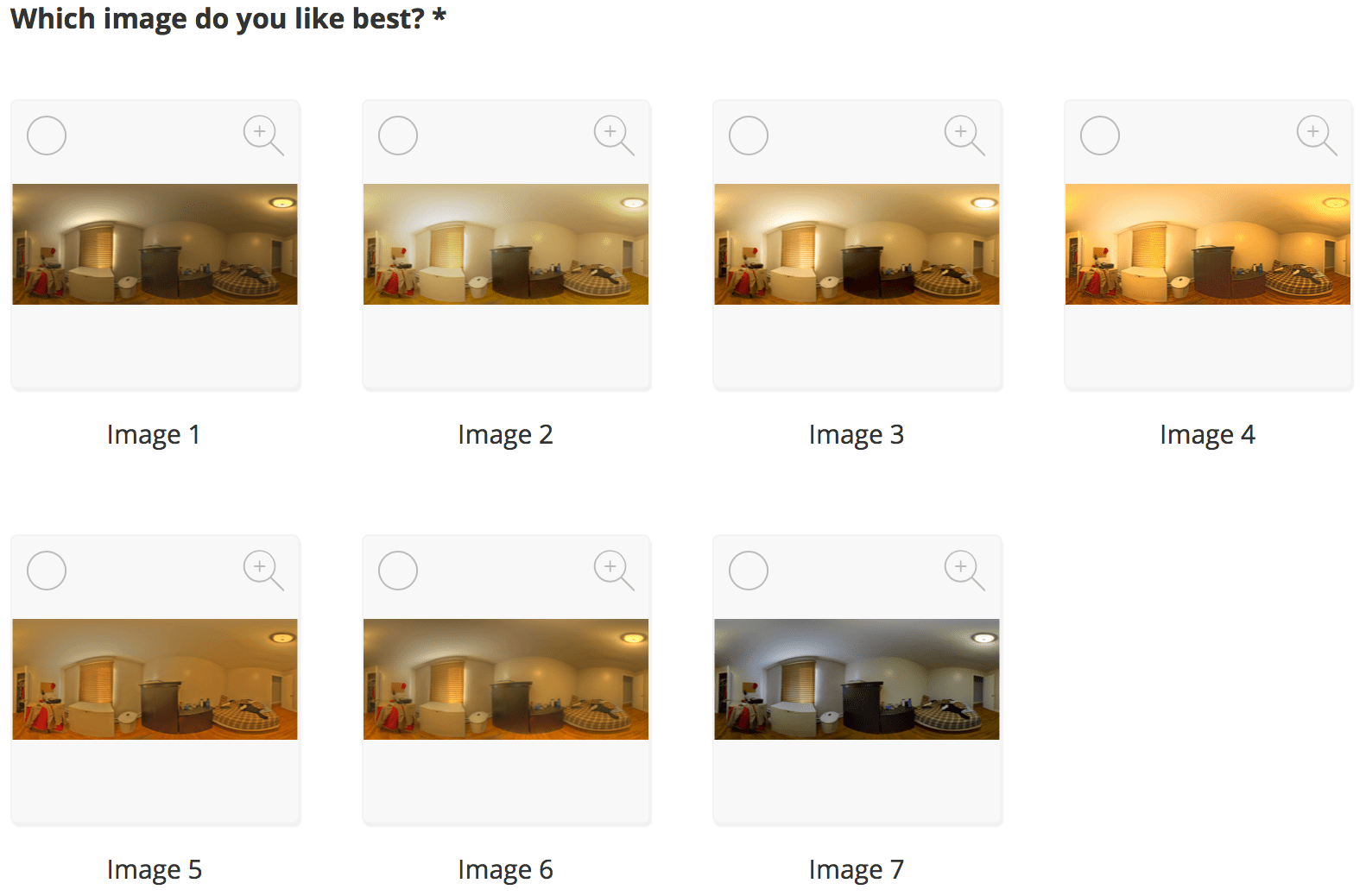}
    \captionof{figure}{A screenshot of a survey question in the Comparative Selection section.}
    \label{fig:survey_1}
\end{figure*}

This section contains 8 questions.  Each question has a group of 7 tone mapped LDR images in random order.  Fig. \ref{fig:survey_1} shows an example of a question.  Participants can click on each image to view the full size. During the survey, participants need to choose an image with the best visual preference.

\subsection{Image Quality Rating}
\begin{figure*}[!b]
    \centering
    \includegraphics[scale = 0.32]{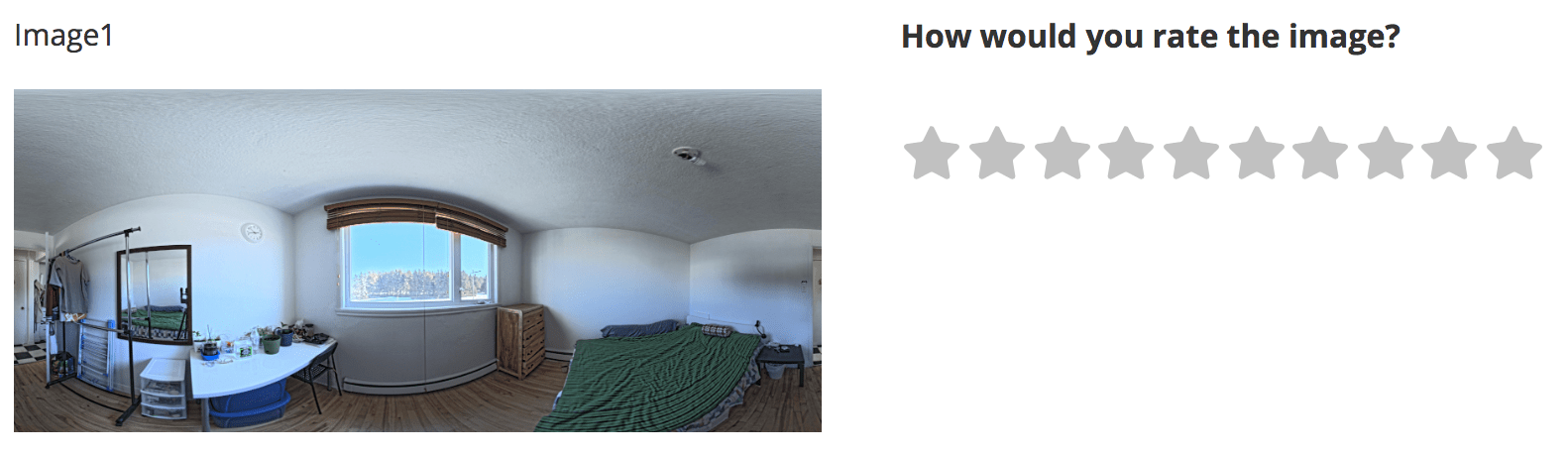}
    \captionof{figure}{A screenshot of a survey question in the Image Quality Rating section.}
    \label{fig:survey_2}
\end{figure*}

This section contains 6 questions, 7 tone mapped LDR images in each question.  Fig. \ref{fig:survey_2} shows an example of a question. Participants are asked to rate each LDR image from \textbf{1} to \textbf{10} based on image brightness, image contrast, and the extent to which overexposure and underexposure details are revealed. \textbf{1} represents "dislike" or "fuzzy details" and \textbf{10} stands for "most favorite" or "clear and rich details".  

\subsection{Assessment Process and Results}
We send out our survey\footnote{https://surveyhero.com/c/7040d7c6} via WeChat (the Chinese social media and multipurpose application) and email. Each participant is only allowed to complete the survey once. Participants can also choose to abandon the test at any time. Until the submission of the paper, there were a total of 71 people participated in the survey.  Table \ref{table:survey_1} and Table \ref{table:survey_2} summarizes result of Comparative Selection and Image Quality Rating respectively.

We perform a simple averaging function to obtain $Ave_{selection}$ in the Comparative Selection section by the equation
\begin{equation}
    Ave_{selection} = \frac{\sum_{i=1}^{N_s}s_i}{N_s}
\end{equation}
where s denotes the number of LDR images being selected from each algorithm in each question.  $N_s$ represents the total number of questions in the Comparative Selection section.

To acquire the $Ave_{rating}$ in the Image Quality Rating section, we first computed the weighted average score $w$ of each TMO by the equation
\begin{equation}
    w = \frac{\sum_{i=1}^{10}r_in_i}{10}
\end{equation}
where $r_i$ denote each rating score (from 1 to 10).  $n_i$ is the number of $r_i$ score that participants rated to the LDR image.

Then we used $w$ to calculate the $Ave_{rating}$ 
\begin{equation}
    Ave_{rating} = \frac{\sum_{i=1}^{N_r}w_i}{N_r}
\end{equation}
where $N_r$ represents the total number of questions in the Image Quality Rating section.

\begin{table*}[t]
\footnotesize
\begin{center}
\caption{The summary of the result in the Comparative Selection section. The number from 1 to 8 in the first row indicate each survey question.  $Sum$ denotes the total number of each TMO being selected.  $Ave_{selection}$ denotes the average number of each TMO being selected.}
\begin{tabular}{l|c|c|c|c|c|c|c|c|c|c}\hline
  Methods & 1 & 2 & 3 & 4 & 5 & 6 & 7 & 8 & $Sum$ & $Ave_{selection}$ \\ \hline
  Ferradans \cite{ferradans2011analysis} & 8 & 4 & 7 & 10 & 8 & 14 & 9 & 9 & 69 & 8.625 \\
   Gu \cite{gu2013local} & 16 & 25 & 9 & 13 & 18 & 8 & 11 & 11 & 111 & 13.875 \\
  Mai \cite{mai2011optimizing} & 15 & 7 & 8 & 8 & 5 & 7 & 9 & 6 & 65 & 8.125 \\
  Mantiuk \cite{mantiuk2008display} & 3 & 6 & 10 & 11 & 7 & 9 & 4 & 4 & 54 & 6.75 \\
  Paris \cite{paris2015local} & 8 & 4 & 17 & 9 & 5 & 3 & 2 & 4 & 52 & 6.5 \\
  Photomatix \cite{photomatrix} & 2 & 2 & 10 & 3 & 4 & 4 & 8 & 6 & 39 & 4.875 \\
  Ours & 19 & 23 & 7 & 13 & 11 & 11 & 13 & 15 & \textbf{112} & \textbf{14} \\
  \hline
\end{tabular}
\label{table:survey_1}
\end{center}
\end{table*}

\begin{table*}[t]
\footnotesize
\begin{center}
\caption{The summary of the results in the Image Quality Rating section. The number from 1 to 6 in the first row indicate each survey question.  $Ave_{rating}$ represents the average rating score of each TMO.}
\begin{tabular}{l|c|c|c|c|c|c|c}\hline
  Methods & 1 & 2 & 3 & 4 & 5 & 6 & $Ave_{rating}$ \\ \hline
  Ferradans \cite{ferradans2011analysis} & 6.45 & 5.59 & 5.47 & 6.50 & 5.87 & 7.23 & 6.18 \\
Gu \cite{gu2013local} & 6.00 & 6.39 & 6.06 & 7.00 & 6.10 & 6.87 & 6.40 \\
Mai \cite{mai2011optimizing} & 6.64 & 6.29 & 6.27 & 6.73 & 6.19 & 7.31 & 6.57 \\
Mantiuk \cite{mantiuk2008display} & 5.88 & 6.11 & 5.73 & 6.72 & 5.52 & 6.52 & 6.08 \\
Paris \cite{paris2015local} & 5.47 & 5.93 & 5.76 & 5.29 & 5.16 & 5.90 & 5.59 \\
Photomatix \cite{photomatrix} & 6.00 & 5.52 & 4.55 & 6.21 & 6.10 & 5.48 & 5.64 \\
Our & 6.56 & 6.25 & 6.40 & 6.43 & 6.84 & 7.00 & \textbf{6.58} \\

  \hline
\end{tabular}
\label{table:survey_2}
\end{center}
\end{table*}

\subsection{Color Recovery}
Our approach operates on luminance WDR images.  We employed an additional color recovery step to assign a color to the pixels of compressed dynamic range images using the method described in \cite{fattal2002gradient}

\begin{equation}
    \hat{x}_{c} = (\frac{\hat{x}_{l}}{h})^sl
\end{equation}
where $\hat{x}$ is the final WDR-LDR output after the fine tune network. $h$ and $l$ denote the WDR image of the output and the one after dynamic range compression, respectively.  We set the color saturation controller s=0.6 as \cite{fattal2002gradient} found to produce satisfactory results.

\subsection{Additional Qualitative Comparisons}
\begin{figure*}[h!]
        \centering
        \begin{subfigure}[b]{0.24\textwidth}  
            \centering 
            \includegraphics[width=\textwidth]{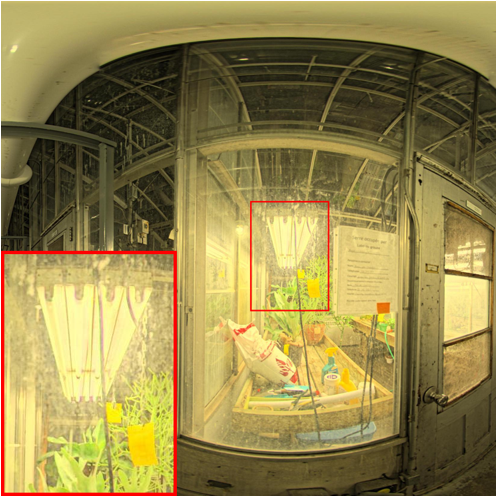}
            \caption[]%
            {{\small Reference }}    
            \label{fig:mean and std of net24}
        \end{subfigure}
        \begin{subfigure}[b]{0.24\textwidth}  
            \centering 
            \includegraphics[width=\textwidth]{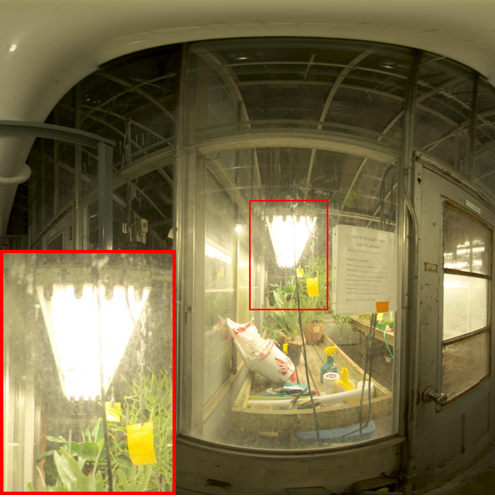}
            \caption[]%
            {{\small Mantiuk \cite{mantiuk2008display} }}    
            \label{fig:mean and std of net24}
        \end{subfigure}
        \begin{subfigure}[b]{0.24\textwidth}   
            \centering 
            \includegraphics[width=\textwidth]{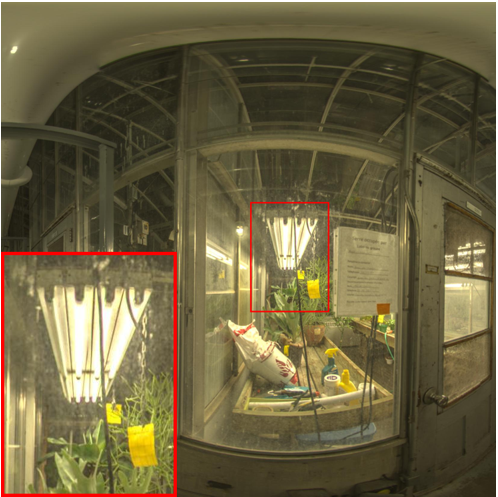}
            \caption[]%
            {{\small Paris \cite{paris2015local} }}    
            \label{fig:mean and std of net34}
        \end{subfigure}
        \begin{subfigure}[b]{0.24\textwidth}   
            \centering 
            \includegraphics[width=\textwidth]{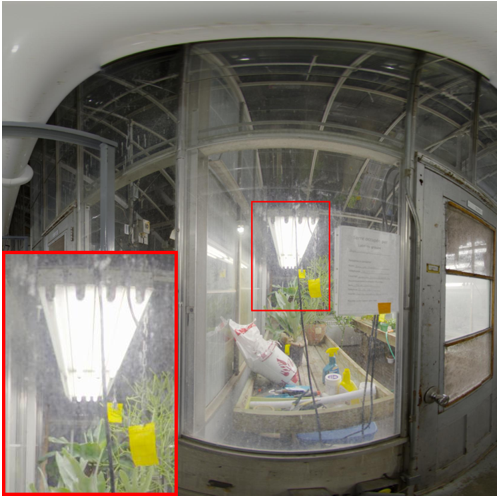}
            \caption[]%
            {{\small Ferradans \cite{ferradans2011analysis}     }}    
            \label{fig:mean and std of net44}
        \end{subfigure}
        \label{fig:mean and std of nets}
        \centering
        \begin{subfigure}[b]{0.24\textwidth}
            \centering
            \includegraphics[width=\textwidth]{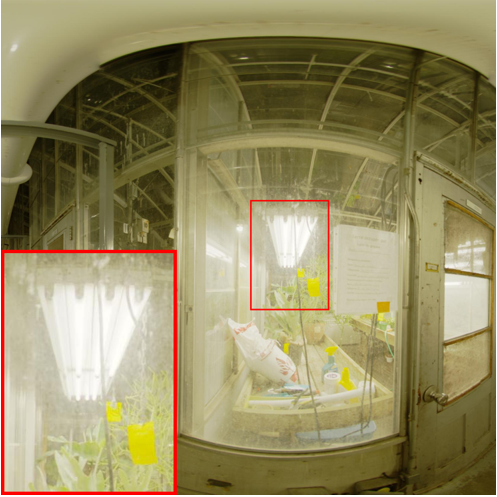}
            \caption[]%
            {{\small Mai \cite{mai2011optimizing} }}    
            \label{fig:mean and std of net14}
        \end{subfigure}
        \begin{subfigure}[b]{0.24\textwidth}
            \centering
            \includegraphics[width=\textwidth]{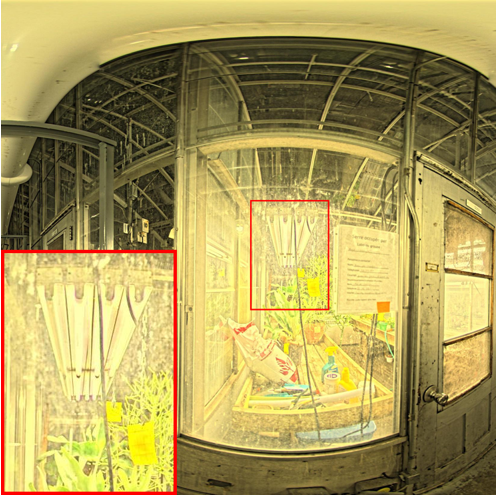}
            \caption[]%
            {{\small Gu \cite{gu2013local} }}    
            \label{fig:mean and std of net14}
        \end{subfigure}
        \begin{subfigure}[b]{0.24\textwidth}   
            \centering 
            \includegraphics[width=\textwidth]{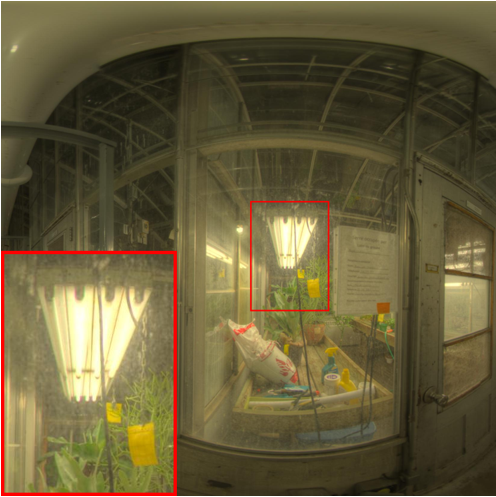}
            \caption[]%
            {{\small Photomatrix \cite{photomatrix} }}    
            \label{fig:mean and std of net34}
        \end{subfigure}
        \begin{subfigure}[b]{0.24\textwidth}   
            \centering 
            \includegraphics[width=\textwidth]{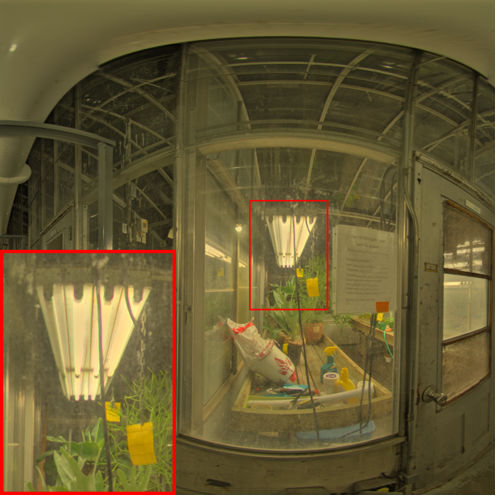}
            \caption[]%
            {{\small Proposed}}    
            \label{fig:mean and std of net44}
        \end{subfigure}
        \caption{Qualitative comparison on Laval data test set. The proposed method is able to recover local details in the saturated region.}
        \label{fig:Laval_9C4A0006_5133111e97}
\end{figure*}

\begin{figure*}[t]
        \centering
        \begin{subfigure}[b]{0.24\textwidth}  
            \centering 
            \includegraphics[width=\textwidth]{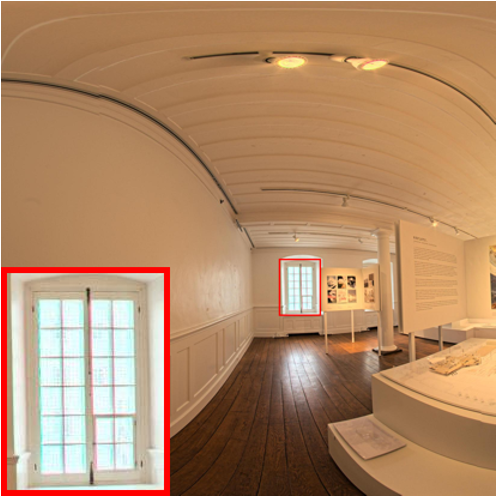}
            \caption[]%
            {{\small Reference }}    
            \label{fig:mean and std of net24}
        \end{subfigure}
        \begin{subfigure}[b]{0.24\textwidth}  
            \centering 
            \includegraphics[width=\textwidth]{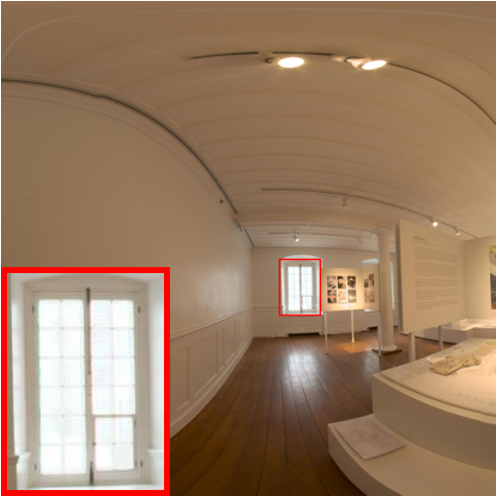}
            \caption[]%
            {{\small Mantiuk \cite{mantiuk2008display} }}    
            \label{fig:mean and std of net24}
        \end{subfigure}
        \begin{subfigure}[b]{0.24\textwidth}   
            \centering 
            \includegraphics[width=\textwidth]{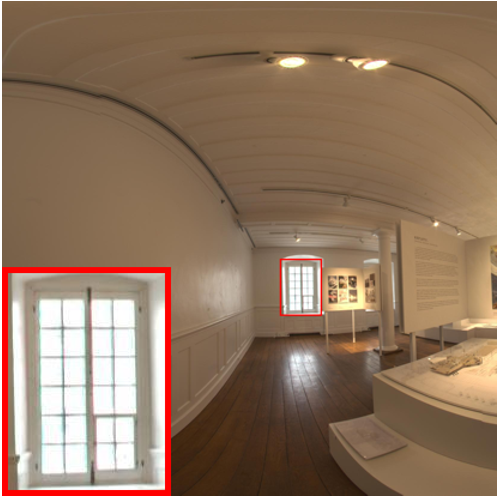}
            \caption[]%
            {{\small Paris \cite{paris2015local} }}    
            \label{fig:mean and std of net34}
        \end{subfigure}
        \begin{subfigure}[b]{0.24\textwidth}   
            \centering 
            \includegraphics[width=\textwidth]{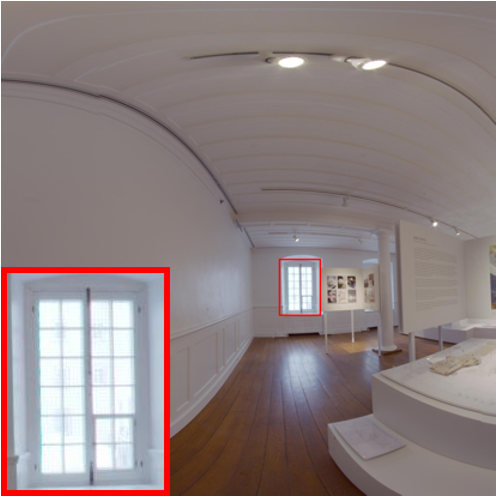}
            \caption[]%
            {{\small Ferradans \cite{ferradans2011analysis}     }}    
            \label{fig:mean and std of net44}
        \end{subfigure}
        \label{fig:mean and std of nets}
        \centering
        \begin{subfigure}[b]{0.24\textwidth}
            \centering
            \includegraphics[width=\textwidth]{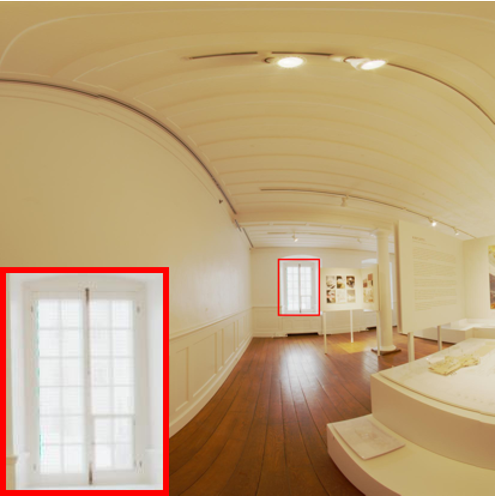}
            \caption[]%
            {{\small Mai \cite{mai2011optimizing} }}    
            \label{fig:mean and std of net14}
        \end{subfigure}
        \begin{subfigure}[b]{0.24\textwidth}
            \centering
            \includegraphics[width=\textwidth]{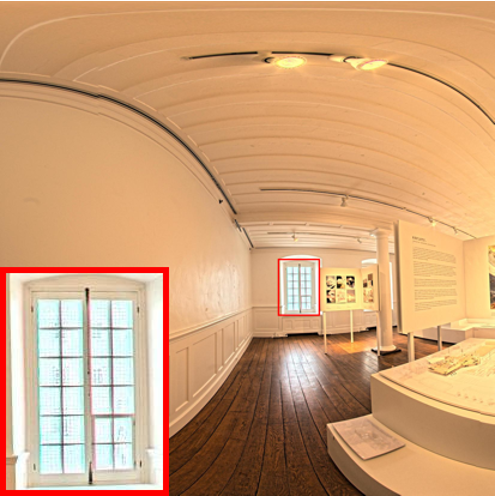}
            \caption[]%
            {{\small Gu \cite{gu2013local} }}    
            \label{fig:mean and std of net14}
        \end{subfigure}
        \begin{subfigure}[b]{0.24\textwidth}   
            \centering 
            \includegraphics[width=\textwidth]{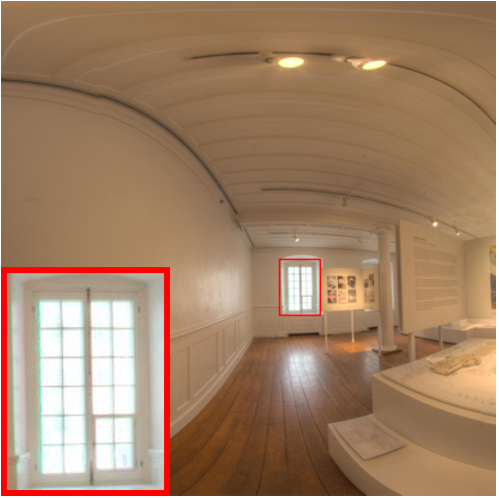}
            \caption[]%
            {{\small Photomatrix \cite{photomatrix} }}    
            \label{fig:mean and std of net34}
        \end{subfigure}
        \begin{subfigure}[b]{0.24\textwidth}   
            \centering 
            \includegraphics[width=\textwidth]{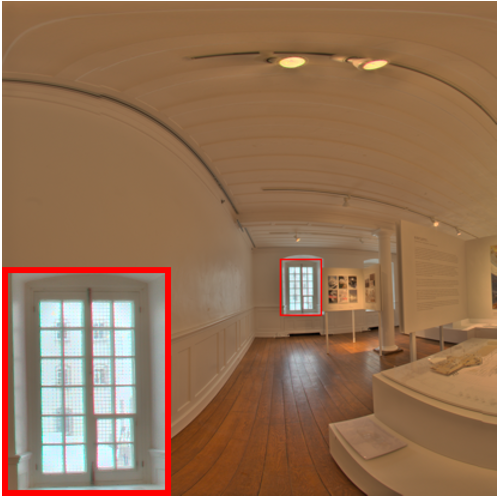}
            \caption[]%
            {{\small Proposed TMO}}    
            \label{fig:mean and std of net44}
        \end{subfigure}
        \caption{Qualitative comparison on Laval data test set. The proposed method is able to recover local details in the saturated region.}
        \label{fig:Laval_9C4A0227_29aa066f77}
\end{figure*}

\begin{figure*}[t]
        \centering
        \begin{subfigure}[b]{0.24\textwidth}  
            \centering 
            \includegraphics[width=\textwidth]{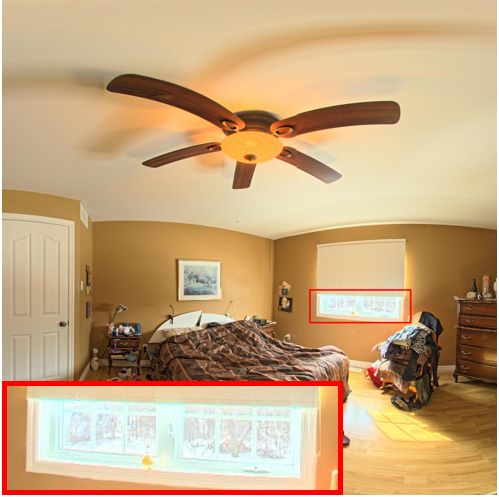}
            \caption[]%
            {{\small Reference }}    
            \label{fig:mean and std of net24}
        \end{subfigure}
        \begin{subfigure}[b]{0.24\textwidth}  
            \centering 
            \includegraphics[width=\textwidth]{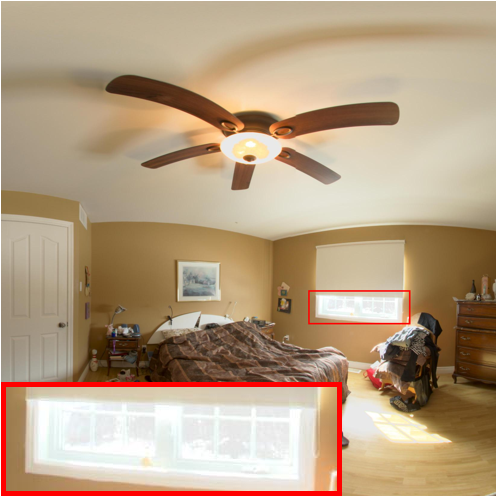}
            \caption[]%
            {{\small Mantiuk \cite{mantiuk2008display} }}    
            \label{fig:mean and std of net24}
        \end{subfigure}
        \begin{subfigure}[b]{0.24\textwidth}   
            \centering 
            \includegraphics[width=\textwidth]{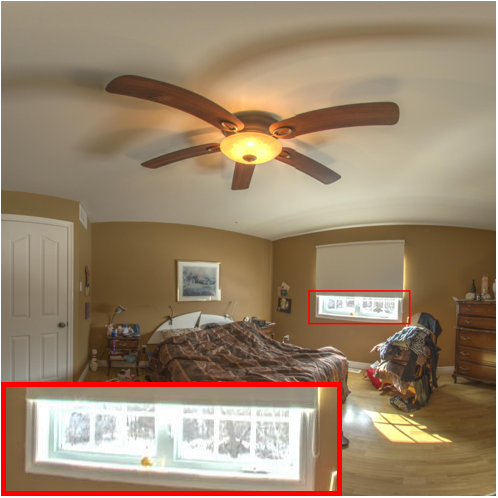}
            \caption[]%
            {{\small Paris \cite{paris2015local} }}    
            \label{fig:mean and std of net34}
        \end{subfigure}
        \begin{subfigure}[b]{0.24\textwidth}   
            \centering 
            \includegraphics[width=\textwidth]{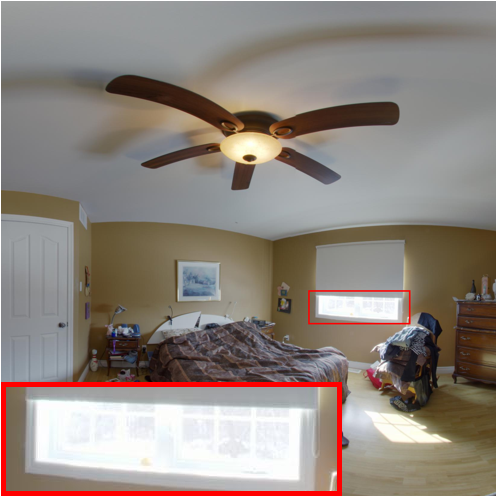}
            \caption[]%
            {{\small Ferradans \cite{ferradans2011analysis}     }}    
            \label{fig:mean and std of net44}
        \end{subfigure}
        \label{fig:mean and std of nets}
        \centering
        \begin{subfigure}[b]{0.24\textwidth}
            \centering
            \includegraphics[width=\textwidth]{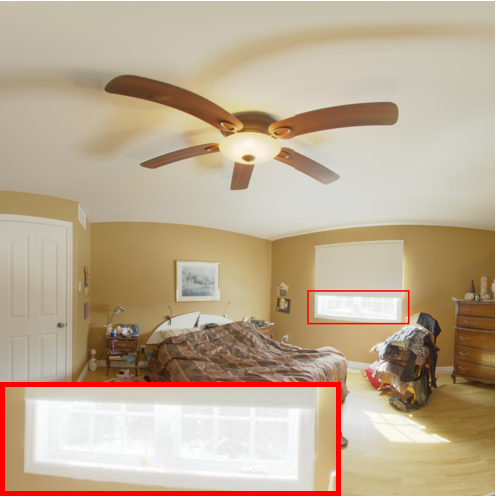}
            \caption[]%
            {{\small Mai \cite{mai2011optimizing} }}    
            \label{fig:mean and std of net14}
        \end{subfigure}
        \begin{subfigure}[b]{0.24\textwidth}
            \centering
            \includegraphics[width=\textwidth]{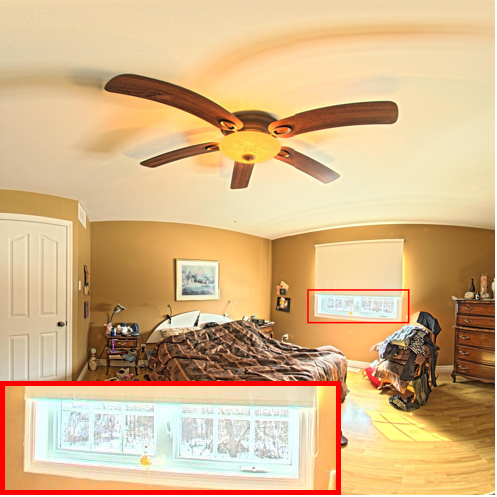}
            \caption[]%
            {{\small Gu \cite{gu2013local} }}    
            \label{fig:mean and std of net14}
        \end{subfigure}
        \begin{subfigure}[b]{0.24\textwidth}   
            \centering 
            \includegraphics[width=\textwidth]{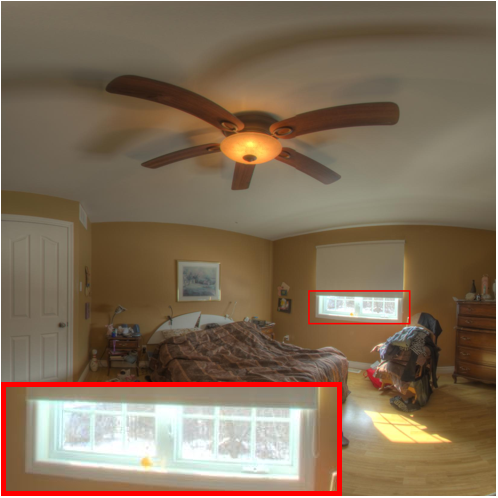}
            \caption[]%
            {{\small Photomatrix \cite{photomatrix} }}    
            \label{fig:mean and std of net34}
        \end{subfigure}
        \begin{subfigure}[b]{0.24\textwidth}   
            \centering 
            \includegraphics[width=\textwidth]{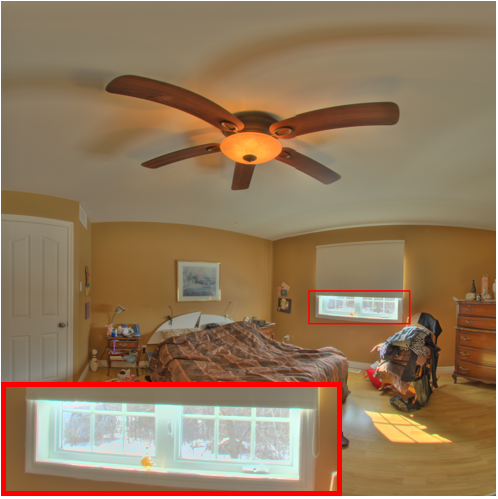}
            \caption[]%
            {{\small Proposed TMO}}    
            \label{fig:mean and std of net44}
        \end{subfigure}
        \caption{Qualitative comparison on Laval data test set. The proposed method is able to recover local details in the saturated region.}
        \label{fig:Laval_9C4A0632_aba46c619b}
\end{figure*}

\begin{figure*}[t]
        \centering
        \begin{subfigure}[b]{0.24\textwidth}  
            \centering 
            \includegraphics[width=\textwidth]{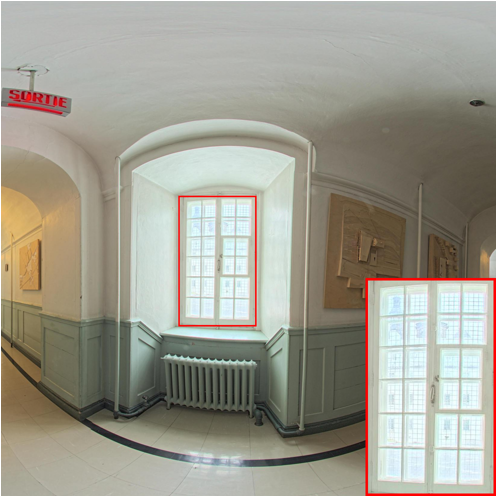}
            \caption[]%
            {{\small Reference }}    
            \label{fig:mean and std of net24}
        \end{subfigure}
        \begin{subfigure}[b]{0.24\textwidth}  
            \centering 
            \includegraphics[width=\textwidth]{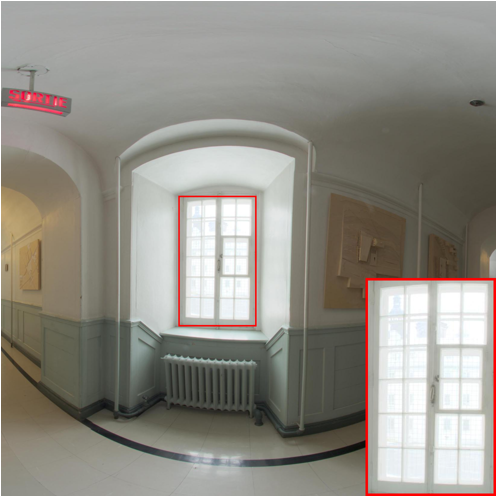}
            \caption[]%
            {{\small Mantiuk \cite{mantiuk2008display} }}    
            \label{fig:mean and std of net24}
        \end{subfigure}
        \begin{subfigure}[b]{0.24\textwidth}   
            \centering 
            \includegraphics[width=\textwidth]{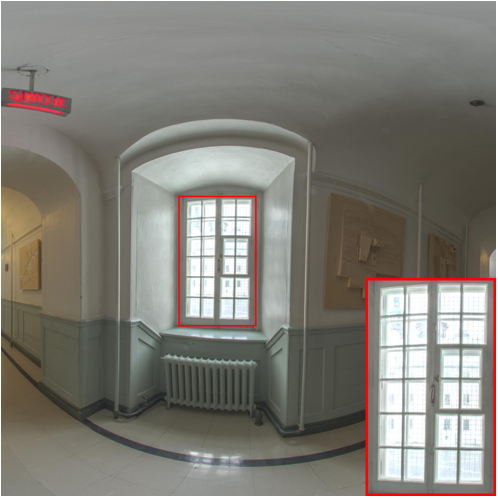}
            \caption[]%
            {{\small Paris \cite{paris2015local} }}    
            \label{fig:mean and std of net34}
        \end{subfigure}
        \begin{subfigure}[b]{0.24\textwidth}   
            \centering 
            \includegraphics[width=\textwidth]{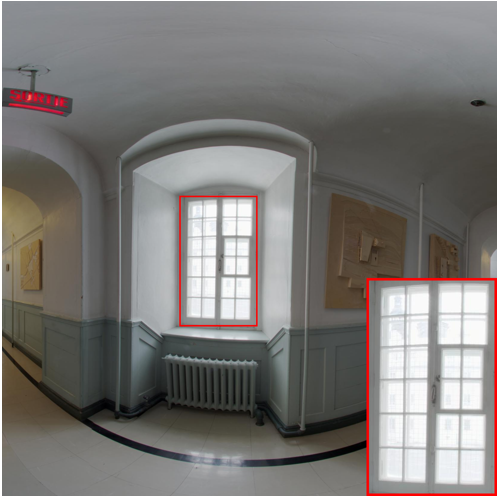}
            \caption[]%
            {{\small Ferradans \cite{ferradans2011analysis}     }}    
            \label{fig:mean and std of net44}
        \end{subfigure}
        \label{fig:mean and std of nets}
        \centering
        \begin{subfigure}[b]{0.24\textwidth}
            \centering
            \includegraphics[width=\textwidth]{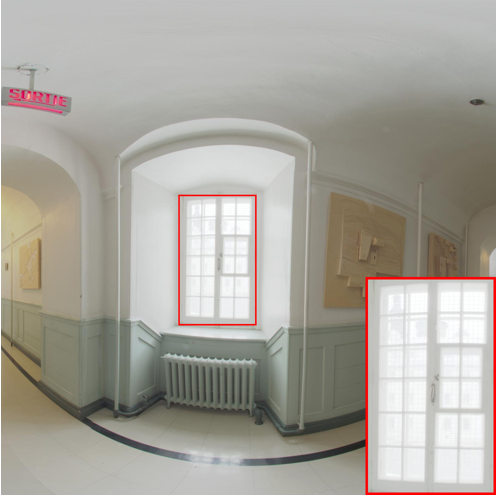}
            \caption[]%
            {{\small Mai \cite{mai2011optimizing} }}    
            \label{fig:mean and std of net14}
        \end{subfigure}
        \begin{subfigure}[b]{0.24\textwidth}
            \centering
            \includegraphics[width=\textwidth]{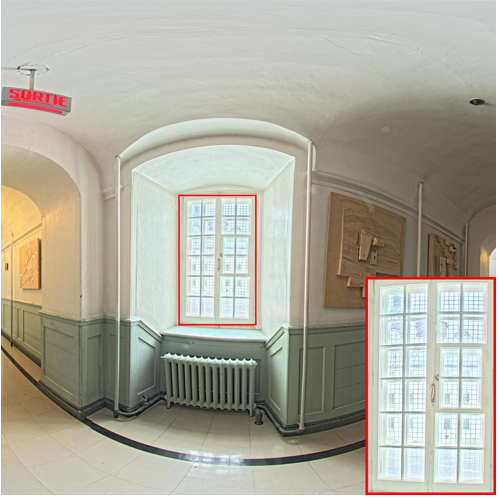}
            \caption[]%
            {{\small Gu \cite{gu2013local} }}    
            \label{fig:mean and std of net14}
        \end{subfigure}
        \begin{subfigure}[b]{0.24\textwidth}   
            \centering 
            \includegraphics[width=\textwidth]{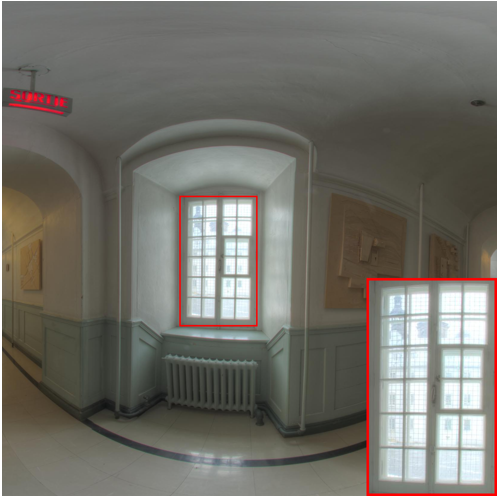}
            \caption[]%
            {{\small Photomatrix \cite{photomatrix} }}    
            \label{fig:mean and std of net34}
        \end{subfigure}
        \begin{subfigure}[b]{0.24\textwidth}   
            \centering 
            \includegraphics[width=\textwidth]{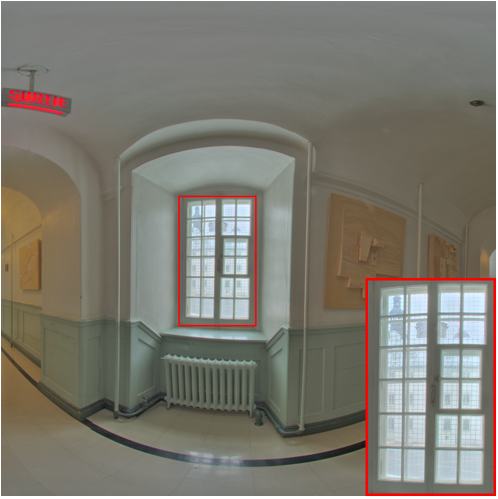}
            \caption[]%
            {{\small Proposed TMO}}    
            \label{fig:mean and std of net44}
        \end{subfigure}
        \caption{Qualitative comparison on Laval data test set. The proposed method is able to recover local details in the saturated region.}
        \label{fig:Laval_9C4A1578_457ab482d7}
\end{figure*}

\begin{figure*}[t]
        \centering
        \begin{subfigure}[b]{0.24\textwidth}  
            \centering 
            \includegraphics[width=\textwidth]{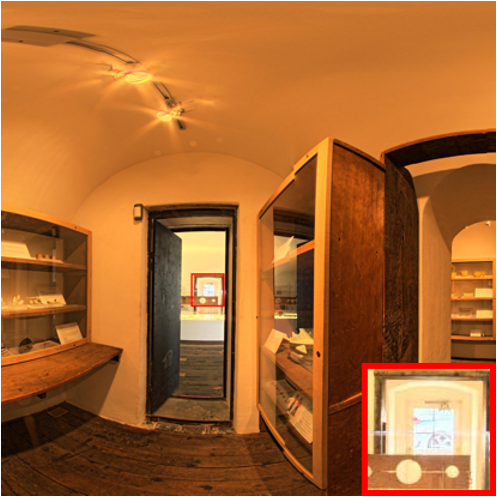}
            \caption[]%
            {{\small Reference }}    
            \label{fig:mean and std of net24}
        \end{subfigure}
        \begin{subfigure}[b]{0.24\textwidth}  
            \centering 
            \includegraphics[width=\textwidth]{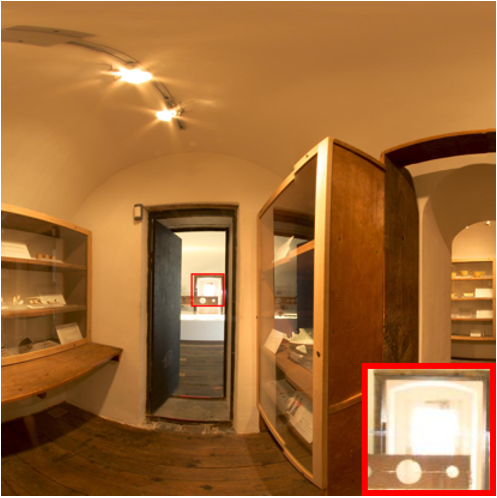}
            \caption[]%
            {{\small Mantiuk \cite{mantiuk2008display} }}    
            \label{fig:mean and std of net24}
        \end{subfigure}
        \begin{subfigure}[b]{0.24\textwidth}   
            \centering 
            \includegraphics[width=\textwidth]{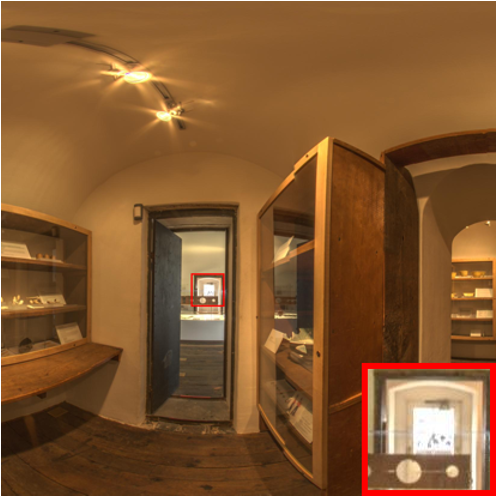}
            \caption[]%
            {{\small Paris \cite{paris2015local} }}    
            \label{fig:mean and std of net34}
        \end{subfigure}
        \begin{subfigure}[b]{0.24\textwidth}   
            \centering 
            \includegraphics[width=\textwidth]{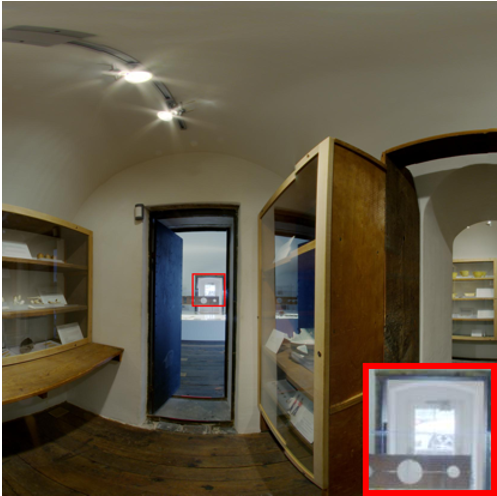}
            \caption[]%
            {{\small Ferradans \cite{ferradans2011analysis}     }}    
            \label{fig:mean and std of net44}
        \end{subfigure}
        \label{fig:mean and std of nets}
        \centering
        \begin{subfigure}[b]{0.24\textwidth}
            \centering
            \includegraphics[width=\textwidth]{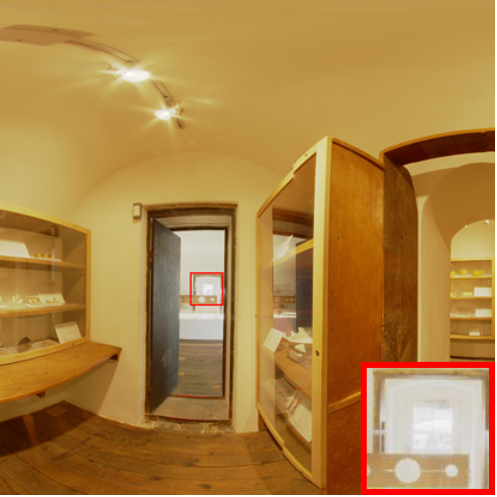}
            \caption[]%
            {{\small Mai \cite{mai2011optimizing} }}    
            \label{fig:mean and std of net14}
        \end{subfigure}
        \begin{subfigure}[b]{0.24\textwidth}
            \centering
            \includegraphics[width=\textwidth]{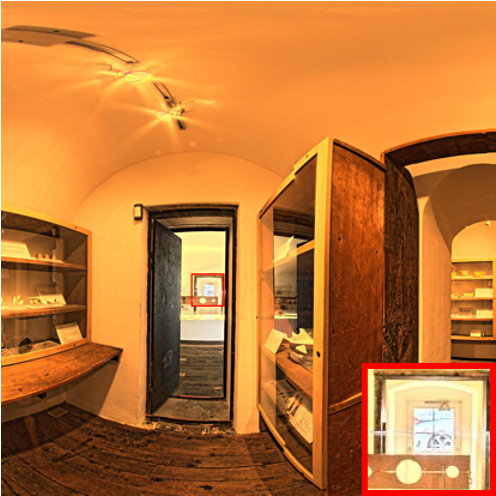}
            \caption[]%
            {{\small Gu \cite{gu2013local} }}    
            \label{fig:mean and std of net14}
        \end{subfigure}
        \begin{subfigure}[b]{0.24\textwidth}   
            \centering 
            \includegraphics[width=\textwidth]{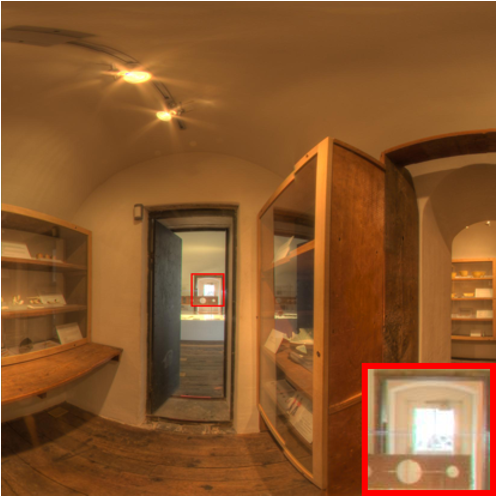}
            \caption[]%
            {{\small Photomatrix \cite{photomatrix} }}    
            \label{fig:mean and std of net34}
        \end{subfigure}
        \begin{subfigure}[b]{0.24\textwidth}   
            \centering 
            \includegraphics[width=\textwidth]{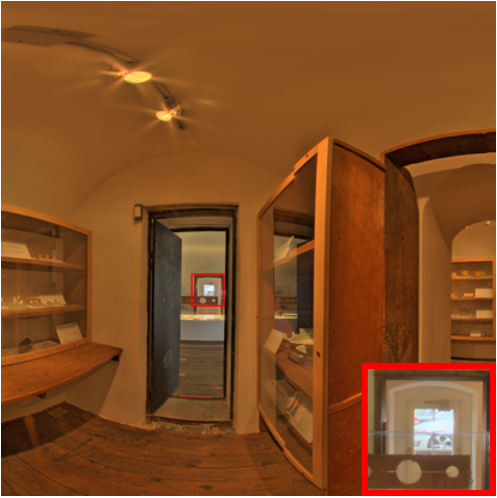}
            \caption[]%
            {{\small Proposed TMO}}    
            \label{fig:mean and std of net44}
        \end{subfigure}
        \caption{Qualitative comparison on Laval data test set. The proposed method is able to enhance local details in the saturated region.}
        \label{fig:Laval_9C4A3237-2aa4d412ea}
\end{figure*}

\begin{figure*}[t]
        \centering
        \begin{subfigure}[b]{0.24\textwidth}  
            \centering 
            \includegraphics[width=\textwidth]{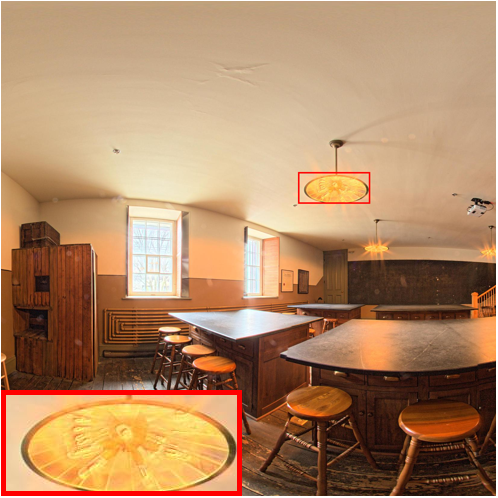}
            \caption[]%
            {{\small Reference }}    
            \label{fig:mean and std of net24}
        \end{subfigure}
        \begin{subfigure}[b]{0.24\textwidth}  
            \centering 
            \includegraphics[width=\textwidth]{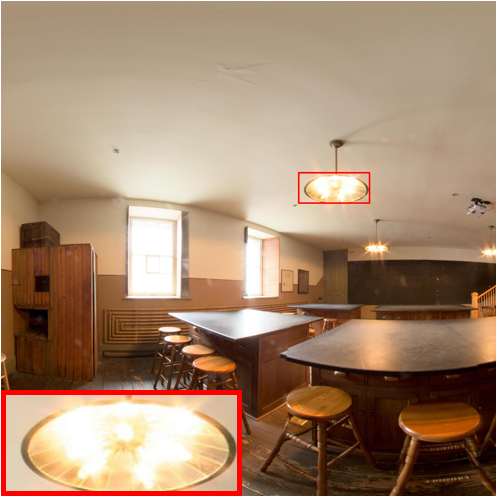}
            \caption[]%
            {{\small Mantiuk \cite{mantiuk2008display} }}    
            \label{fig:mean and std of net24}
        \end{subfigure}
        \begin{subfigure}[b]{0.24\textwidth}   
            \centering 
            \includegraphics[width=\textwidth]{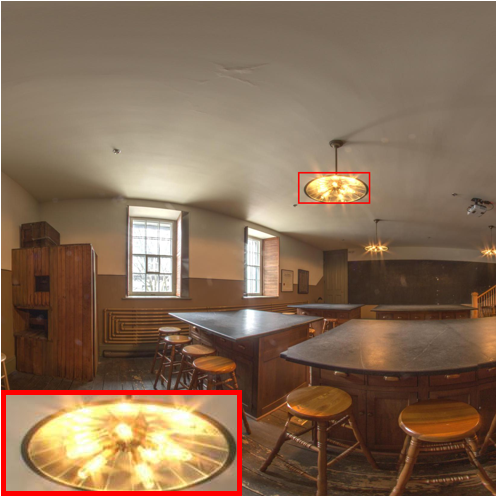}
            \caption[]%
            {{\small Paris \cite{paris2015local} }}    
            \label{fig:mean and std of net34}
        \end{subfigure}
        \begin{subfigure}[b]{0.24\textwidth}   
            \centering 
            \includegraphics[width=\textwidth]{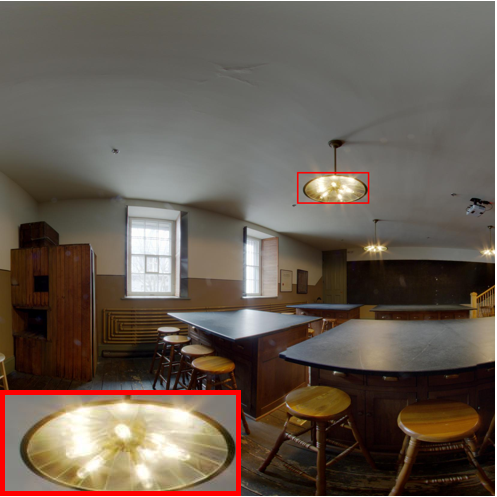}
            \caption[]%
            {{\small Ferradans \cite{ferradans2011analysis}     }}    
            \label{fig:mean and std of net44}
        \end{subfigure}
        \label{fig:mean and std of nets}
        \centering
        \begin{subfigure}[b]{0.24\textwidth}
            \centering
            \includegraphics[width=\textwidth]{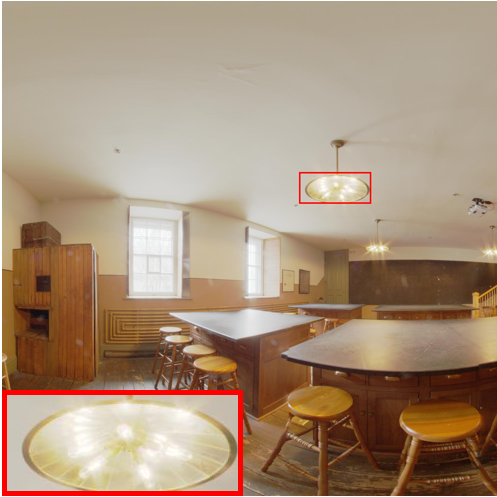}
            \caption[]%
            {{\small Mai \cite{mai2011optimizing} }}    
            \label{fig:mean and std of net14}
        \end{subfigure}
        \begin{subfigure}[b]{0.24\textwidth}
            \centering
            \includegraphics[width=\textwidth]{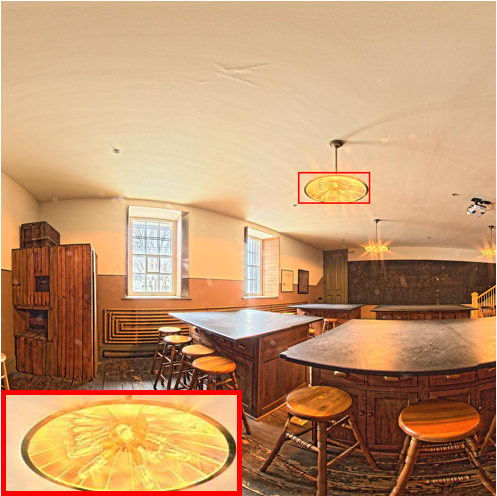}
            \caption[]%
            {{\small Gu \cite{gu2013local} }}    
            \label{fig:mean and std of net14}
        \end{subfigure}
        \begin{subfigure}[b]{0.24\textwidth}   
            \centering 
            \includegraphics[width=\textwidth]{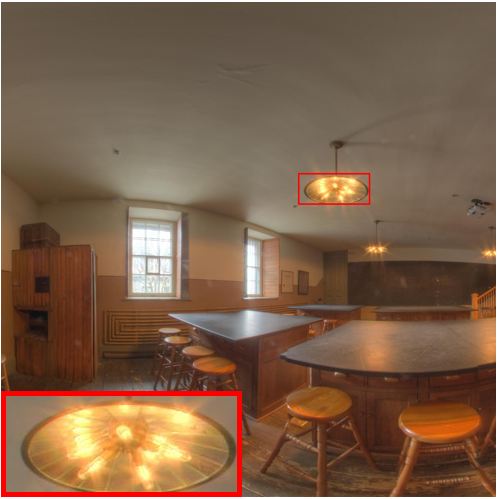}
            \caption[]%
            {{\small Photomatrix \cite{photomatrix} }}    
            \label{fig:mean and std of net34}
        \end{subfigure}
        \begin{subfigure}[b]{0.24\textwidth}   
            \centering 
            \includegraphics[width=\textwidth]{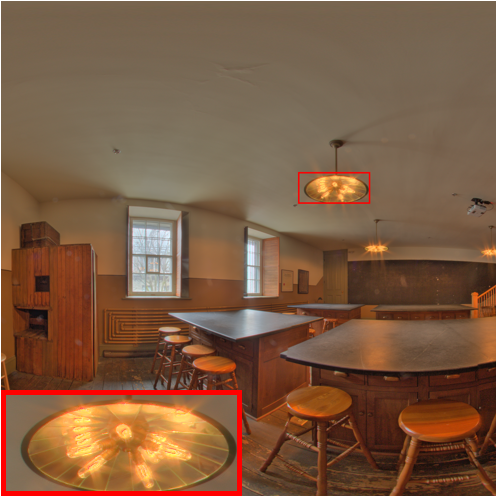}
            \caption[]%
            {{\small Proposed TMO}}    
            \label{fig:mean and std of net44}
        \end{subfigure}
        \caption{Qualitative comparison on Laval data test set. The proposed method is able to enhance local details in the saturated region.}
        \label{fig:Laval_9C4A3335-edf32a8ffe}
\end{figure*}

\begin{figure*}[t]
{\begin{minipage}{\textwidth}
\centering
            \includegraphics[width=\columnwidth]{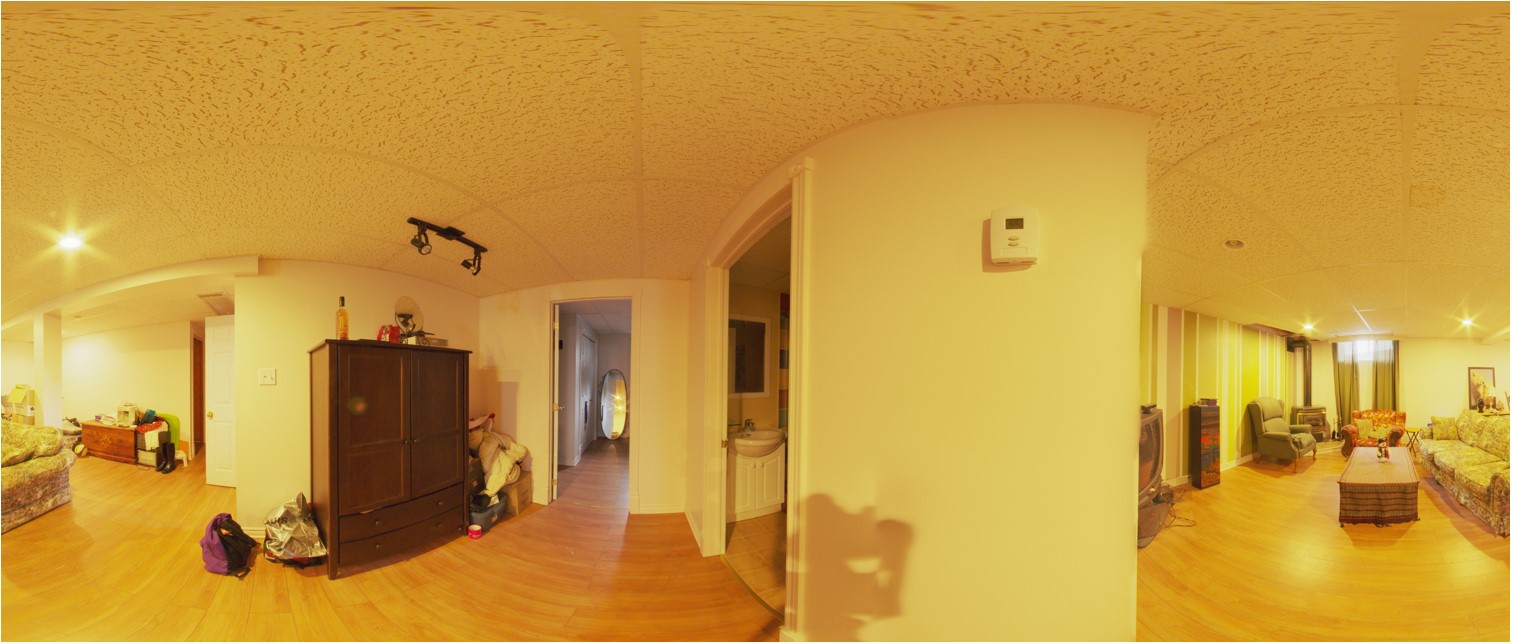}

            \includegraphics[width=\columnwidth]{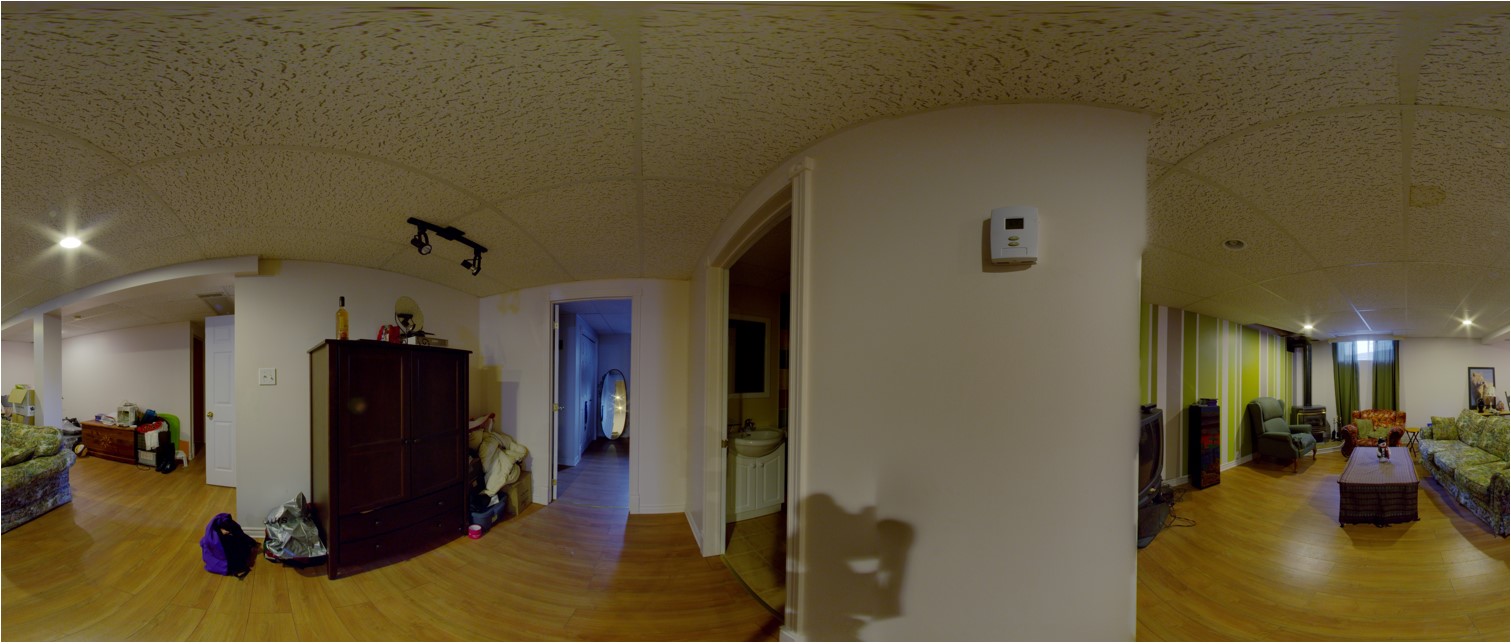}
            
\caption{top: Mai TMO \cite{mai2011optimizing}, bottom: Ferradans TMO \cite{ferradans2011analysis}}
\label{9C4A3187-9c8f0e2b1b-1}
\end{minipage}}
\end{figure*}

\begin{figure*}[t]
{\begin{minipage}{\textwidth}
\centering
            \includegraphics[width=\columnwidth]{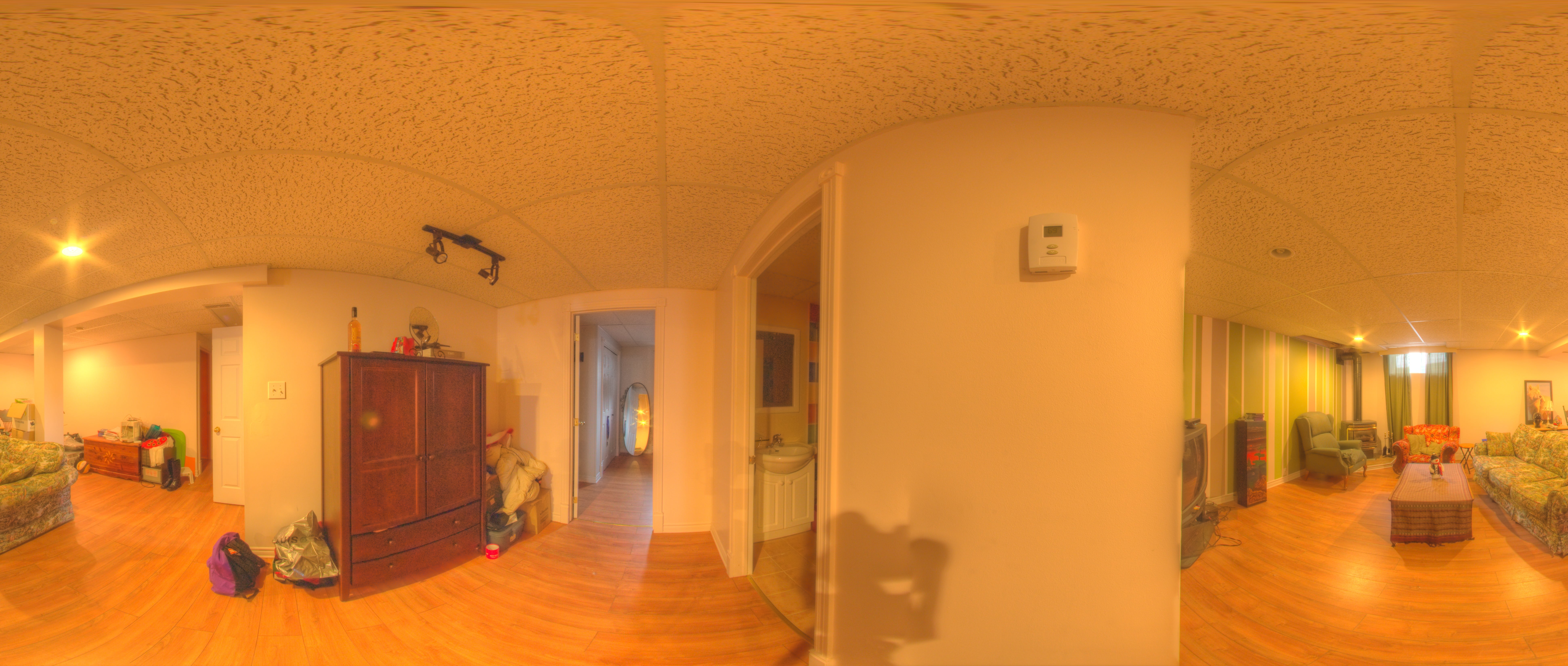}

            \includegraphics[width=\columnwidth]{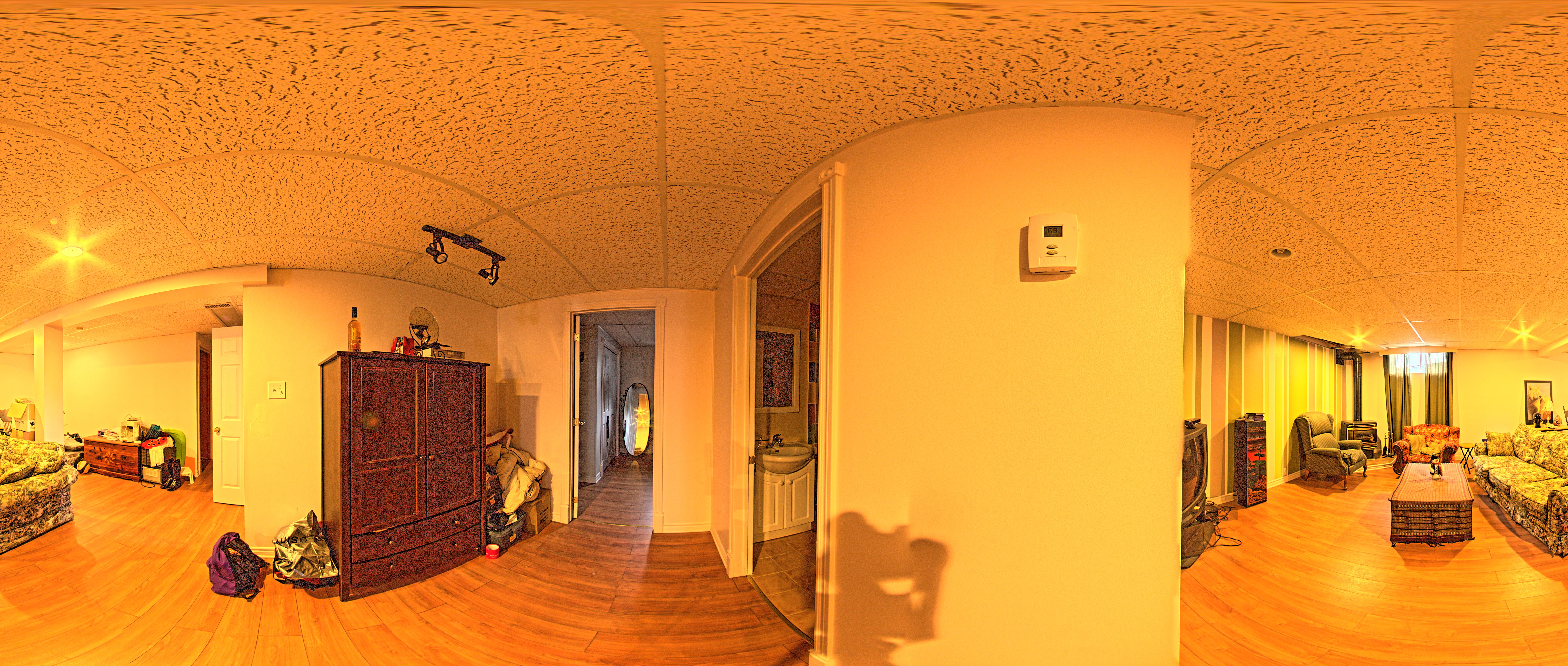}
            
\caption{top: Photomatix TMO \cite{photomatrix}, bottom: Gu TMO \cite{gu2013local}}
\label{9C4A3187-9c8f0e2b1b-2}
\end{minipage}}
\end{figure*}

\begin{figure*}[t]
{\begin{minipage}{\textwidth}
\centering
            \includegraphics[width=\columnwidth]{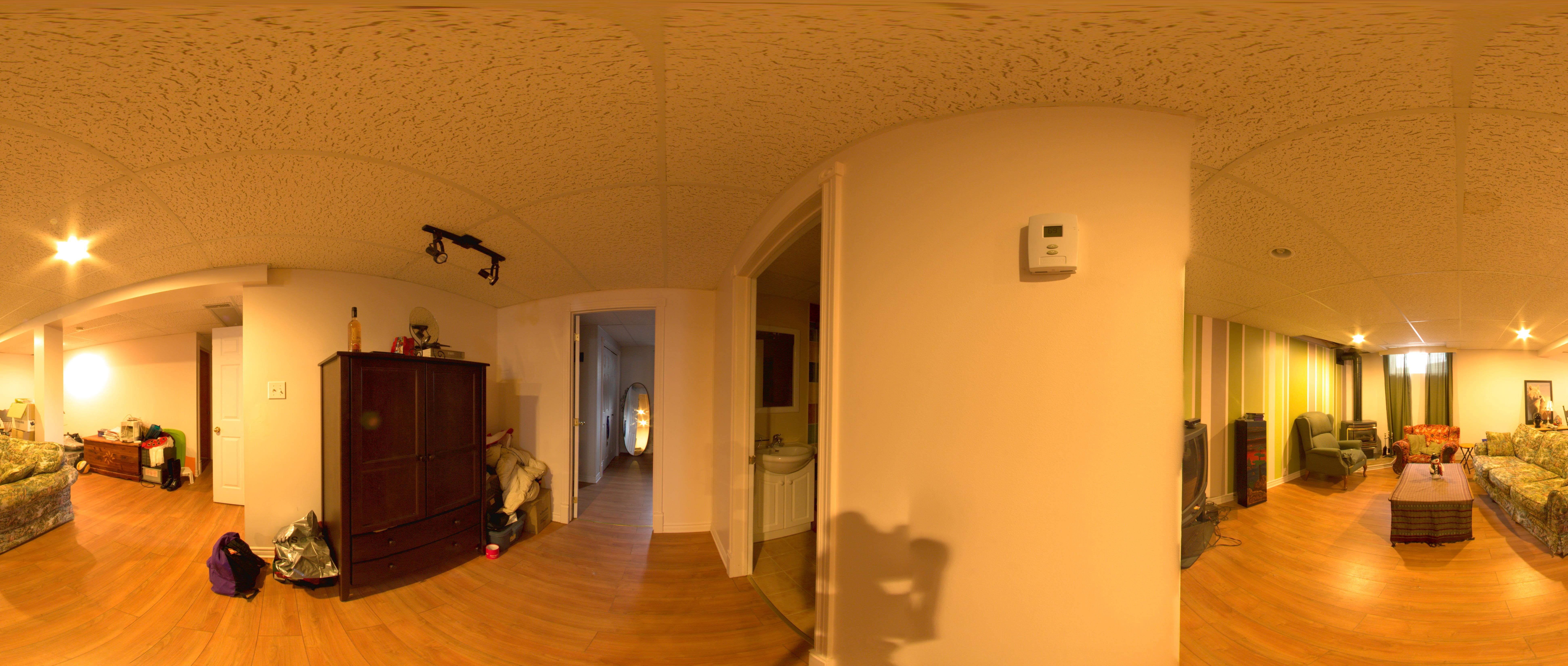}

            \includegraphics[width=\columnwidth]{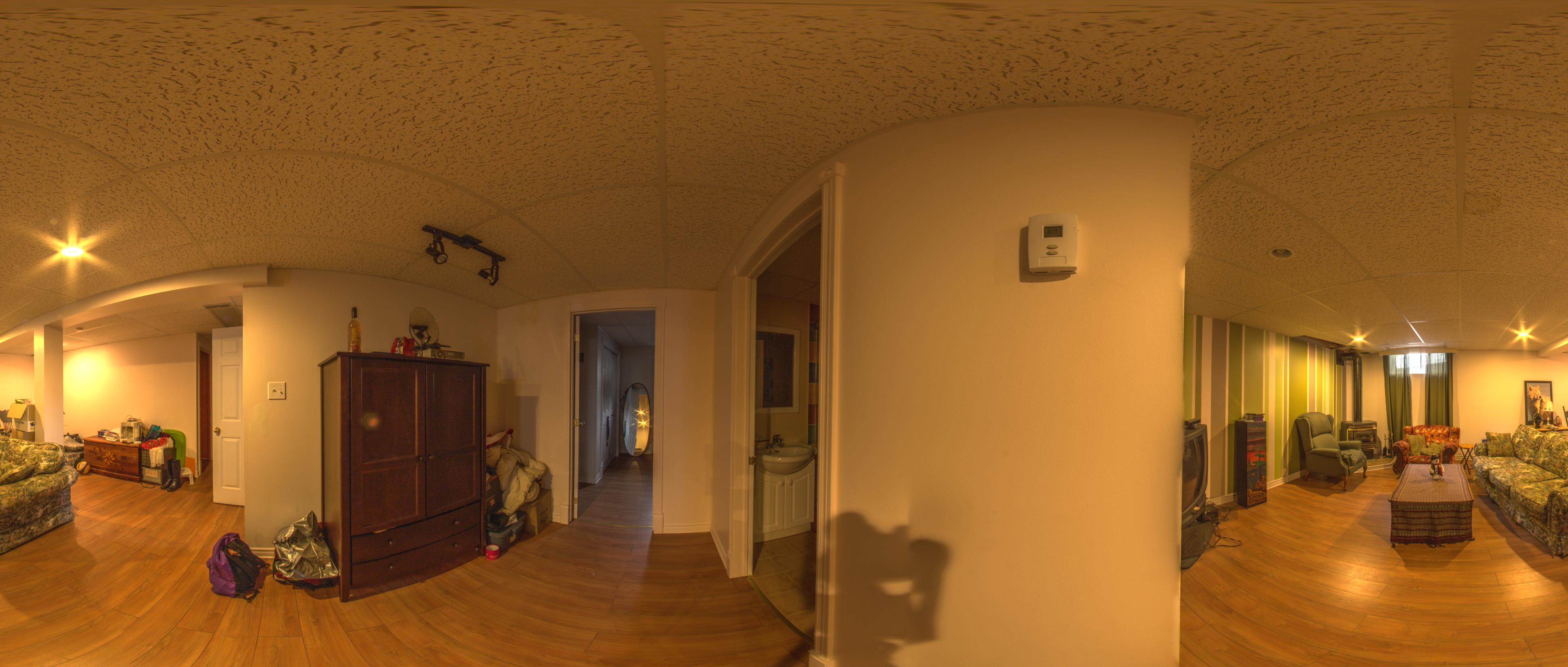}
            
\caption{top: Mantiuk TMO \cite{mantiuk2008display}, bottom: Paris TMO \cite{paris2015local}}
\label{9C4A3187-9c8f0e2b1b-3}
\end{minipage}}
\end{figure*}

\begin{figure*}[t]
{\begin{minipage}{\textwidth}
\centering
            \includegraphics[width=\columnwidth]{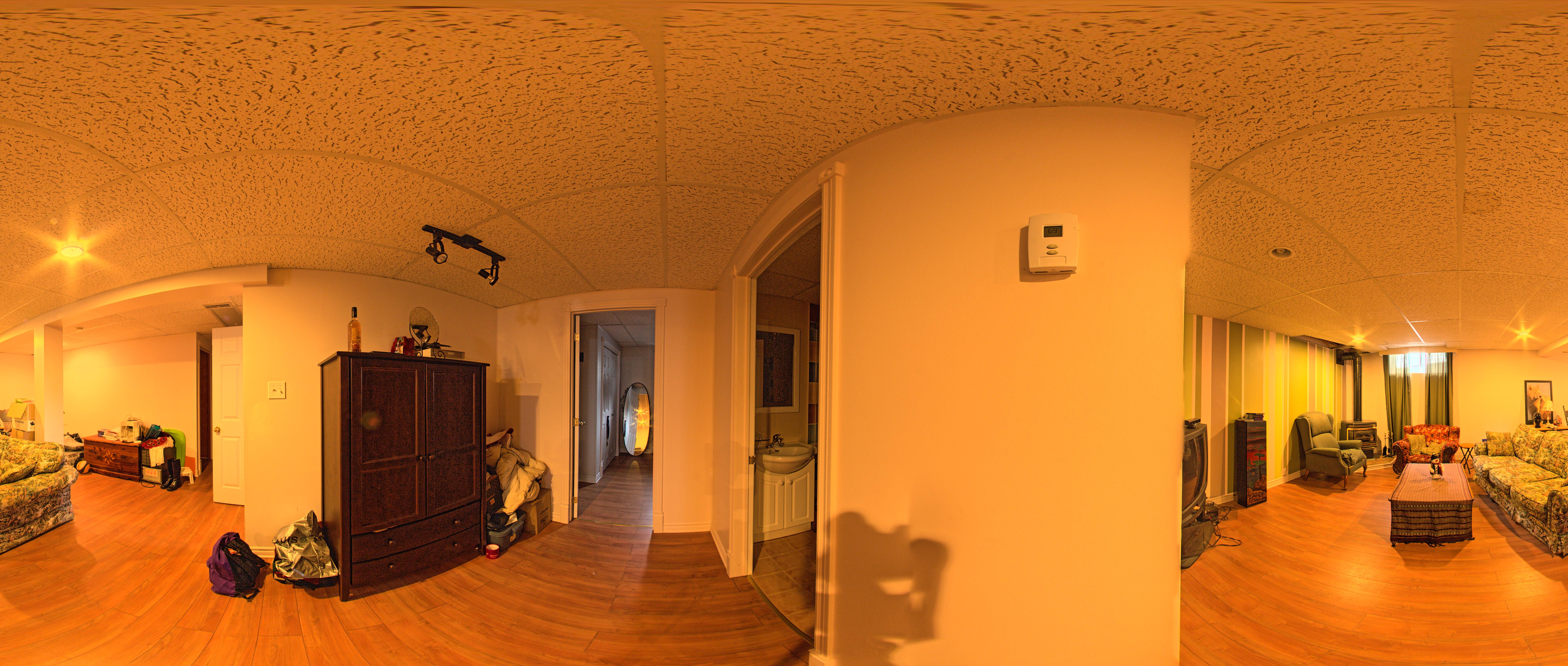}

            \includegraphics[width=\columnwidth]{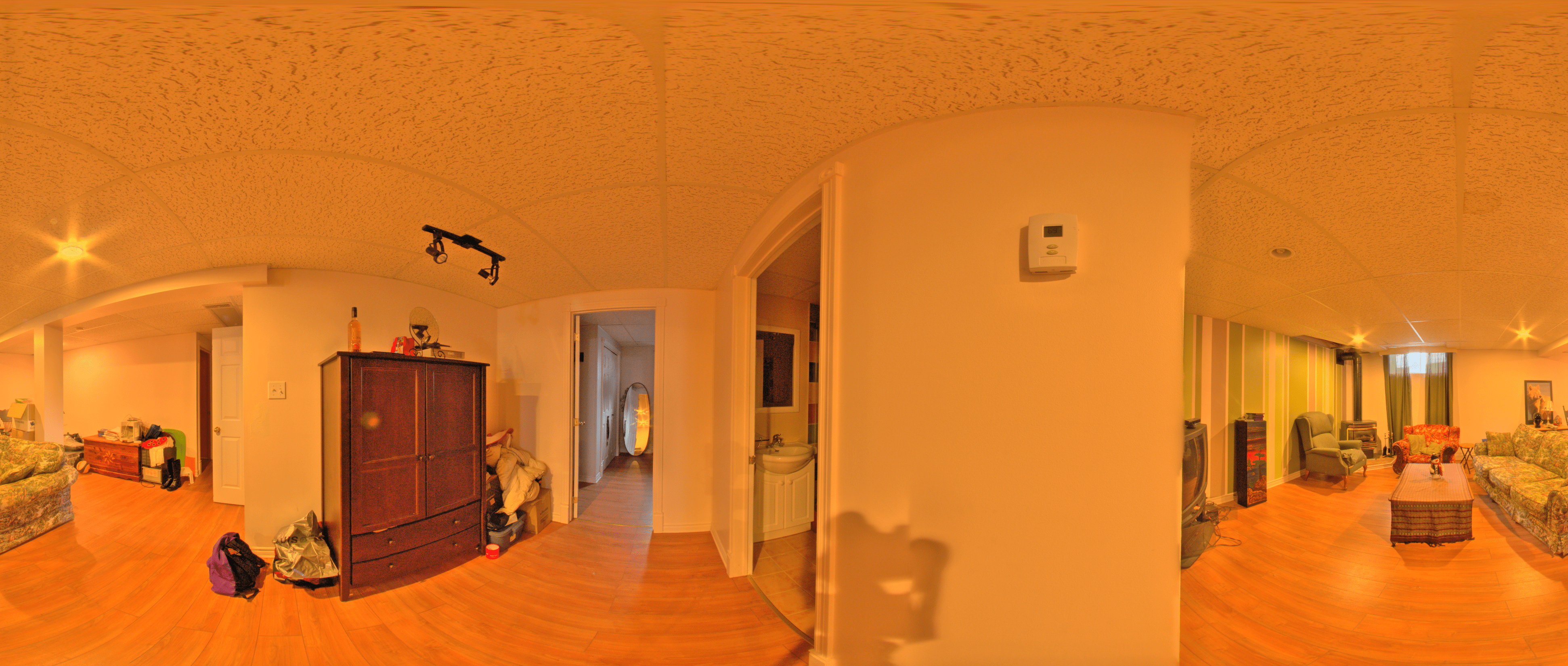}
            
\caption{top: Reference, bottom: Proposed TMO}
\label{9C4A3187-9c8f0e2b1b-4}
\end{minipage}}
\end{figure*}

\begin{figure*}
    \centering
    \includegraphics[scale = 0.25]{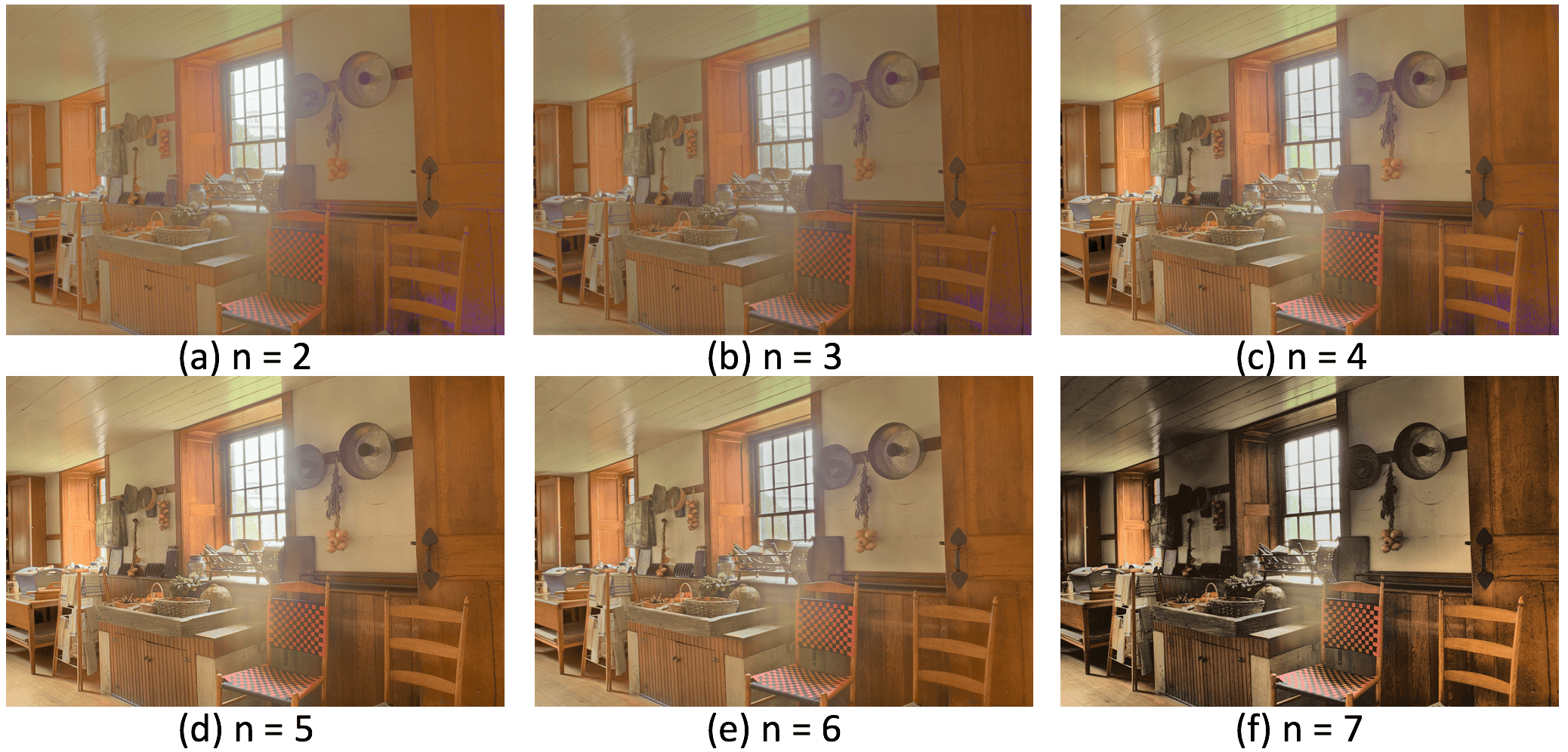}
    \captionof{figure}{Visual comparison of the resulting images in different frequency bands. (a), (b), (c), (d), (e) and (f) are the images with the frequency band $n=2$, $3$, $4$, $5$, $6$ and $7$, respectively.}
    \label{fig:fairchild_lv_compare_1}
\end{figure*}

\begin{figure*}
    \centering
    \includegraphics[scale = 0.24]{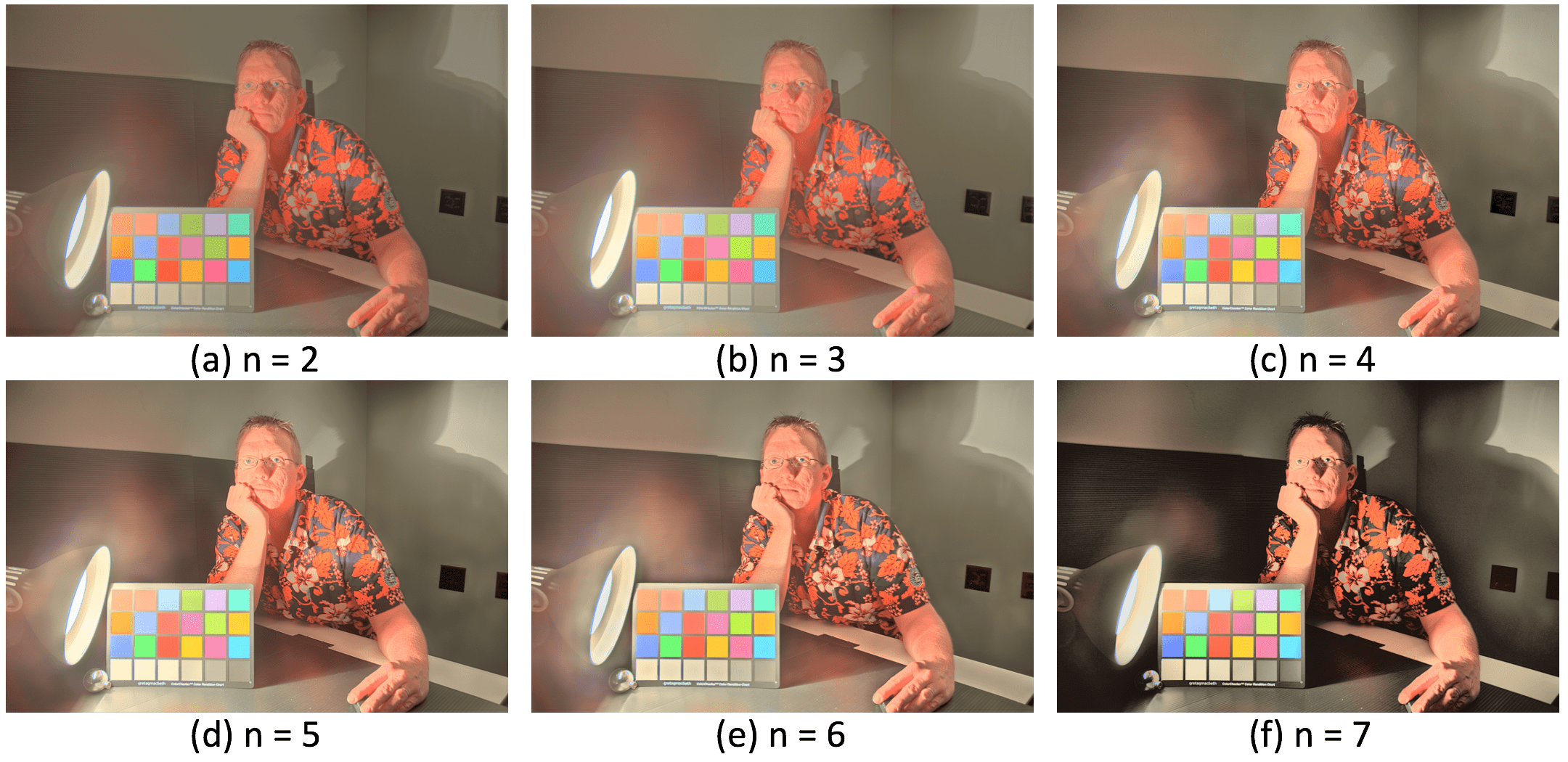}
    \captionof{figure}{Visual comparison of the resulting images in different frequency bands. (a), (b), (c), (d), (e) and (f) are the images with the frequency band $n=2$, $3$, $4$, $5$, $6$ and $7$, respectively.}
    \label{fig:fairchild_lv_compare_2}
\end{figure*}

\begin{figure*}
    \centering
    \includegraphics[scale = 0.24]{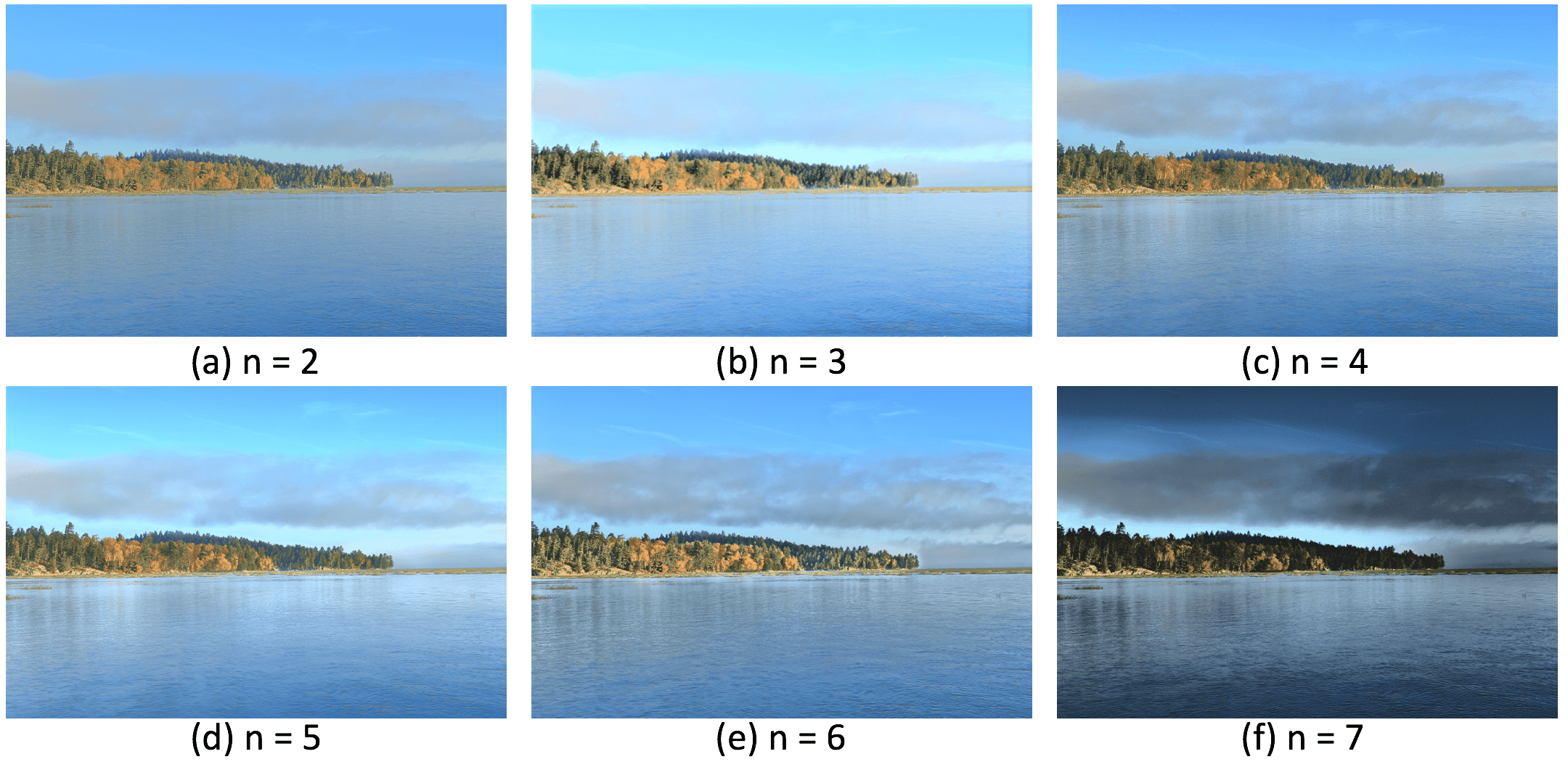}
    \captionof{figure}{Visual comparison of the resulting images in different frequency bands. (a), (b), (c), (d), (e) and (f) are the images with the frequency band $n=2$, $3$, $4$, $5$, $6$ and $7$, respectively.}
    \label{fig:fairchild_lv_compare_3}
\end{figure*}

\begin{figure*}
    \centering
    \includegraphics[scale = 0.22]{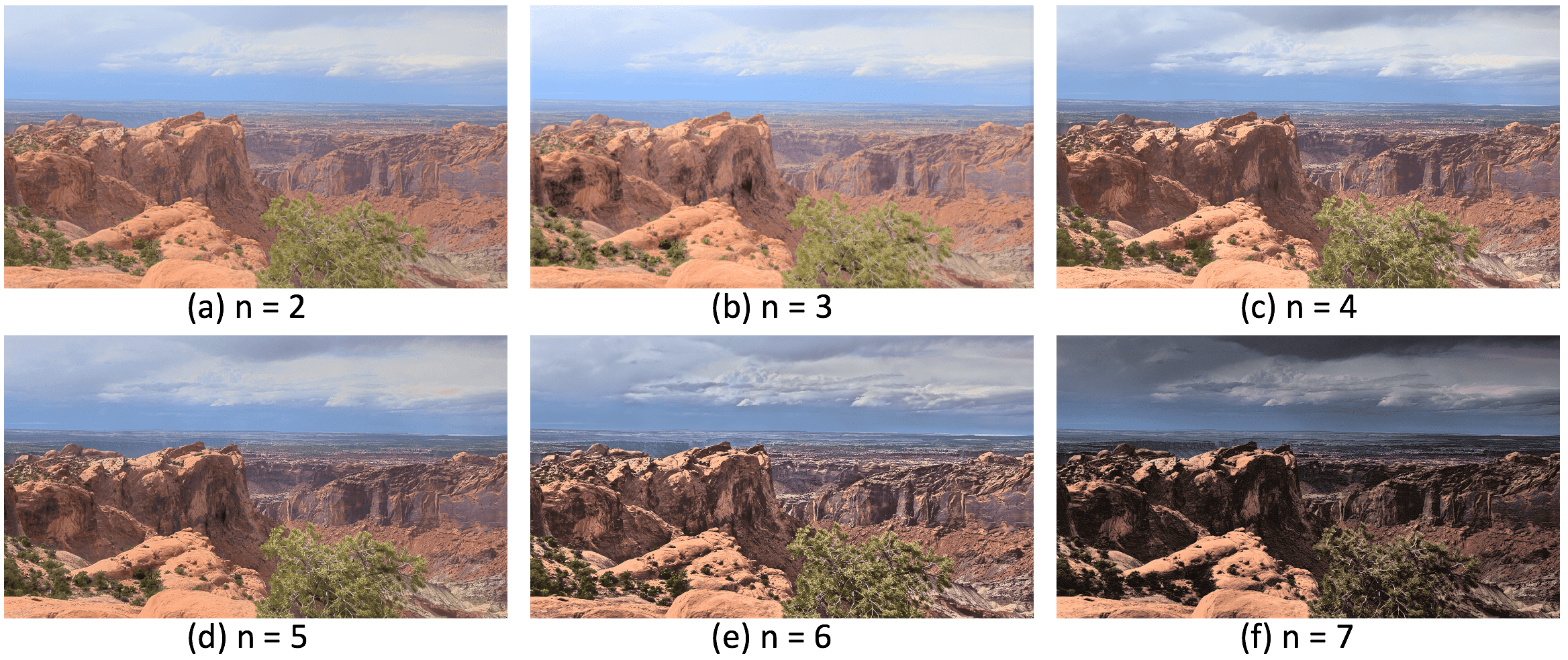}
    \captionof{figure}{Visual comparison of the resulting images in different frequency bands. (a), (b), (c), (d), (e) and (f) are the images with the frequency band $n=2$, $3$, $4$, $5$, $6$ and $7$, respectively.}
    \label{fig:fairchild_lv_compare_4}
\end{figure*}

Figs. \ref{fig:Laval_9C4A0006_5133111e97} - \ref{fig:Laval_9C4A3335-edf32a8ffe} show additional results of qualitative comparison with the state-of-the-art algorithms \cite{mantiuk2008display, paris2015local, ferradans2011analysis, mai2011optimizing, gu2013local, photomatrix}. The images were randomly chosen from the test set of Laval dataset \cite{gardner2017learning}.  The dataset contains uniformly panorama indoor WDR images in various scenes with the aspect ratio of around 1.0 to 2.0.  To better demonstrate our model's performance in preserving local details, we crop the full size images to half size and keep the overexposed regions. Our model is able to recover more details in these regions when compared with other state-of-the-art methods.  Figs. \ref{9C4A3187-9c8f0e2b1b-1} - \ref{9C4A3187-9c8f0e2b1b-4} show full size output of different methods.  It also yields visually comparable or better WDR-LDR images when compared with other methods.  Figs. \ref{fig:fairchild_lv_compare_1} - \ref{fig:fairchild_lv_compare_4} show visual comparison with different choices of the number of the frequency band $n$ in the Fairchild dataset \cite{fairchild2007hdr}.  Although our neural network was trained with an indoor dataset which the dynamic range inevitably much smaller than the outdoors, our method still be able to yield pleasing WDR-LDR images.

From these exhaustive predicted outputs, we show the proposed method generates output not only effectively compress global dynamic range like other compared methods, but also preserves and enhances the details of saturated region better.


\end{document}